\documentclass[%
  aps,
  superscriptaddress,
  longbibliography,
  12pt,
  onecolumn,
  a4paper,
  showpacs,
  showkeys,
  amsfonts, amssymb, amsmath]{revtex4-2}

\usepackage[utf8]{inputenc}
\usepackage[caption=false]{subfig}
\usepackage{hyperref} 

\hypersetup{pdfpagemode={UseOutlines},
bookmarksopen=true,
bookmarksopenlevel=0,
hypertexnames=false,
colorlinks=true, 
citecolor=blue, 
linkcolor=red, 
urlcolor=black, 
pdfstartview={FitV},
unicode,
breaklinks=true,
}
\usepackage{graphicx,amssymb,amsmath}
\usepackage{grffile}
\usepackage{color}
\usepackage{array}
\usepackage{hhline}

\usepackage{ulem}

\usepackage{siunitx}
\usepackage{listings}
\newlength{\figwidth}
\newcommand{\R}{\mathcal{R}}

\newcommand{\epsK}{{\varepsilon_{\!\scriptscriptstyle K}}}
\newcommand{\epsKK}{{\varepsilon_{\!\scriptscriptstyle K 2}}}
\newcommand{\epsKKKK}{{\varepsilon_{\!\scriptscriptstyle K 4}}}
\newcommand{\epsA}{{\varepsilon_{\!\scriptscriptstyle A}}}

\newcommand{\xx}{\boldsymbol{x}}
\newcommand{\kk}{\boldsymbol{k}}

\newcommand{\eeh}{\boldsymbol{e}_h}
\newcommand{\eetheta}{\boldsymbol{e}_\theta}
\newcommand{\eez}{\boldsymbol{e}_z}

\newcommand{\vv}{\boldsymbol{v}}
\newcommand{\ff}{\boldsymbol{f}}
\newcommand{\bomega}{\boldsymbol{\omega}}
\newcommand{\bnabla}{\boldsymbol{\nabla}}

\newcommand{\p}{\partial}
\newcommand{\mean}[1]{\langle #1 \rangle}
\newcommand{\CKA}{C_{K\rightarrow A}}

\newcommand{\diff}{\text{d}}
\newcommand{\bv}{Brunt-V\"ais\"al\"a}
\newcommand{\kmax}{k_{\max}}

\begin{document}

\title{Numerical study of experimentally inspired stratified turbulence forced by waves}

\author{Jason Reneuve}
\author{Clément Savaro}
\author{Géraldine Davis}
\author{Costanza Rodda}
\author{Nicolas Mordant}
\author{Pierre Augier}

\email[]{pierre.augier@univ-grenoble-alpes.fr}
\affiliation{Laboratoire des Ecoulements G\'eophysiques et Industriels, Universit\'e
Grenoble Alpes, CNRS, Grenoble-INP, F-38000 Grenoble, France}

\begin{abstract}

Stratified flows forced by internal waves similar to those obtained in the Coriolis
platform (LEGI, Grenoble, France) \cite{Savaro2020} are studied by pseudospectral
triply-periodic simulations. The experimental forcing mechanism consisting in large
oscillating vertical panels is mimicked by a penalization method. The analysis of
temporal and spatiotemporal spectra reveals that the flow for the strongest forcing in
the experiments is composed of two superposed large and quasi-steady horizontal
vortices, of internal waves in box modes and of much weaker waves outside the modes.
Spatial spectra and spectral energy budget confirm that the flow is in an intermediate
regime for very small horizontal Froude number $F_h$ and buoyancy Reynolds number $\R$
close to unity. Since the forcing frequency $\omega_f$ is just slightly smaller than
the Brunt-V\"ais\"al\"a frequency $N$, there are energy transfers towards slower waves
and large vortices, which correspond to an upscale energy flux over the horizontal.

Two other experimentally feasible sets of parameters are investigated. A larger
amplitude forcing shows that it would indeed be possible to produce in huge apparatus
like the Coriolis platform stratified turbulence forced by waves for small $F_h$ and
buoyancy Reynolds number $\R$ of order 10. Forcing slower waves for $\omega_f = 0.40 N$
leaves space between $\omega_f$ and $N$ for ``down-time-scale" transfers through weakly
nonlinear interactions with temporal spectra consistent with $\omega^{-2}$ slope.
However, for this set of parameters, the large scales of the flow are strongly
dissipative and there is no downscale energy cascade.

\end{abstract}


\maketitle


\section{Introduction}

Statistical results obtained from oceanic measurements are usually interpreted as the
signature of an internal wave field. Gravito-inertial waves can propagate into the
oceans because of the stable density stratification and the Earth rotation. These two
effects are characterized by two frequencies, the Brunt-Väisälä frequency $N$ and the
Coriolis frequency $f$, respectively. In the oceans, $N$ is usually much larger than
$f$. The velocity and temperature spectra measured at different times and locations
were observed to present some similarities. Temporal spectra scale as $\omega^{-2}$
between the Coriolis frequency $f$ and the Brunt-Väisälä frequency $N$. Vertical
spectra tend to scale as ${k_z}^{-2}$ at large scales and as $N^2{k_z}^{-3}$
(``saturated spectra") at intermediate scales. A turbulent $k^{-5/3}$ slope is observed
only at small scales along the vertical (dropped spectra) and at much larger scales
along the horizontal (towed spectra). The ${k_z}^{-5/3}$ spectrum at small scales
corresponds to weakly stratified isotropic turbulence. In contrast, the ${k_h}^{-5/3}$
spectrum at much larger scales together with the steeper vertical spectra cannot be due
to isotropic turbulence and another interpretation should be found.

Garrett \& Munk (GM) \cite{Garrett1979} showed that a simple model of a continuous
superposition of internal waves is consistent with different types of measurements.
Even though the GM model is fully empirical and not based on a physical understanding
of the underlying dynamics, it is a remarkable result that the oceanic spectra can be
modeled with only internal waves, i.e. that the different spectra are consistent with
the dispersion and polarization relations. These spectra are often called ``wave
spectra" but they are actually the footprint of the full dynamics, which can also
involve non-wavy flows.

Indeed, flows influenced by stable stratification and system rotation can also contain
a non-wavy ``balanced" part \cite{mcintyre2003balanced}. For example, in the case
without rotation, the equation for the vertical vorticity $\omega_z$ does not contain
any linear terms, which implies that the so-called "horizontal vortices" (associated
with $\omega_z$ and horizontal velocity) are not linearly coupled with the buoyancy. In
the large majority of recent numerical simulations of stratified turbulence, the
associated "vortical" (more precisely toroidal) energy is not small and horizontal
vortices play a key role. Moreover, wave-vortex interactions appear to be more
efficient than wave-wave interactions to drive a forward cascade
\cite{waite2006stratified}.

There had been a lot of debate about the dynamical regime producing the GM spectra and
different physical explanations have been proposed (for reviews, see
\cite{waite2006stratified, CaillolZeitlin2000}). The success of the GM model seems to
indicate that the oceanic spectra are due to a kind of internal wave turbulence.
However, such spectra have not been reproduced with internal waves, neither with
laboratory experiments nor with numerical simulations. The large-scale $k_h^{-5/3}$
spectrum tends to indicate that there could be an anisotropic turbulent cascade with an
energy flux through the horizontal scales. Recently, the theory of Weak Wave Turbulence
(WWT), which assumes that the flow consists only of weakly interacting waves, has been
used to derive solutions corresponding to the standard GM spectra and to observed
variabilities around this historical model \cite{CaillolZeitlin2000, polzin2011toward}.


An alternative dynamical explanation of the oceanic spectra has been proposed by
Lindborg, Brethouwer and Riley \cite{lindborg2007stratified, riley2008stratified}. They
show that many oceanic measurements could actually be compatible with ``strongly
stratified turbulence" \cite{lindborg2006energy, deBruynKops2019effects}. This regime
is strongly nonlinear and involves, similarly to isotropic turbulence, both toroidal
(associated with $\omega_z$) and poloidal, horizontally divergent, modes. In the
inertial range, the energy is approximately equipartitioned between these two modes. We
will follow \cite{Portwood2016robust} and call this particular regime ``LAST" (for
Layered Anisotropic Stratified Turbulence) to avoid confusion with other somehow
turbulent regimes in stratified fluids. The LAST regime is associated with a downscale
energy cascade and is obtained only when the large scales are simultaneously strongly
stratified (small horizontal Froude number $F_h$) and weakly influenced by viscosity
(large buoyancy Reynolds number $\R = Re {F_h}^2$)
\cite{brethouwer_billant_lindborg_chomaz_2007}.

Flows in the LAST regime have been obtained in high resolution idealized simulations.
In contrast, we are not aware of studies reporting forced dissipative flows composed of
a continuum of internal waves (without vortical modes) associated with spectra similar
to oceanic ones. Let us note that reproducing this regime might require very large
experimental apparatus or very large simulations that were until recently unfeasible.
Moreover, most numerical simulations of stratified turbulence were not designed to
produce a pure internal wave field without vortices, as described in the GM models. We
can mention few studies focussing on this subject.

\begin{itemize}

\item Waite \& Bartello \cite{waite2006stratified} carried out simulations of
stratified turbulence forced in waves with $k_z \sim k_h$ (fast waves for which
$\omega \simeq 0.7N$). They concluded that they have been unsuccessful at
reproducing the observed spectra and that the effects of adding vortices can be
dramatic.

\item Lindborg \& Brethouwer \cite{lindborg2006energy} completed this study by
forcing in waves with $k_z \gg k_h$ (slow waves for which $\omega \ll N$) such
that the vertical Froude number $F_v = UN/l_v$ is of order unity. For some
parameters, they observed a clear downscale energy cascade. Their spatial and
spatiotemporal spectra show that the dynamics of the inertial range corresponds
to the LAST regime and that inertial waves dominate only for the very large
horizontal scales.

\item Le Reun {\it et al.} \cite{LeReun2018parametric} simulate some forced
dissipative flows made of internal gravity waves but the buoyancy Reynolds
number is quite small. Therefore, these waves should be dissipated at large
horizontal scales and there is no need for a downscale energy cascade bringing
energy at small horizontal scales.

\item Calpe Linares {\it et al.} \cite{linares2020numerical} studied
two-dimensional stratified turbulence forced by fast internal waves. They
verified that the conditions on $F_h$ and $\R$ also determined the regime for
2d stratified turbulence. For values corresponding to the oceanic dynamics
($F_h \ll 1$ and $\R \gg 1$), the dynamics corresponds to a strongly nonlinear
regime which cannot be interpreted in the framework of the WWT theory.

\end{itemize}

Recent experiments \cite{Savaro2020, Rodda2022experimental} have managed to generate
strongly stratified turbulent flows by forcing large scale internal waves into a large
scale tank (the Coriolis platform in Grenoble) filled with stratified salty water. A
continuum of waves was observed, and key elements of WWT phenomenology were identified.
However, some discrepancies with the phenomenology were noticed, such as a flat
frequency spectra or finite size effects in the form of resonant modes. Frequency
spectra qualitatively consistent with GM spectra were observed at frequencies greater
than the forcing and extending at frequencies greater than $N$. This observation
suggests the occurence of strongly non linear wave turbulence
\cite[]{Rodda2022experimental}.

In this article, we present the results of numerical simulations that were designed in
direct inspiration of the experiments of \cite{Savaro2020}. In particular, we model
with a immersed boundary method the experimental forcing mechanism which uses
oscillating panels on the boundaries of the fluid domain to generate waves. This
forcing scheme is uncommon for numerical simulations, as it is local in space and
forces only waves, as opposed to most numerical forcing schemes used in stratified
turbulence. We focus is this study on sets of physical parameters that were or could be
obtained in real experiments in huge apparatus like the Coriolis platform. The paper is
not aimed at simulating the details of the experiment but only its major
characteristics.

This article is organized as follow. The numerical methods are presented in section 2.
Section 3 is dedicated to the analysis of simulations corresponding to a set of
parameters considered in \cite{Savaro2020}. We then extend in section 4 the experiments
of \cite{Savaro2020} by considering two other sets of parameters that could be obtained
in the Coriolis platform.


\section{Numerical setup}
\label{sec:num}

The numerical simulations presented in this article are performed using the
pseudospectral solver \lstinline{ns3d.strat} from the FluidSim Python package
\cite{fluiddyn,fluidfft,fluidsim}. The simulations and the analysis should be
reproducible with Fluidsim version 0.6.1
\footnote{\url{https://pypi.org/project/fluidsim/0.6.1/}}. Using this solver, we
integrate the three-dimensional Navier-Stokes equations under the Boussinesq
approximation with an added fourth-order hyperviscosity term:
\begin{align}
\p_t\vv + (\vv \cdot \bnabla)\vv = b\boldsymbol{e}_z - \frac{1}{\rho_0}\bnabla p +
\nu_2\nabla^2\vv + \nu_4\nabla^4\vv + \ff_h,\label{ns} \\
\p_t{b} + (\vv \cdot \bnabla)b = -N^2v_z + \kappa_2\nabla^2{b} +
\kappa_4\nabla^4{b},\label{buoy}
\end{align}
where $\vv$ is the velocity, $p$ the pressure, $b=-g\delta\rho/\rho_0$, with $\rho_0$
the mean density and $\delta\rho$ the departure from the stable linear density
stratification. For all simulations, the second-order viscosity is set to the value for
water in usual temperature and pressure conditions, $\nu_2=10^{-6}$\,m$^2$/s. In the
following, all physical quantities from simulation data are expressed in SI units.

The fourth-order viscosity $\nu_4$ is left as a free parameter and adapted to the
resolution of simulations in order to ensure that the energy brought at the smallest
simulated scales by the non linear fluxes are dissipated without accumulation.
Similarly, the equation of motion \eqref{buoy} for the buoyancy field $b$ presents both
second and fourth order diffusive terms, with corresponding diffusion coefficients
$\kappa_2$ and $\kappa_4$. We can then build two different Prandtl numbers
$\text{Pr}_i=\nu_i/\kappa_i$ for $i=2,4$. In all simulations, both those Prandtl
numbers are set to unity, such that $\kappa_2=\nu_2$ and $\kappa_4=\nu_4$. The use of
both normal and hyperviscosity is an important tool for the comparison with
experiments. It allows us to carry out simulations with a well defined physical
Reynolds number at a coarse resolution and to be able to quantify the difference with a
proper DNS. More specifically, we use the measure of the turbulent kinetic dissipations
$\epsKK$ and $\epsKKKK$ based on both viscosities, and the ratio $\epsKK/\epsK$ where
$\epsK=\epsKK+\epsKKKK$ is the total kinetic energy dissipation, as an indicator of how
close the simulations we perform are to proper DNS of the true Navier-Stokes equations
with only water viscosity. For a set of physical parameters, the needed hyperviscosity
decreases when the resolution is increased and the ratio $\epsKK/\epsK$ grows towards
unity.

\begin{figure}
\centering
\includegraphics[width=0.95\textwidth]{%
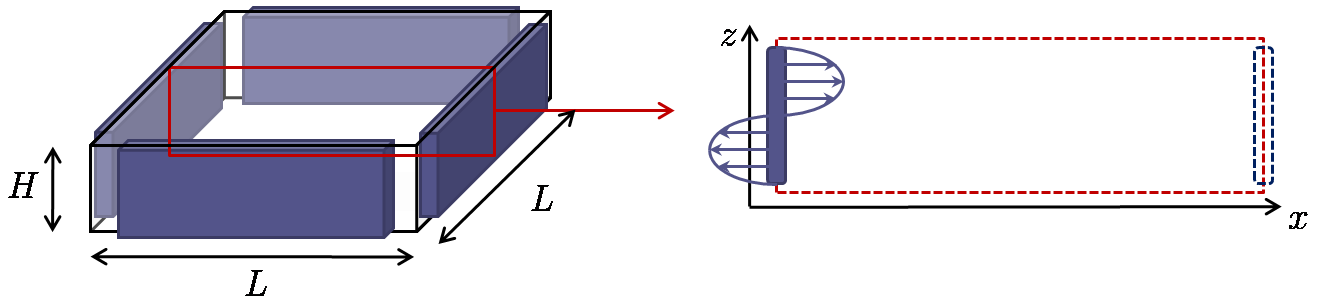}
\caption{Schematic representation of the $L^2$ volume penalization method, shown in the
whole simulation domain (left) and in a vertical cut (right). Wavemakers are modeled by
parallepipedic penalization volumes (in dark blue) close to the vertical boundaries of
the simulation domain. In each of these volumes, we implement a virtual body force
designed to impose a prescribed velocity profile (blue arrows) on the normal component
of the fluid velocity field. The vertical periodicity of the simulation domain is
ensured by choosing a sinusoidal velocity profile with one wavelength over the height
$H$ of the box. Horizontal periodicity is ensured by having the penalization volumes
sit over the vertical boundaries of the domain, so that facing boundaries are imposed
the same velocity profile at all times. The mathematical details of the virtual body
force are given in equation \eqref{forcing}. \label{fig:penalization}}
\end{figure}

The geometry and forcing scheme used for the simulations are inspired by the
configuration of experiments that were performed in the Coriolis facility in LEGI,
Grenoble (France) \cite{Savaro2020}. In the experiments, a parallelepipedic domain of
size $6\times6\times1$\,m$^3$ is isolated inside the tank of the facility, using two
adjacent fixed walls and two oscillating walls acting as wavemakers. The wavemakers are
set to oscillate around their mid-height horizontal axis, with frequency $\omega_f$ and
maximum horizontal excursion $a$ being free parameters of this forcing mechanism. In
the simulations we present here, the wavemakers are modeled using a $L^2$ volume
penalization method \cite{Angot1999, engels2016flusi}, which is schematized in figure
\ref{fig:penalization}.

This penalization methods works by introducing a forcing term of the following shape in
the momentum equation \eqref{ns}:
\begin{equation}
\ff_h = \frac{1}{\tau} \Theta(\xx) (\vv_h^*(\xx,t) - \vv_h), \label{forcing}
\end{equation}
where $\vv_h^*(\xx,t)$ is the target velocity profile imposed by the wavemakers,
$\Theta(\xx)$ is an activation function ensuring that the forcing acts only locally in
place of the wavemakers, and $\tau$ is a typical timescale at which the forcing acts.
The forcing term described by equation \eqref{forcing} ensures that in the region where
$\Theta(\xx)=1$, the horizontal velocity field $\vv_h$ follows the target velocity
$\vv_h^*$ over timescales that are large compared to $\tau$. In order to give a
meaningful description of the dynamics imposed by wavemakers, we must then choose
$\tau$ to be small compared to the typical timescales of the flow that we simulate.
Typically, we take $\tau$ to be small compared to the \bv{} period,
i.e.\ $\tau=T_N/20$, with $T_N = 2\pi/N$. The activation function $\Theta(\xx)$ is
zero outside the wavemakers, and unity inside a region representing the oscillating
panels. For the sake of simplicity, we choose this activation region of a single panel
to be a paralleleliped of the same height as the simulation domain, of length slightly
smaller than the horizontal size of the box in order to avoid interpenetration of the
panels in the corners, and with a thickness equal to the forcing amplitude $2a$. In
order to ensure the horizontal periodicity of the forcing term $\ff$ that is required
by the pseudospectral solver, the activation region of each panel is centered on the
corresponding vertical boundary of the box, meaning that contrary to the experimental
setup, the panels are not facing motionless walls but rather a virtual copy of
themselves. Additionally to the periodicity constraint, the use of a pseudospectral
solver requires that the activation function $\Theta$ is a smooth function of space in
order to avoid Gibbs oscillations. To ensure this, we take $\Theta$ to grow from zero
to unity as a hyperbolic tangent when crossing the activation region boundary, over a
small scale set arbitrarily to four grid points. Inside the activation region of each
wavemaker, we apply the forcing term only to the fluid velocity component that is
orthogonal to the corresponding vertical boundary, meaning that we model the panels
with a free-slip boundary condition, and that we neglect their vertical motion. This
last approximation is justified by the fact that the experimental values for $a$ are of
the order of a few centimeters, meaning that the angular excursion of a panel is very
small. We thus only prescribe the orthogonal component of the target velocity field to
be $v^*_\bot=a\sin(2\pi z/L_z)s(t)$, with $L_z$ the height of the simulation box, and
$s(t)$ a prescribed temporal forcing signal. The sinusoidal shape ensures that the
prescribed velocity field is periodic vertically. Let us note that as a consequence,
here the panels force modes with one vertical wavelength, as opposed to the
experimental setup where the wavemakers force modes with half a vertical wavelength. In
order to have modes with the same wavelengths as in the experiments, and to keep the
same aspect ratio $L_x/L_z = L_y/L_z=6$, we then double the size of the numerical
domain, effectively performing simulations in a box of size $12\times12\times2$\,m$^3$.
Finally, the time signal $s(t)$ of the forcing oscillates at a frequency $\omega_f =
FN$, which is randomly modulated in time in a narrow band of width 10\% as was done in
the experiment. At the beginning of every simulation, two realizations of the forcing
time signal are generated, one for each pair of facing wavemakers. By doing so, we make
sure that the phases of adjacent wavemakers stay uncorrelated in time. This forcing
scheme corresponds to the Fluidsim parameter \lstinline{params.forcing.type="watu_coriolis"}
and is implemented in the module
\href{https://foss.heptapod.net/fluiddyn/fluidsim/-/blob/branch/default/fluidsim/solvers/ns3d/forcing/watu.py}{fluidsim/solvers/ns3d/forcing/watu.py}.

For all simulations we take $N=0.6$\,rad/s and $\nu_2=10^{-6}$\,m$^2$/s. The remaining
free parameters are the resolution $n_x\times n_y\times n_z$, the reduced forcing
frequency $F\equiv\omega_f/N$ and the amplitude $a$. The hyperviscosity $\nu_4$ is
adapted to the other parameters so that the dissipative scales are well-resolved. The
simulations were performed on a local cluster at LEGI for resolutions up to
$480\times480\times80$, and on the national CINES cluster Occigen for resolutions up to
$2304\times2304\times384$. Parameters and dimensionless numbers for each simulations
are summarized in table~\ref{table_simul}. The turbulent nondimensional numbers are
computed from the statistically stationary flows as $F_h = \epsK / ({U_h}^2 N)$ and
$\R_2 = \epsK / (\nu_2 N^2)$, where $\epsK$ is the mean kinetic energy dissipation and
$U_h$ the rms horizontal velocity. The results presented in this article are obtained
from periods of the simulation when a steady state has been approximately reached.
Because the time scales of the flows studied here are very long, finding such
steady-state period can be very difficult and computationally costly. In order to reach
an approximately steady state in a reasonable time, we start all the simulations at a
reduced resolution $240\times240\times40$, and increase the resolution step by step
only when a sufficiently stationary state has been reached. When such a state is
observed, specific outputs are turned on and the simulation is ran further for 10 to 20
minutes of equation time in order to produce substantial data to analyze, before
increasing the resolution again if needed.

\begin{figure}
\centerline{
\includegraphics[width=0.48\textwidth]{%
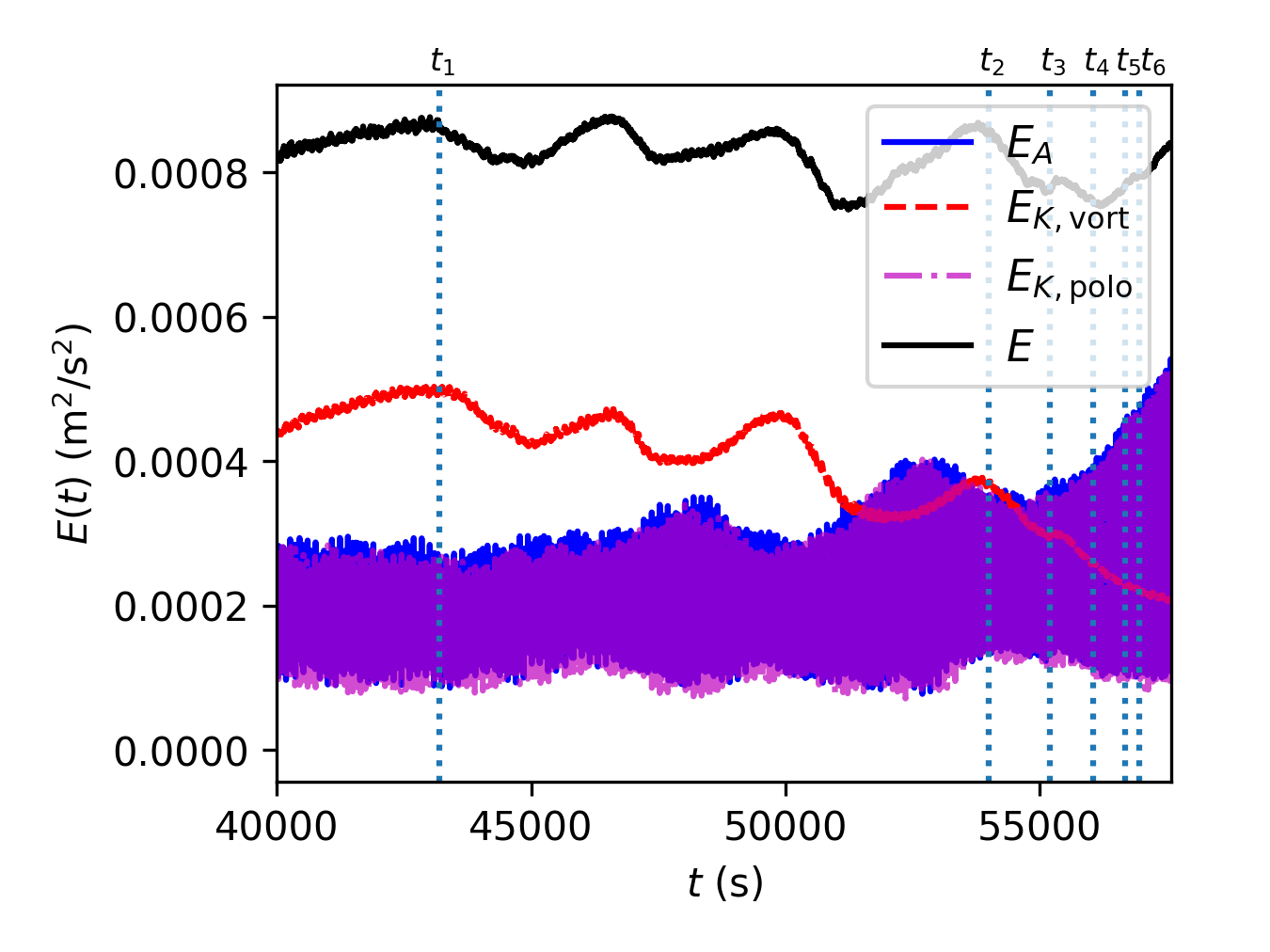}
\includegraphics[width=0.48\textwidth]{%
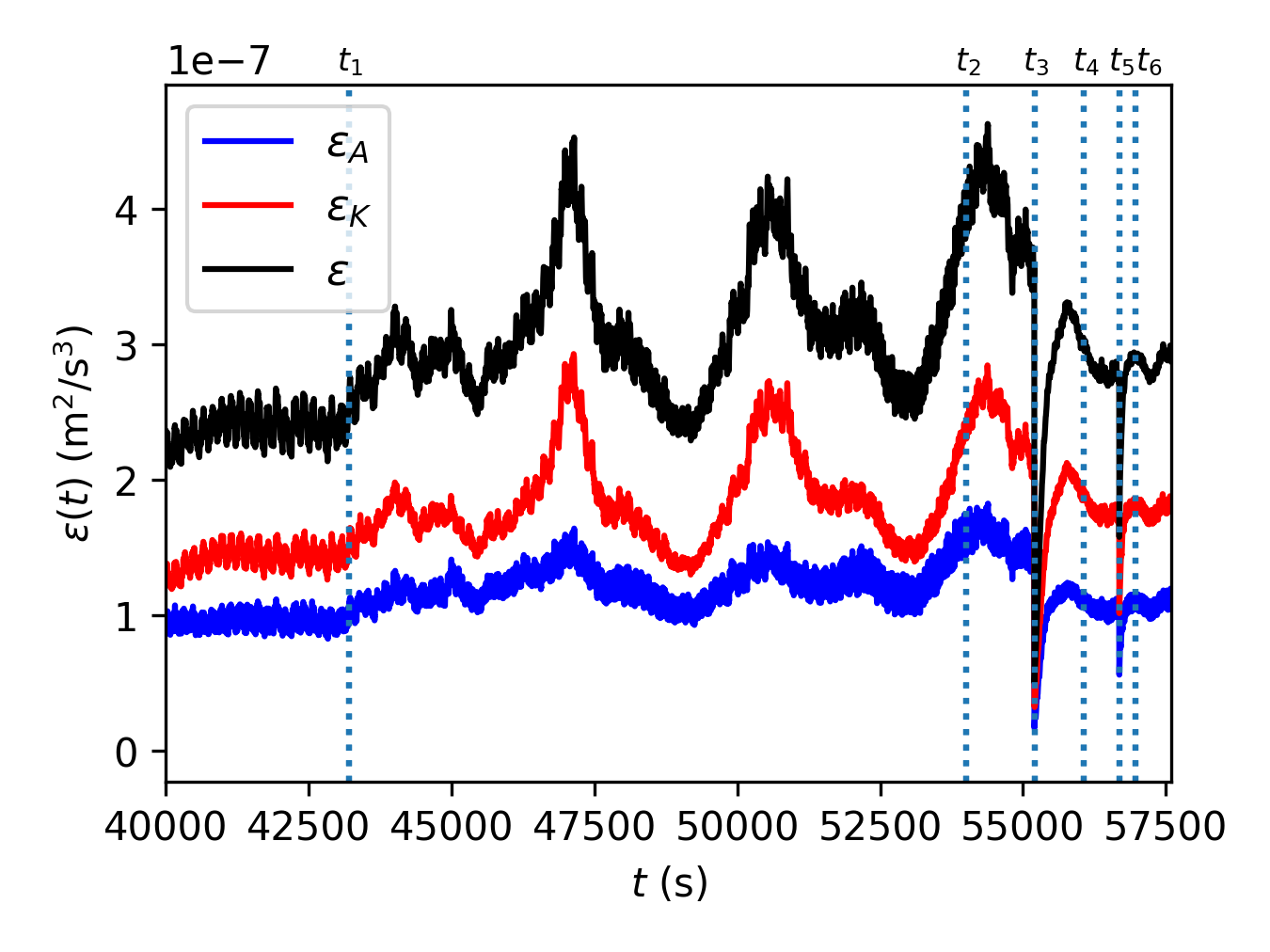}
}
\caption{Time evolution of the energy components (left) and dissipation (right) for
parameters $F =0.73$ and $a=5$\,cm, since the beginning of the initial run. Vertical
dotted lines signal the initial time $t_i$ of each successive run, for $i=0,...,4$. The
simulations first runs at resolution $240\times240\times40$ between $t_0=0$\,s and
$t_1=43200$\,s. At $t=t_1$ the resolution is increased to $480\times480\times80$ and
the simulation runs until $t_2=54000$\,s. At $t=t_2$, temporal and spatiotemporal
outputs are turned on and the simulation runs until $t_3=55200$\,s for analysis. At
$t=t_3$, the resolution is increased to $1152\times1152\times192$ and the simulation
runs with outputs off until $t_4=56058$\,s. At $t=t_4$ outputs are turned back on and
the simulation runs until $t_5=56684$\,s for analysis. At $t=t_5$, outputs are turned
off, the resolution is increased a last time to $2304\times2304\times384$ and the
simulation runs until $t_6=56972$\,s. Finally from $t=t_6$ to $t_\text{end}=57598$\,s
the simulation runs with outputs on for analysis. \label{fig:spatial-mean-history}}
\end{figure}

Figure \ref{fig:spatial-mean-history} summarizes the successive runs and resolution
changes since the beginning of the simulation for $F = \omega_f / N =0.73$ and
$a=5$\,cm. We see that most of the equation time is taken by the initial transient
state from a static initial condition, which justifies the use of smaller resolutions
for the simulation of transient states, as it is faster and less computationally
expensive. Indeed, the first transient between $t_0$ and $t_1$ takes up most of the
equation time but is computed in a few hours on a local cluster at resolution
$240\times240\times40$, whereas the last period of data analysis from $t_4$ to
$t_\text{end}$ represent a small amount of equation time but requires days of
computation on the national cluster Occigen.

\begin{table}
\centering
\begin{tabular}{cccccccccc}
\hline \hline
$F$ & $a$ & $n_x\times n_y\times n_z$ & $\nu_4$ & $F_h$ & $\R_2$ & $\Gamma$ & $\eta \kmax$ & $\epsKK/\epsK$\\
    & cm  &                           & m$^4$/s &       &        &          &              &               \\
\hline
0.73 & 5 & 480$\times$480$\times$80 & 3.86e-08 & 7.87e-04 & 0.68 & 0.64 & 0.12 & 0.13 \\
0.73 & 5 & 1152$\times$1152$\times$192 & 6.39e-10 & 6.70e-04 & 0.49 & 0.59 & 0.44 & 0.47 \\
0.73 & 5 & 2304$\times$2304$\times$384 & 1.2e-11 & 6.52e-04 & 0.49 & 0.59 & 0.88 & 0.87 \\
0.73 & 10 & 480$\times$480$\times$80 & 1.11e-07 & 1.89e-03 & 5.06 & 0.74 & 0.07 & 0.05 \\
0.73 & 10 & 1152$\times$1152$\times$192 & 1.28e-09 & 2.20e-03 & 6.26 & 0.60 & 0.23 & 0.17 \\
0.73 & 10 & 2304$\times$2304$\times$384 & 1.85e-10 & 2.33e-03 & 6.60 & 0.57 & 0.46 & 0.29 \\
0.45 & 10 & 480$\times$480$\times$80 & 2.18e-08 & 5.89e-04 & 0.48 & 0.91 & 0.13 & 0.19 \\
0.40 & 10 & 480$\times$480$\times$80 & 2.18e-08 & 4.10e-04 & 0.22 & 0.54 & 0.16 & 0.23 \\
0.40 & 10 & 1152$\times$1152$\times$192 & 6.18e-10 & 5.81e-04 & 0.31 & 0.52 & 0.49 & 0.48 \\
0.40 & 10 & 2304$\times$2304$\times$384 & 2.65e-11 & 5.85e-04 & 0.33 & 0.54 & 0.96 & 0.76 \\
\hline \hline
\end{tabular}
\caption{Summary of the simulations. For all simulations, $L_z=\lambda_{fz} =
2$\,m, $L_x=L_y=$12\,m, $N = 0.6$\,rad/s and $\nu_2 = 10^{-6}$\,m$^2$/s are
fixed. The first four columns contain the simulations control parameters which
are the normalized forcing frequency $F=\omega_f/N$, the forcing amplitude $a$,
the resolution $n_x\times n_y\times n_z$ and the hyperviscosity $\nu_4$. The
last five columns contain important dimensionless numbers such as the
horizontal Froude number $F_h$, buoyancy Reynolds number $\R_2$,
the mixing coefficient $\Gamma=\epsA/\epsK$ (with $\epsA$ the available
potential energy dissipation rate) and the dissipation ratio $\epsKK/\epsK$.
All dimensionless numbers are computed from turbulent quantities and averaged
in time over a statistically stationary period.}
\label{table_simul}
\end{table}

\begin{figure}
\centering
\includegraphics[width=0.9\textwidth]{%
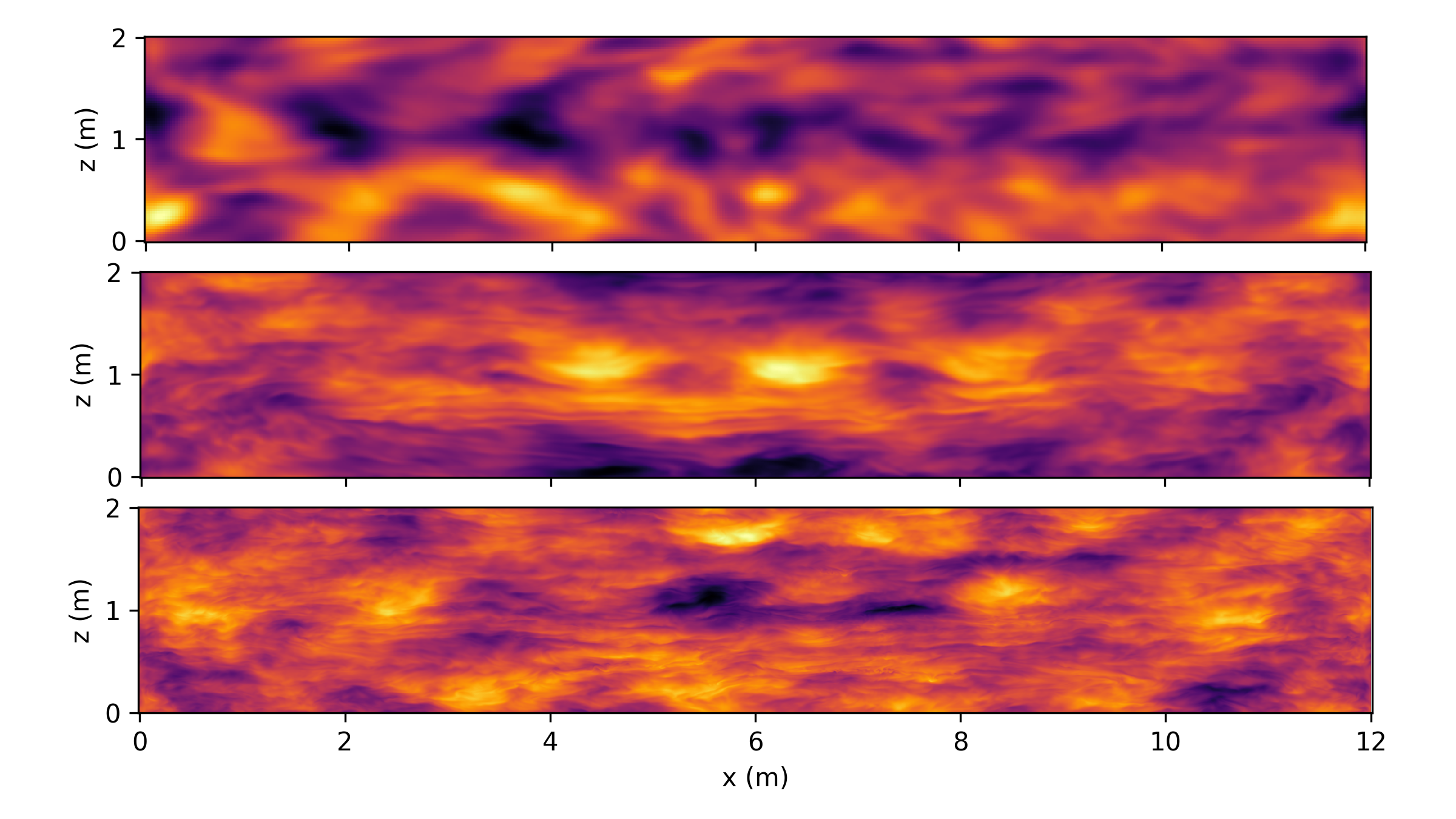}
\caption{Buoyancy field snapshots from three successive runs for $F=0.73$ and
$a=5$\,cm. Top: taken at $t_3=55200$\,s, resolution is $480\times480\times80$. Middle:
taken at $t_5=56684$\,s, resolution is $1152\times1152\times192$. Bottom: taken at
$t_\text{end}=57598$\,s, resolution is $2304\times2304\times384$. All snapshots are
vertical slices in the middle of the numerical domain at $x=6$\,m.
\label{fig:phys-fields}}
\end{figure}

In the following sections, results are presented for the resolutions
$480\times480\times80$, $1152\times1152\times192$ and $2304\times2304\times384$.
Snapshots of the buoyancy field are shown in figure~\ref{fig:phys-fields} for
resolutions $480\times480\times80$, $1152\times1152\times192$ and
$2304\times2304\times384$. The snapshots show clear anisotropy confirming that the flow
is strongly affected by gravity. We see that small scales appear only for the two
largest simulations, which is consistent with the values of $\eta \kmax$ given in the 3
first lines of table~\ref{table_simul}. In the following, the results of the
simulations with the coarser resolution are used only when more statistics are needed
for a better convergence.

Throughout this article, we make an extensive use of the poloidal/toroidal
decomposition \cite{RileyLelong2000, Cambon2001} also known as the Craya-Herring
decomposition \cite{Craya1958, Herring1974}. This decomposition consists in separating
the horizontal velocity field into a horizontally divergent part $\vv_{hd}$ and a part
$\vv_{hr}$ carrying the vertical vorticity:
\begin{align}
\vv = \vv_z + \vv_h = \vv_z + \vv_{hd} + \vv_{hr},
\label{helmholtz}
\end{align}
where $\bnabla\cdot\vv_h = \bnabla\cdot\vv_{hd}$ and
$\eez\cdot(\bnabla\times\vv_h)=\eez\cdot(\bnabla\times\vv_{hr})$. In Fourier space,
this decomposition becomes simple when expressed in the cylindrical basis
$(\eeh,\eetheta,\eez)$ around axis $z$:
\begin{align}
\hat{\vv} = \hat{v}_z\eez + \hat{v}_{hd}\eeh + \hat{v}_{hr}\eetheta.
\label{helmholtz_fourier}
\end{align}
This decomposition is thus conveniently performed by projecting the horizontal velocity
field in Fourier space on the radial and azimuthal unit vectors $\eeh$ and $\eetheta$,
respectively. Since the cylindrical basis is orthogonal, the kinetic energy can also be
decomposed in the same fashion:
\begin{align}
E_K = E_{Kz} + E_{Khd} + E_{Khr}.
\label{helmholtz_nrj}
\end{align}
In technical terms, the velocity components are regrouped into a poloidal field
$\vv_p=\vv_z+\vv_{hd}$, associated with poloidal energy $E_{Kp}=E_{Kz}+E_{Khd}$ and a
toroidal field $\vv_t=\vv_{hr}$ accounting for toroidal energy $E_{Khr}$. To help
nonspecialist readers, in the following we will not use the word toroidal and we will
prefer "vortical" (associated with vertical vorticity). This rearrangement of the
velocity components is meaningful in stratified turbulence. First, one can see by
considering the inviscid equation for the vertical vorticity
\begin{equation}
\partial_t \omega_z + \vv \cdot \bnabla \omega_z = \bomega \cdot \bnabla v_z
\end{equation}
that this part of the flow is not directly influenced by a linear term related to
buoyancy and has therefore no wavy behavior. Second, for weak nonlinearity levels, the
poloidal component can be identified as being mostly waves.

A typical aspect of triply periodic numerical simulations of stratified turbulence is
the uncontrolled growth of so-called shear modes, which are modes of the horizontal
velocity with no horizontal structures
\cite{smith2002generation,lindborg2006energy,augier2015stratified}. In Fourier space,
these modes correspond to $\hat{\vv}_{hr}(k_x=0,k_y=0,k_z)$. Although these modes can
play an important role in the overall flow dynamics \cite{Maffioli2020}, they are
absent from the experiments in \cite{Savaro2020} as half of the vertical walls of the
tank are immobile, thus imposing a hard boundary condition on the horizontal velocity
field and preventing any horizontally-averaged horizontal flow. In order to stay as
reasonably close as possible to the experiments, we decided to prevent shear modes
altogether by imposing $\hat{\vv}_{hr}(0,0,k_z)=0$ at all times \footnote{This is done
with the parameter \lstinline{params.oper.NO_SHEAR_MODES} of the ns2d.strat
and ns3d Fluidsim solvers, which extends the truncation used for dealiasing to the
shear modes.}.

Similarly to shear modes, fast waves oscillating at $\omega=N$ are absent from the
experiments. These waves oscillate vertically with no vertical structure,
i.e.\ they correspond in Fourier space to $\hat{v}_z(k_x,k_y,0)$. As a consequence,
they are prevented from the experiments by the hard boundary condition at the bottom of
the tank where velocity must be zero at all times. In our periodic simulations, those
waves have no reason to vanish and can actually be observed with significant energy
levels compared to waves in the frequency range just below $N$. For the sake of the
comparison with experiments, we proceed as for shear modes and impose
$\hat{v}_z(k_x,k_y,k_z=0)=0$ in all the following simulations \footnote{Parameter
\lstinline{params.no_vz_kz0} in Fluidsim.}.

\section{Simulations for $N=0.6$\,rad/s, $F = 0.73$ and $a=5$\,cm}
\label{sec:like-exp}

We present in this section a first set of simulations for parameters ($N=0.6$\,rad/s,
$F= \omega_f / N =0.73$ and $a=5$\,cm) corresponding to the experiment with the largest
amplitude described in \cite{Savaro2020}. For this experiment, the strong mixing of
salt in the tank prevented the full usage of imaging techniques. By starting with this
experimental set of parameter, we thus intend to give a complementary vision to what
was done in \cite{Savaro2020}.

\subsection{Time evolution of energy and dissipation terms}

\begin{figure}
\centerline{
\includegraphics[width=0.6\textwidth]{%
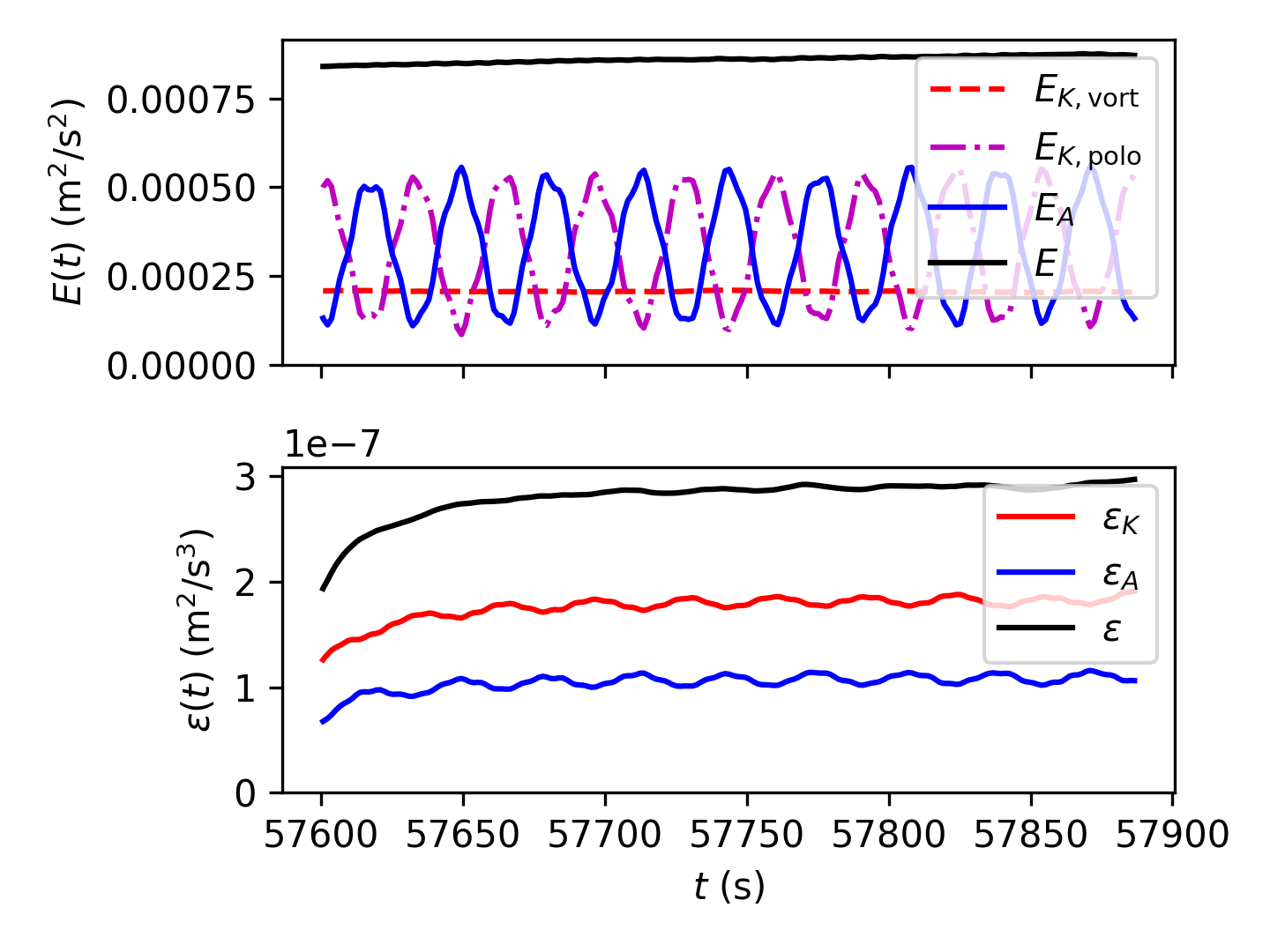}
}
\caption{Time evolution over a statistically stationary period of the energy components
(top row) and dissipation (bottom row) for parameters $F = 0.73$, $a = 5$\,cm and for
$2304\times2304\times384$. \label{fig:spatial-mean}}
\end{figure}

Figure \ref{fig:spatial-mean} shows the evolution of energy components and dissipation
over the period of time that was selected for data analysis for the resolution
$2304\times2304\times384$. It corresponds to 10 minutes of equation time. For the
coarser resolutions $1152\times1152\times192$ and $480\times480\times80$, we average on
10 and 20 minutes, respectively. In comparison, the experiments described in
\cite{Savaro2020} ran for five hours, and temporal analysis such as temporal spectra
were performed using a Welch method with a 20 minutes long temporal window. We see that
over the selected periods the statistical stationarity of the system is not perfect, as
long-time drifts are still observable, for example in the total energy and dissipations
(black lines). However, as suggested by figure \ref{fig:spatial-mean-history} and
because of the length of the time scales that are at play, obtaining a more meaningful
statistical stationarity would require to run the simulation and output data on a
period of equation time that is not computationally reasonable. Therefore, the size of
the time window that we selected for further analysis is the result of a compromise
between statistical stationarity, time resolution of the specific data outputs, and
computational time. Looking into the detail of energy components in figure
\ref{fig:spatial-mean} (top row), we see that kinetic poloidal and potential energy
present well-defined, antiphase oscillations around the same average level, suggesting
the presence of stationary waves in the simulation box as were seen in the experiment
\cite[]{Savaro2020}. The remaining kinetic energy is vortical energy (associated with
vortices with purely horizontal velocity), which is slightly smaller than the averaged
poloidal energy.

\subsection{Representations of the flow in physical space}


\begin{figure}
\centering
\includegraphics[width=0.65\textwidth]{%
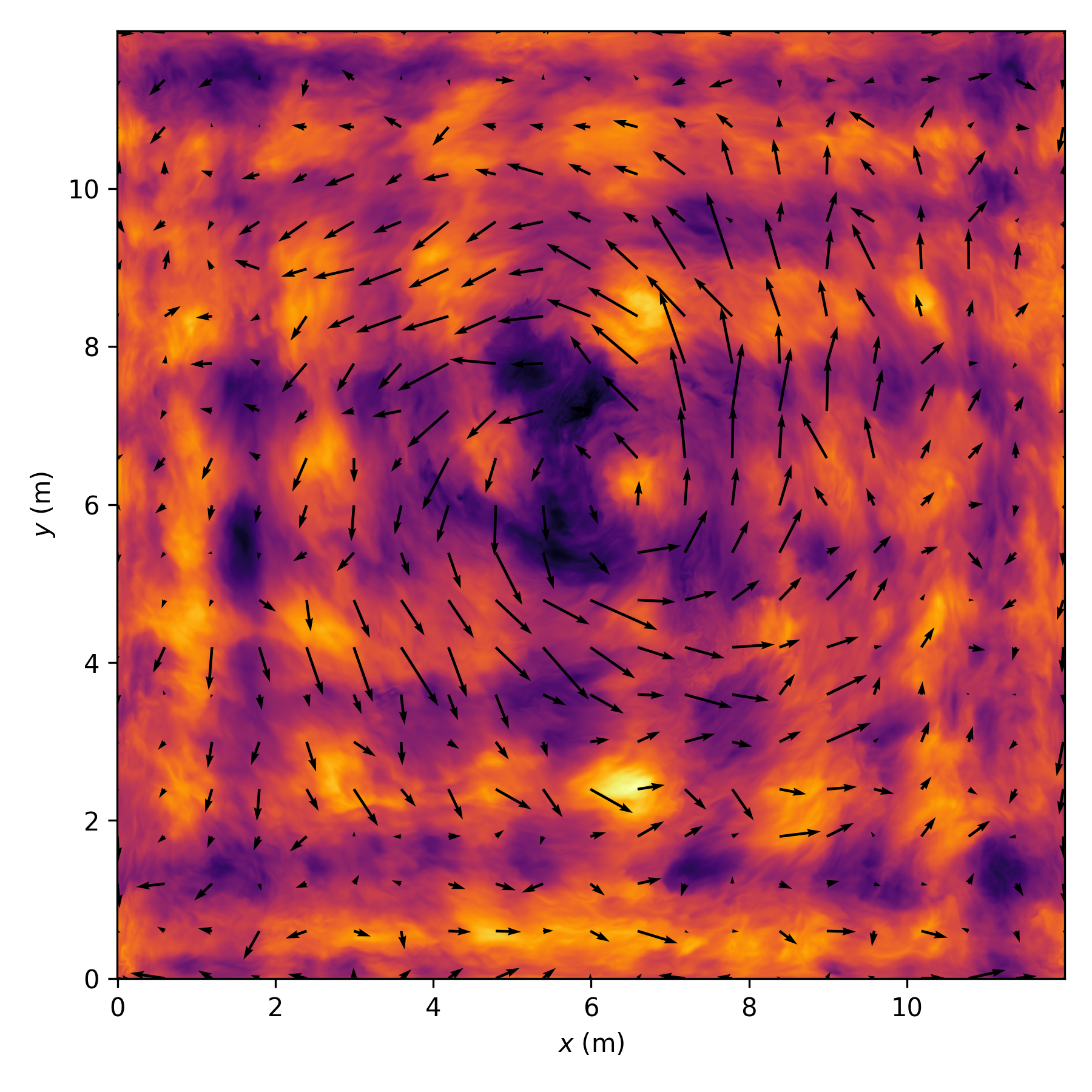}
\caption{Horizontal cross-session for $F=0.73$ and $a=5$\,cm
($2304\times2304\times384$).  The colors and the vector represent the buoyancy field
and the velocity field, respectively. \label{fig:phys-fields-horiz}}
\end{figure}

A snapshot of the flow is displayed in figure~\ref{fig:phys-fields-horiz}. We clearly
see a check pattern in the buoyancy field represented by the colors. This corresponds
to the large scale forced waves. However, the pattern is deformed and there are weak
smaller scale structures in some regions. In contrast, the horizontal velocity field
represented by the arrows is dominated by a very large vortex.


\begin{figure}
\centerline{
\includegraphics[width=0.48\textwidth]{%
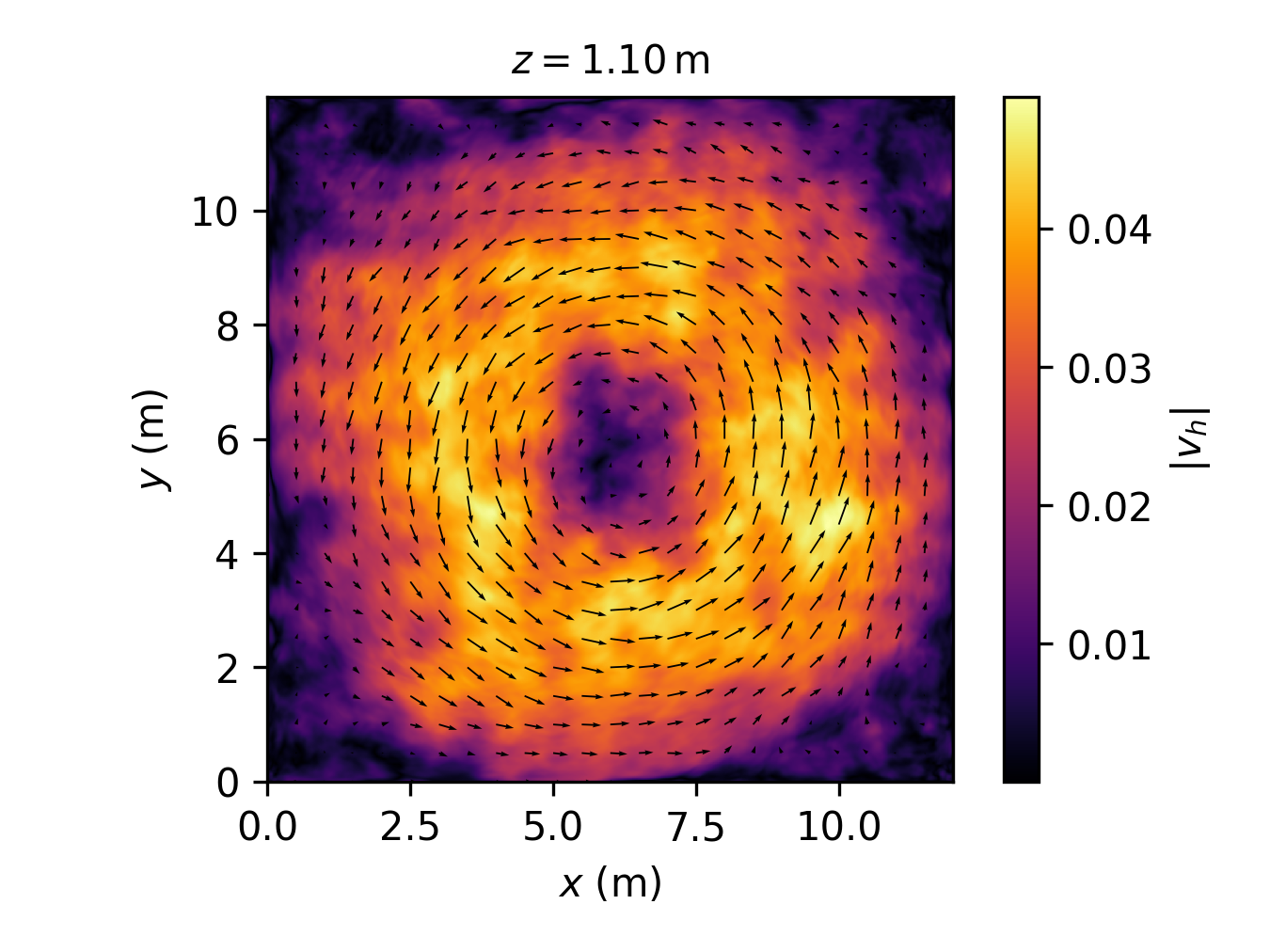}
\includegraphics[width=0.48\textwidth]{%
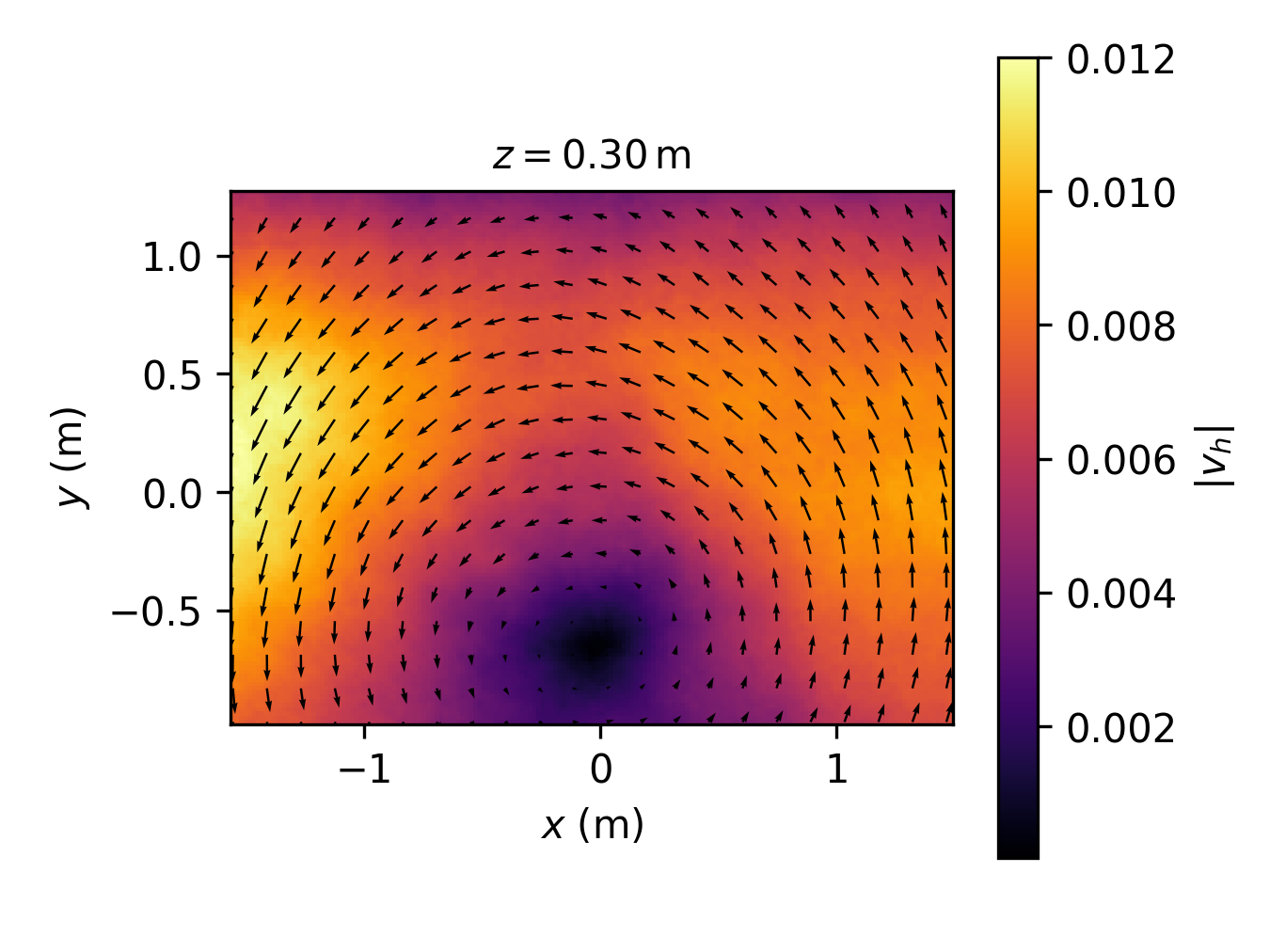}
}
\caption{Time-averaged, horizontal cross-sections of the velocity field in the
simulation (left) and the experiment (right), for parameters $F = 0.73$ and $a=5$\,cm.
The arrows represent the horizontal component of the velocity field, and the colors are
mapped to the magnitude of the total velocity in m/s. The resolution for the simulation
is $1152\times1152\times192$. Data is average over 10 minutes for the simulation and 33
minutes for the experiment. \label{fig:vortex}}
\end{figure}

A time-averaged, horizontal cross-section of the velocity field from the run in
$1152\times1152\times192$ (figure \ref{fig:vortex}, left) shows a large, horizontal
vortex centered in the middle of the simulation box. It is worth noting that a similar
vortex can be observed in experiments, as shown on the right panel of figure
\ref{fig:vortex}. We see that velocity magnitudes differ by approximately a factor 3
between simulation and experiment, though it should be noted first that the
experimental field was acquired one hour after the beginning of the experiment,
compared to around sixteen hours of equation time for the simulation, meaning that the
growth of the vortex mode in the experiment may not have reached saturation. In
addition, because the simulation box is twice as big as the experimental domain in all
directions, the vortex we observe can be twice as large as what is possible in the
experiments.

\begin{figure}
\centerline{
\includegraphics[width=0.75\textwidth]{%
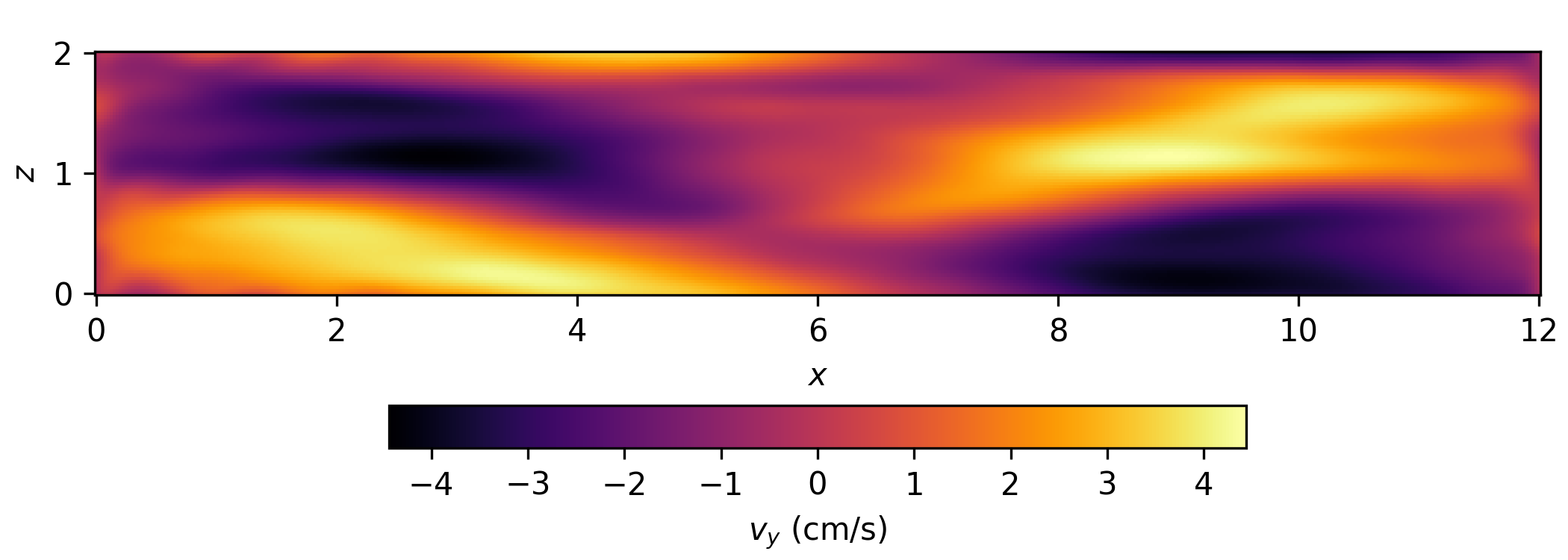}
}
\caption{Vertical cross-section going through the center of the tank ($y = L_y/2$) for
$F = 0.73$, $a=5$\,cm and resolution $480\times480\times80$. The colors show the
time-averaged out-of-plane velocity. \label{fig:vortex_vertical}}
\end{figure}

The vertical structure of the mean flow is shown in figure~\ref{fig:vortex_vertical}.
The colors show the time-averaged out-of-plane velocity for a vertical cross-section
going through the center of the tank ($y = L_y/2$). We see that there are actually two
superposed vortices at two different levels (one vertical wavelength). This very large
horizontal scale structure analogous to shear modes can most likely be interpreted as a
partial condensate as defined in \cite{vanKan2019Condensates} and due to the flux loop
identified in \cite{kumar2017, boffetta2011, linares2020numerical}.

\subsection{Spatial spectra}

\begin{figure}
\centerline{
\includegraphics[width=0.5\textwidth]{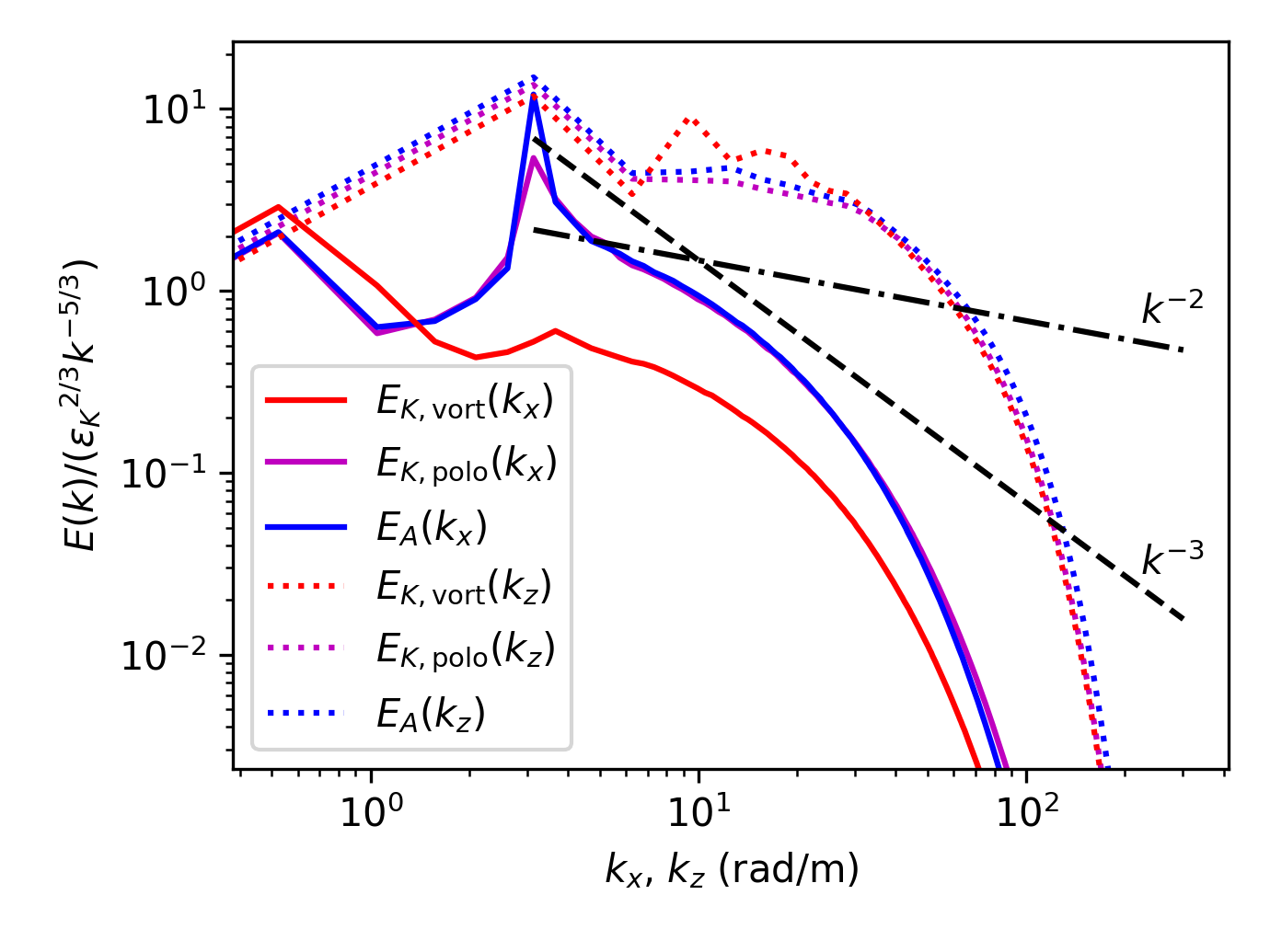}
\includegraphics[width=0.5\textwidth]{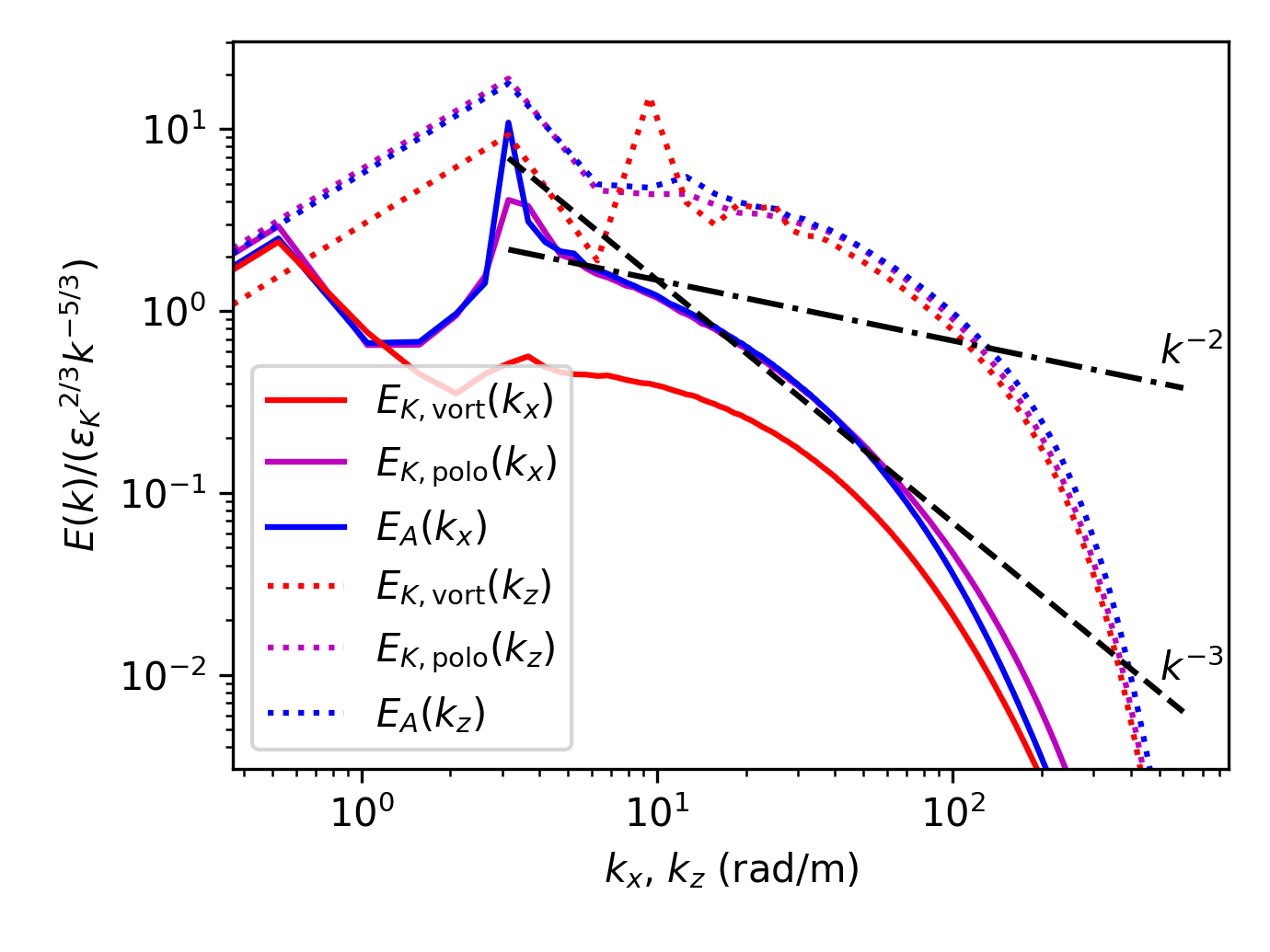}
}
\caption{One-dimensional spatial spectra for parameters $F = 0.73$ and $a=5$\,cm, at
resolutions $1152\times1152\times192$ (left) and $2304\times2304\times384$ (right). All
spectra are compensated by $\epsK^{2/3}k^{-5/3}$. Solid lines correspond to spectra
along the $x$ direction, dashed lines represent spectra along the $z$ direction. Red
curves are kinetic energy spectra, blue curves are available potential energy spectra,
and purple curves represent the poloidal part of kinetic energy spectra. In black
dashed and dash-dotted lines are represented power laws corresponding before
compensation to $k^{-3}$ and $k^{-2}$, respectively. \label{fig:spatial-spectra}}
\end{figure}

One-dimensional spatial energy spectra are presented in figure
\ref{fig:spatial-spectra}. We first note that horizontal (solid lines) and vertical
(dotted lines) spectra are well separated for all scales. Such strong anisotropy for
all scales is typical of strongly stratified flows affected by viscosity at all
horizontal scales \cite{brethouwer_billant_lindborg_chomaz_2007}. Such flows are
characterized by small buoyancy Reynolds numbers, which is consistent here as
table~\ref{table_simul} shows that the buoyancy Reynolds numbers measured in our
simulations are typically of order unity or less. The Helmholtz decomposition of
kinetic energy shows that poloidal and potential energies are equiparted at all scales,
suggesting the presence of internal gravity waves. This is also indicated by the strong
peak at $k_x=k_z=3.14$\,rad/m, which corresponds to the resonant 2D modes with six
horizontal wavelengths in the $x$ direction and one wavelength in the vertical
direction. The frequency of this mode is $\omega_{6,0,1}=0.71N$, which is close to the
forcing frequency $\omega_f=0.73N$. In addition to those waves, we note that vortical
energy levels are non-negligible. In fact for the vertical spectra (dotted lines),
vortical energy is typically of the same order of magnitude as poloidal and potential
energy, at all scales of the simulation. For the horizontal spectra, vortical energy
dominates only at large scales, typically larger than the scale of the principal
horizontal forcing mode $k_x=3.14$\,rad/m, but is negligible at smaller scales.

This analysis suggests that even though waves are present at all vertical and
horizontal scales, vortices still play an important role at all vertical scales, and
even dominate at very large horizontal scales, which is compatible with the picture
given by the horizontal mean flow in figure \ref{fig:vortex}. Finally, it should be
noted that no clear scaling law appear in those spectra, whether horizontal or
vertical, as shown by eye guides for $k^{-2}$ and $k^{-3}$ behaviors (black dashed and
dashed-dotted lines). The vertical poloidal and potential energy spectra are compatible
with a ${k_z}^{-2}$ scaling, but on a range shorter than a decade. This absence of
scaling law is probably due to the limitation of spatial resolution, as
high-resolution, high-$\R$ 2D simulations have shown such scalings to be possible
\cite{Linares2020}.

\subsection{Temporal spectra and linear waves relations}

\begin{figure}
\centering
\includegraphics[width=0.8\figwidth]{%
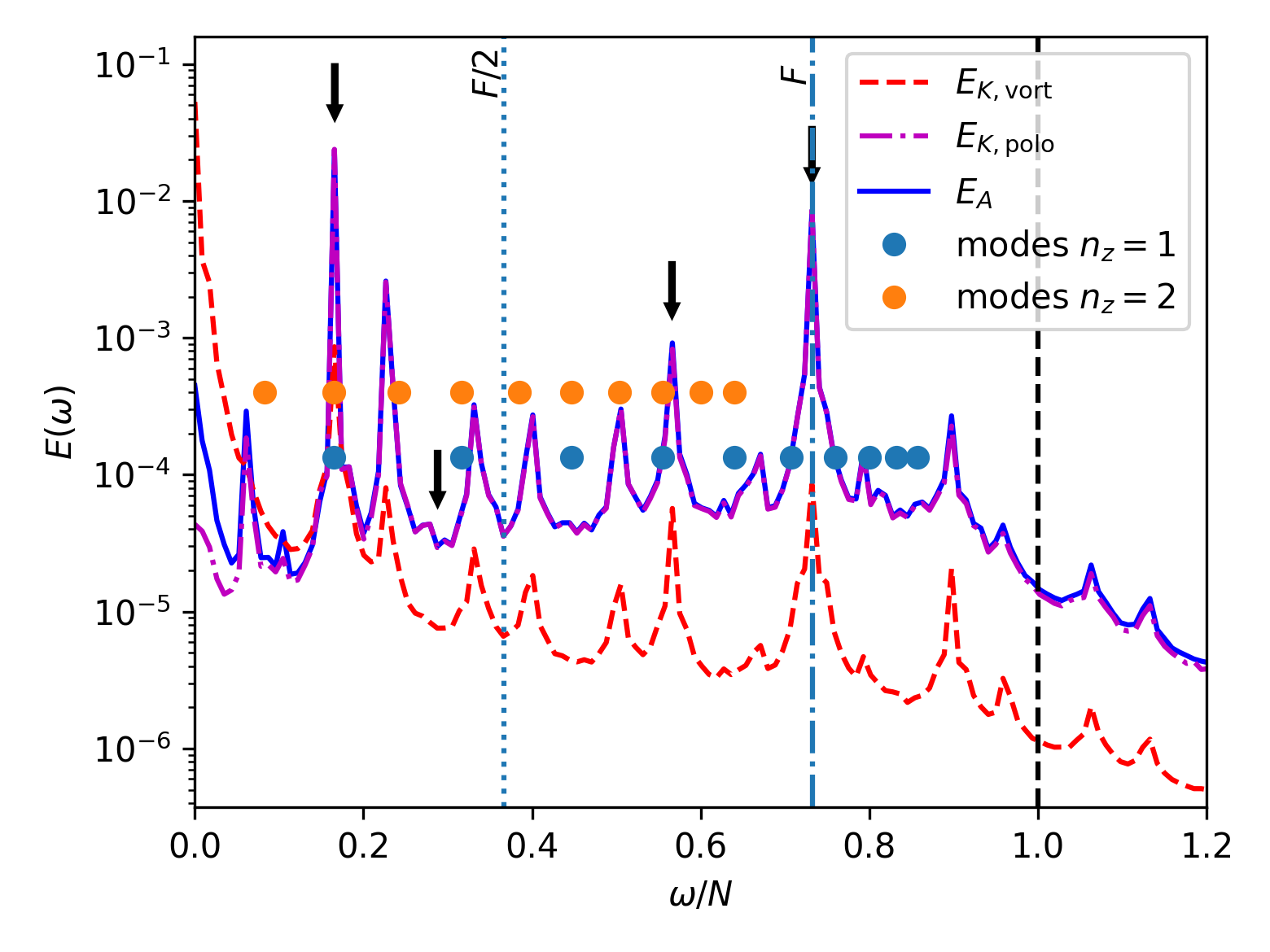}
\caption{Temporal spectra as a function of the normalized frequency $\omega/N$ for
parameters $F = 0.73$ and $a=5$\,cm, at resolution $480\times480\times80$. The kinetic
energy spectrum is represented by the solid red line, with poloidal and vortical parts
of the kinetic energy being plotted in dashed and dashed-dotted red respectively. The
available potential energy spectrum is shown in solid blue. Vertical lines signal
respectively the \bv{} frequency (dashed), the forcing frequency (dashed-dotted) and
half the forcing frequency (dotted). Blue dots represent the frequencies of resonant
modes with $n_z=1$, $n_y=0$ for successive values of $n_x$. Orange dots represent the
frequencies of resonant modes with $n_z=2$, $n_y=0$ for successive values of $n_x$.
Arrows show the frequencies at which a spatio-temporal analysis is performed in figure
\ref{fig:spatiotemporal-spectra-omega}. \label{fig:temporal-spectra}}
\end{figure}

Similarly to what was done in \cite{Savaro2020}, we present the results of a frequency
analysis of the flow. All physical fields are recorded in a reduced, small-wavenumber
region of Fourier space with sufficient time resolution to resolve frequencies as high
as $2N$. Frequency spectra are then obtained by computing the Fourier transform in the
temporal dimension and averaging over all recorded wavenumbers. We thus take advantage
of the numerical setup, which allows us to give a more detailed frequency analysis,
e.g. by computing the poloidal, vortical and potential energy spectra. The results of
this analysis are presented in figure \ref{fig:temporal-spectra}.

We see that all spectra show a marked peak at the forcing frequency $F$, signaled by a
vertical, blue dashed-dotted line. Around this frequency, vortical energy is negligible
and poloidal energy is equiparted with potential energy, suggesting internal gravity
waves. Going down from $F$, the poloidal and potential energy spectra stay equal and
rather flat, with the exceptions of numerous peaks at well-defined frequencies. This
shape of the spectra, though quite different from the $\omega^{-2}$ behaviour suggested
by Garrett and Munk \cite{Garrett1979}, is consistent with the frequency analysis of
\cite{Savaro2020}. In the experimental study, peaks mostly aligned with frequencies of
two-dimensional resonant modes of the tank, i.e.\ modes with either $n_y=0$ or
$n_x=0$. On figure \ref{fig:temporal-spectra}, the frequencies of the ten first modes
with $n_y=0$ are indicated, for $n_z=1$ (blue dots) and $n_z=2$ (orange dots). We see
that only one peak is aligned clearly with modes frequencies, at $\omega=0.17N$ which
corresponds to modes such as $n_z=n_x$ (i.e.\ $k_z = (L_x/L_z)k_x=6k_x$). Other
peaks fall close to other modes frequencies, but not clearly as aligned as the first
peak at $\omega=0.17N$, which is also the highest peaks in the spectra. In addition, we
can note that all peaks below $F$ can be associated in frequency pairs
$(\omega_a,\omega_b)$ such that $\omega_a+\omega_b=\omega_f$, suggesting triadic
resonances with the forcing frequency. This can be intuited graphically in the
representation of figure \ref{fig:temporal-spectra} with a linear scale for the
frequency axis : all pairs of peaks that are symmetric with respect to $F/2$ (signaled
by a vertical, blue dotted line) verify a triadic resonance relation with the forcing.
This graphical hint is then verified numerically by the calculation of the resonance
relations. Such triadic resonance relations were also observed with some approximation
in \cite{Savaro2020}. Finally, we note that vortical energy stays negligible until very
low frequencies under $0.1N$ where it gets larger than poloidal energy by three orders
of magnitude. This is consistent with the large horizontal vortex observed in the mean
flow shown in figure \ref{fig:vortex}. We also note that at those frequencies,
potential and poloidal energies are not equiparted anymore, suggesting that waves are
absent from these scales.


\begin{figure}
\centerline{
\includegraphics[width=0.48\textwidth]{%
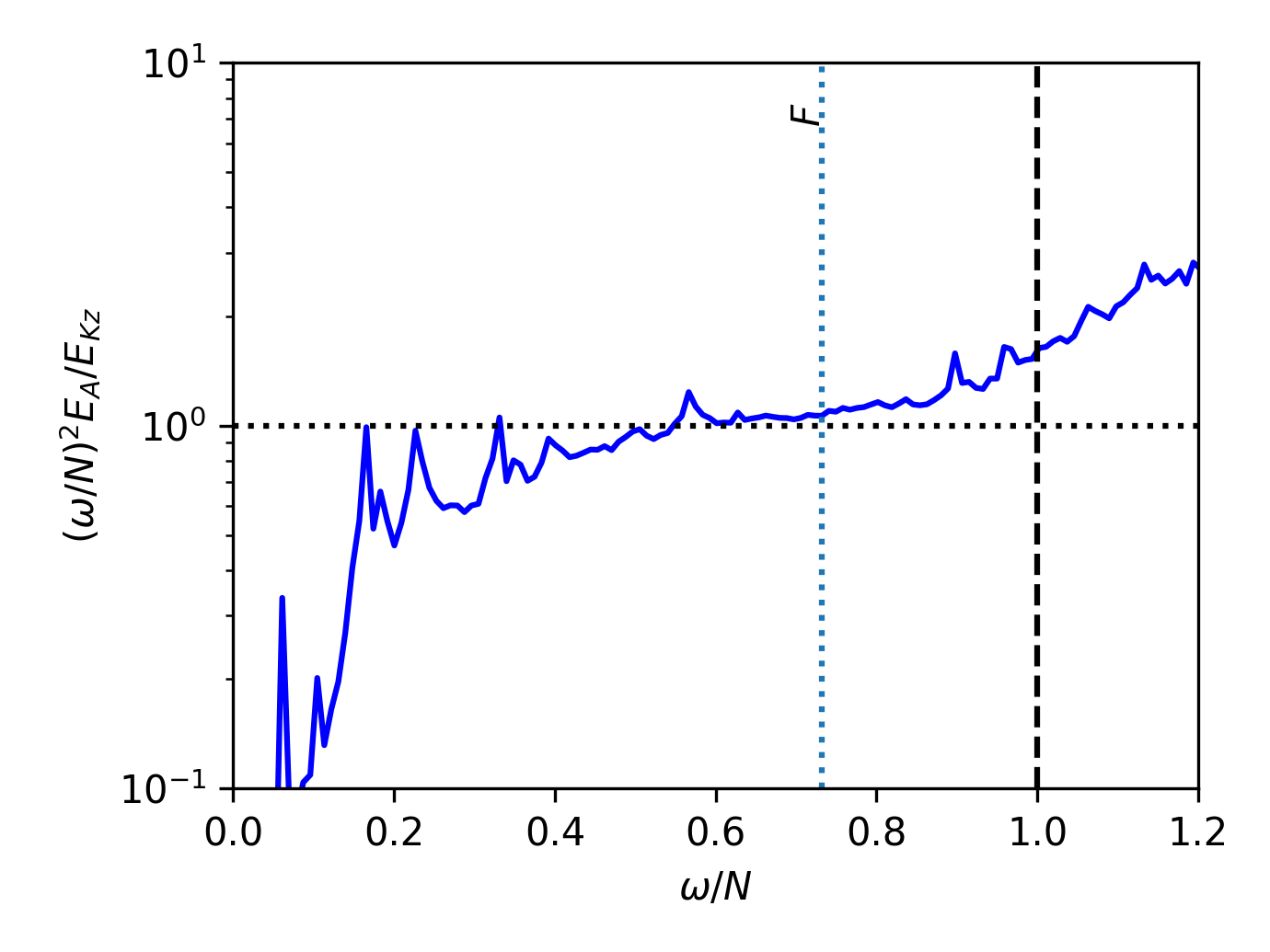}
\includegraphics[width=0.48\textwidth]{%
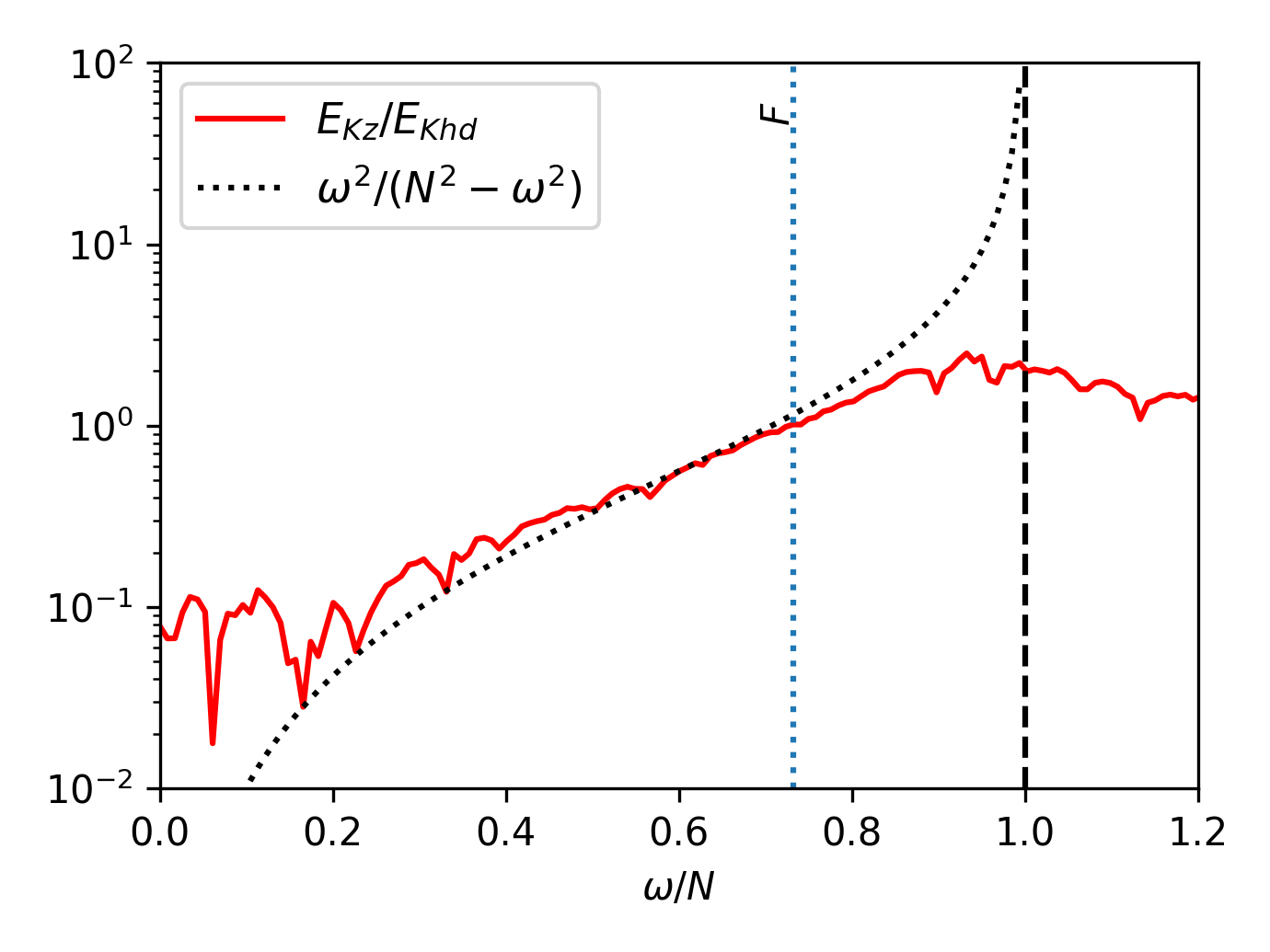}
}
\caption{Linear waves relations in $\omega$-space for $F = 0.73$ and $a=5$\,cm, at
resolution $480\times480\times80$. Left: relation between the vertical kinetic energy
and the available potential energy spectra. The horizontal dotted line represents
unity. Right: relation between the vertical kinetic energy and the poloidal part of the
horizontal kinetic energy spectra. The dashed curve represents the right-hand-side of
equation \eqref{ratio_Kzh}. On both figures, vertical lines signal respectively the
\bv{} frequency (dashed) and the forcing frequency (dotted). \label{fig:linear-waves}}
\end{figure}

The presence of internal gravity waves can be investigated further by looking for
relations between spectral quantities that are verified for linear waves. We note that
the linearized version of \eqref{buoy} reads:
\begin{equation}
\p_t b = N^2 v_z,
\label{buoylin}
\end{equation}
which can be rewritten after taking the time Fourier transform and square modulus of
each side of the equation:
\begin{equation}
\frac{\omega^2}{N^2}E_A(\omega) = E_{Kz}(\omega),
\label{ratio_AKz}
\end{equation}
where $E_A(\omega)=|\hat{b}(\omega)|/(2N^2)$ and $E_{Kz}(\omega)=|\hat{v}_z(\omega)|/2$
are the potential energy and vertical kinetic energy spectra respectively. This
relation is tested in figure \ref{fig:linear-waves} (left), where the ratio of both
sides of the equation is plotted and compared to unity. We see that relation
\eqref{ratio_AKz} is noticeably far from being verified, though by a small amount in a
short frequency range below $F$. Such a discrepancy should be nuanced by the fact that
the set of parameters tested here corresponds to the strongest forcing used in
\cite{Savaro2020}, where strong mixing occurred, suggesting substantial overturning and
a fortiori a regime where nonlinearities are important. In contrast, \eqref{ratio_AKz}
is well verified at weaker forcing, for example in the experiments for $a=2$\,cm or in
simulations with different parameters presented in \ref{sec:slower}.

Another test for the presence of linear waves would use the geometric properties of
internal gravity waves. If we consider one linear wave propagating along
$\kk=(\sin\theta\cos\phi,\sin\theta\sin\phi,\cos\theta)$, in spherical coordinates of
axis $z$, then the wave velocity is orthogonal to $\kk$ and in a vertical plane that
contains $\kk$. We can then decompose the velocity as an horizontal and a vertical
component $\vv = v_h \eeh + v_z \eez$, such that
$\hat{v}_h(\kk,\omega)=a(\kk,\omega)\cos\theta$ and
$\hat{v}_z(\kk,\omega)=a(\kk,\omega)\sin\theta$. Now if we assume that the flow is made
of a random superposition of independent linear plane waves that are statistically
axisymmetric around axis $z$, then the statistics of $a$ only depend on $k=|\kk|$ and
$\omega$, and we can write by taking the average power spectra of the velocity
components:
\begin{equation}
\mean{|\hat{v}_z(\kk,\omega)|^2}=A(k,\omega)\sin^2\theta
\label{axsym_vz}
\end{equation}
and
\begin{equation}
\mean{|\hat{v}_h(\kk,\omega)|^2}=A(k,\omega)\cos^2\theta,
\label{axsym_vh}
\end{equation}
where $A(k,\omega)=\mean{|a(\kk,\omega)|^2}$. By summing over the azimuthal angle
$\phi$ and the wavenumber norm $k$ we get the vertical and horizontal kinetic frequency
spectra $E_{Kz}(\omega)$ and $E_{Kh}(\omega)$, which share the same multiplicative
factor so that their ratio reads:
\begin{equation}
\frac{E_{Kz}(\omega)}{E_{Kh}(\omega)} = \frac{\sin^2\theta}{\cos^2\theta} =
\frac{\omega^2}{N^2 - \omega^2}.
\label{ratio_Kzh}
\end{equation}
Relation \eqref{ratio_Kzh} is tested in figure \ref{fig:linear-waves} (right). In order
to probe the wave component of the flow, the ratio $E_{Kz}/E_{Khd}$ is shown (solid red
line), where $E_{Khd}$ is the kinetic energy associated with the divergent part of the
horizontal velocity. The ratio is compared to the right hand side of equation
\eqref{ratio_Kzh}. Like with equation \eqref{ratio_AKz}, we see that the relation
\eqref{ratio_Kzh} is close to be verified in a short frequency range below the forcing,
but with a significant deviation outside of this range. We note again that this
relative agreement is consistent with the fact that the forcing amplitude $a=5$\,cm
corresponds to a regime where nonlinearities where observed to have a stronger effect
in experiments \cite{Savaro2020}.

\subsection{Spatiotemporal correlations}

\begin{figure}
\centerline{
\includegraphics[width=0.48\textwidth]{%
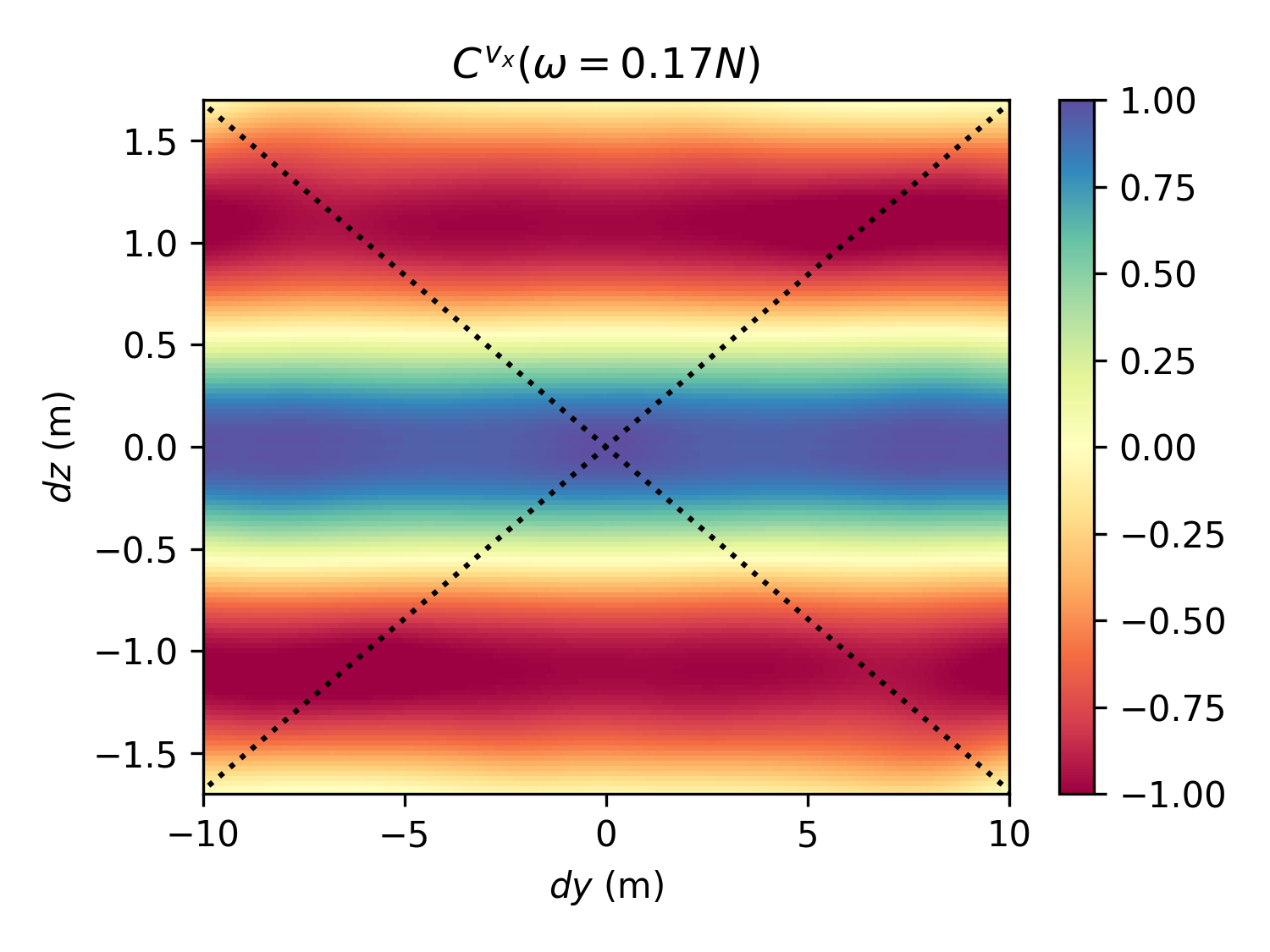}
\includegraphics[width=0.48\textwidth]{
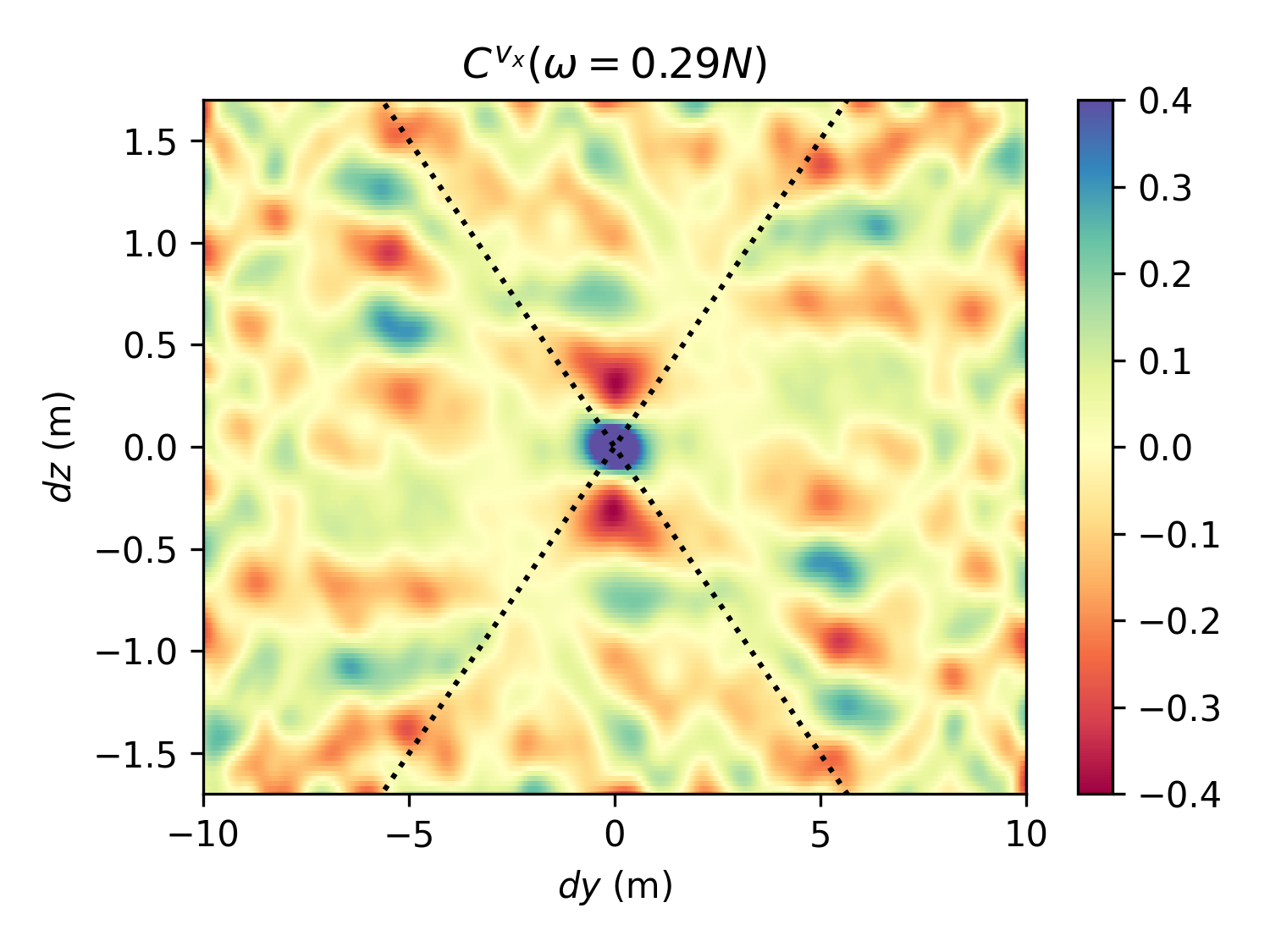}
}
\centerline{
\includegraphics[width=0.48\textwidth]{%
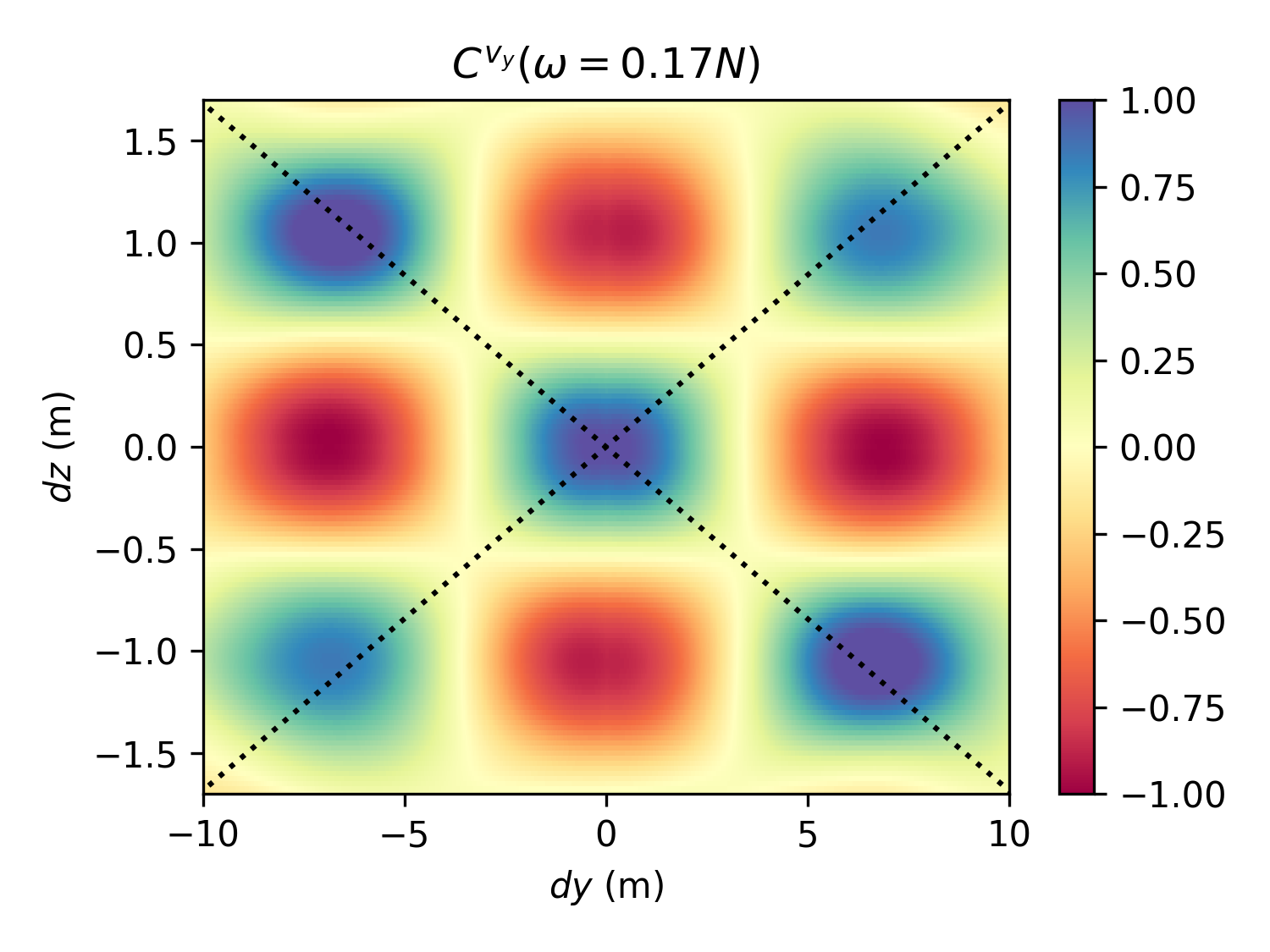}
\includegraphics[width=0.48\textwidth]{%
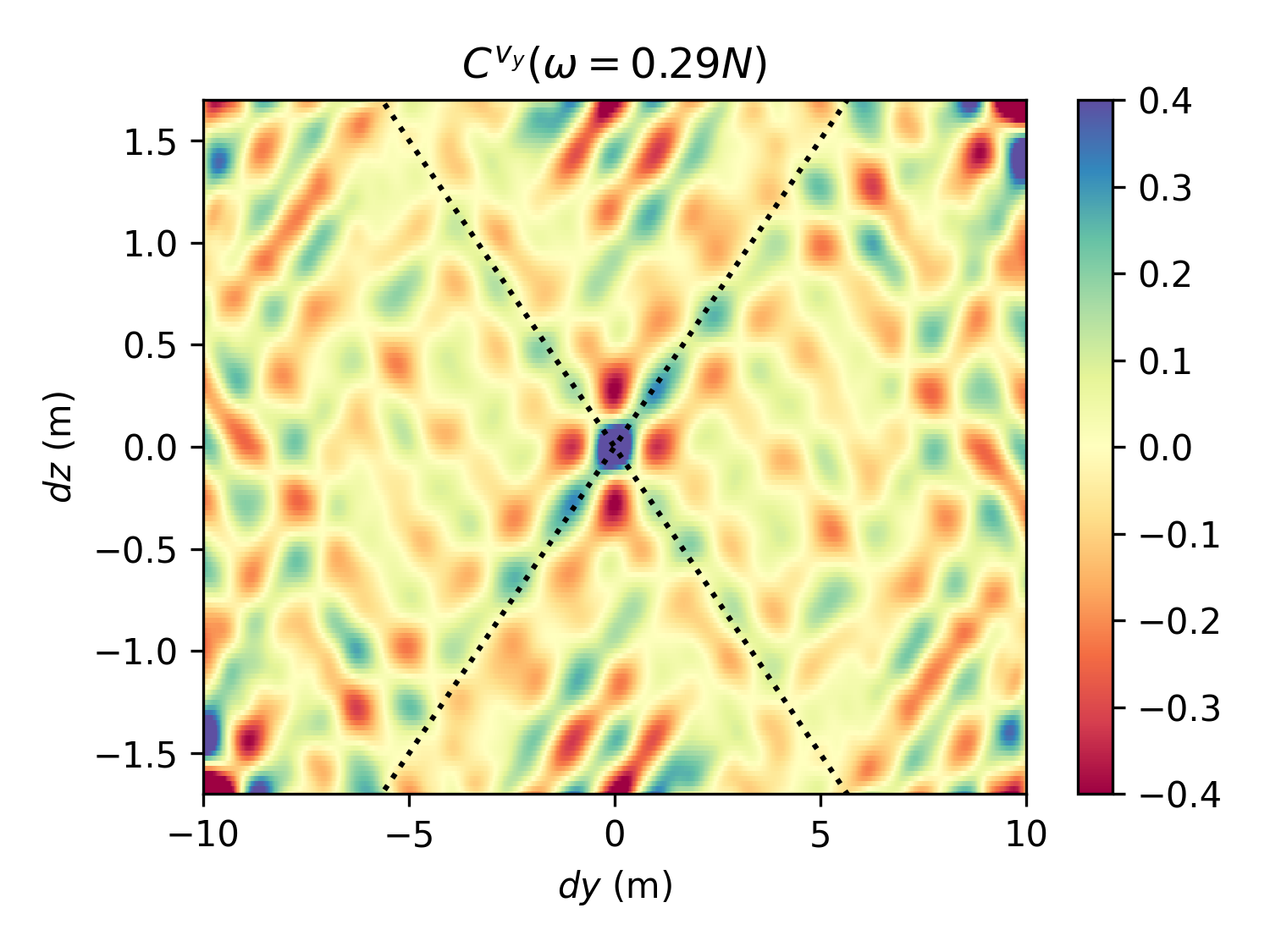}
}
\caption{Two-dimensional spatial correlations at fixed $\omega$ of $v_x$ (top row) and
$v_y$ (bottom row), for $F = 0.73$ and $a=5$\,cm, at resolution $480\times480\times80$.
Correlations are shown at $\omega=0.17N$ (left column) and $\omega=0.29N$ (right
column), which correspond respectively to the first and second arrows from the left on
figure \ref{fig:temporal-spectra}. \label{fig:spatiotemporal-correlations}}
\end{figure}

In order to probe more deeply the structure of the wave field, we perform a
spatiotemporal analysis inspired by \cite{Savaro2020}. We compute the spatial
correlation of velocity component $v_i$ at fixed frequency $\omega$ as:
\begin{align}
C^{v_i}(\xx,\omega) = \frac{\mean{{\hat{v}_i}^*(\xx_0+\xx,\omega)\hat{v}_i(\xx_0,\omega)
+ \text{c.c.}}}{2\mean{|\hat{v}_i(\xx_0,\omega)|^2}},
\label{corr}
\end{align}
where $\mean{\cdot}$ represents an average over $\xx_0$ and c.c. stands for the complex
conjugate of the expression it follows. We study specifically here the vertical
correlations of the horizontal velocity field, i.e.\ for $v_i=v_x,v_y$ and
$\xx=(0,\diff{y},\diff{z})$ lying in a vertical plane. Figure
\ref{fig:spatiotemporal-correlations} show the results of this analysis in the
$\diff{y}-\diff{z}$ plane for two distinct frequencies, corresponding to the first two
arrows in figure \ref{fig:temporal-spectra}. The correlations for $\omega=0.17N$ (left
column), which corresponds to a bright peak in the temporal spectra, show a large band
structure with no $\diff{y}$ dependence for $v_x$, and a large check pattern for $v_y$.
The oscillations appear with one wavelength in the vertical direction for both
components, and one wavelength in the $y$ direction for $v_y$. These structures are
thus compatible with the mode $(n_x=0, n_y=1, n_z=1)$.

In contrast, for a frequency between peaks of the temporal spectra such as
$\omega=0.29N$ (right column), no large-scale correlation structure emerges, but rather
a cross-like pattern is visible in the correlations of $v_y$. The cross is highlighted
by two dotted black lines with an angle $2\theta$ between each other, with $\theta$
being the angle associated with $\omega=0.29N$ through the dispersion relation. It is
to be noted that this cross is observed for any frequency outside peaks of the spectra,
with its angle varying accordingly with the dispersion relation. Both structures,
whether on or off the peaks of the temporal spectra, are observed in experiments
\cite{Savaro2020}. The on-peak correlation structure is interpreted as the presence of
a two-dimensional mode in the $y-z$ plane, whereas the off-peak cross can be
interpreted as the result of the superposition of random, broadband axisymmetric waves
around the vertical axis. The cross pattern we observe is less bright than in
\cite{Savaro2020}, however this difference could be explained by the lack of
statistical convergence in our analysis, due to the shorter time over which the
temporal Fourier transform is computed. Indeed, the results presented in figure
\ref{fig:spatiotemporal-correlations} come from a single time window of twenty minutes
of equation time, whereas in \cite{Savaro2020} a Welch method is used to average over
several windows of around twenty minutes long. In addition to this lesser statistical
convergence, it should again be kept in mind that the present simulation is forced with
a large amplitude compared to experiments, which was associated with stronger
nonlinearities and thus a less visible weak wave turbulence signature. Finally, let us
note that we investigate a larger spatial window in the correlation plane, as we are
able to probe a full $12\times2$\,m$^2$ window whereas the experimental setup of
\cite{Savaro2020} only allows for a $2.5\times2.1$\,m$^2$ window, so the cross pattern
is not observed over the same spatial scales.

\subsection{Spatiotemporal spectra}

\begin{figure}
\centerline{
\includegraphics[width=0.5\textwidth]{%
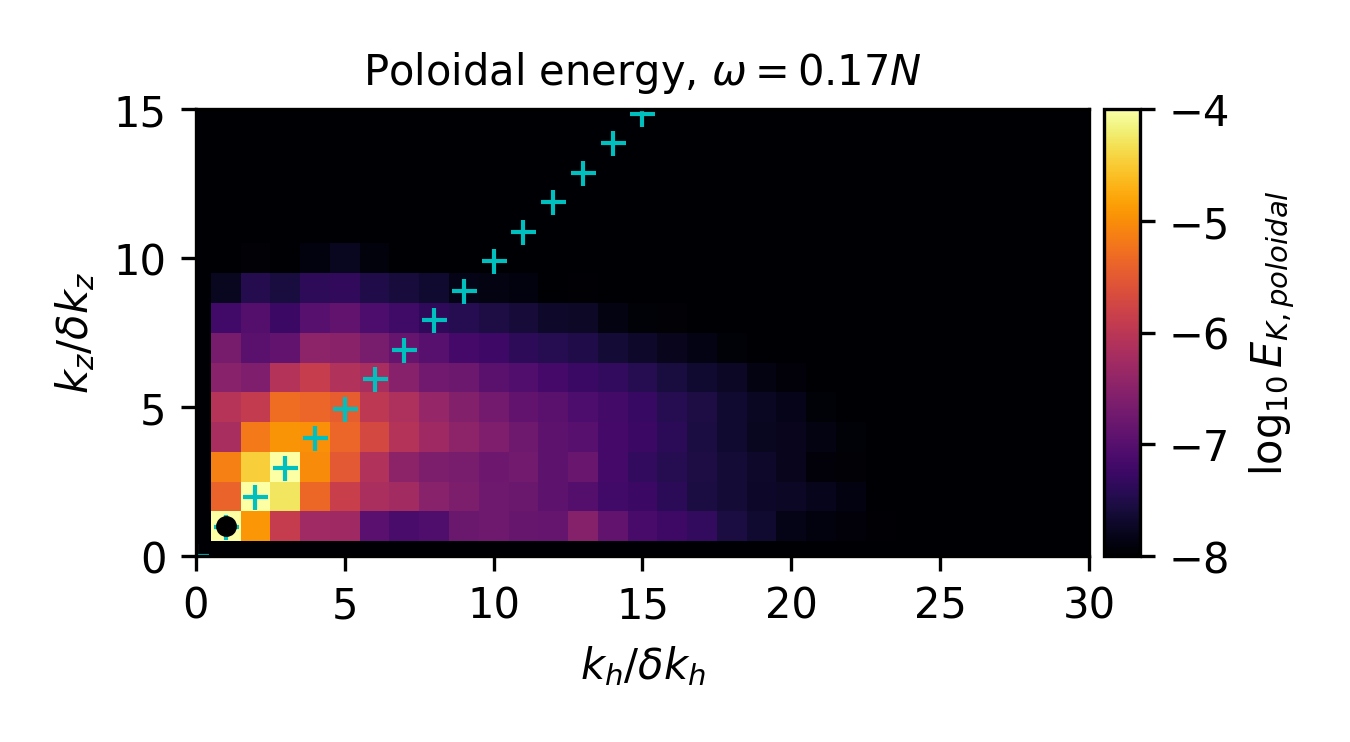}
\includegraphics[width=0.5\textwidth]{%
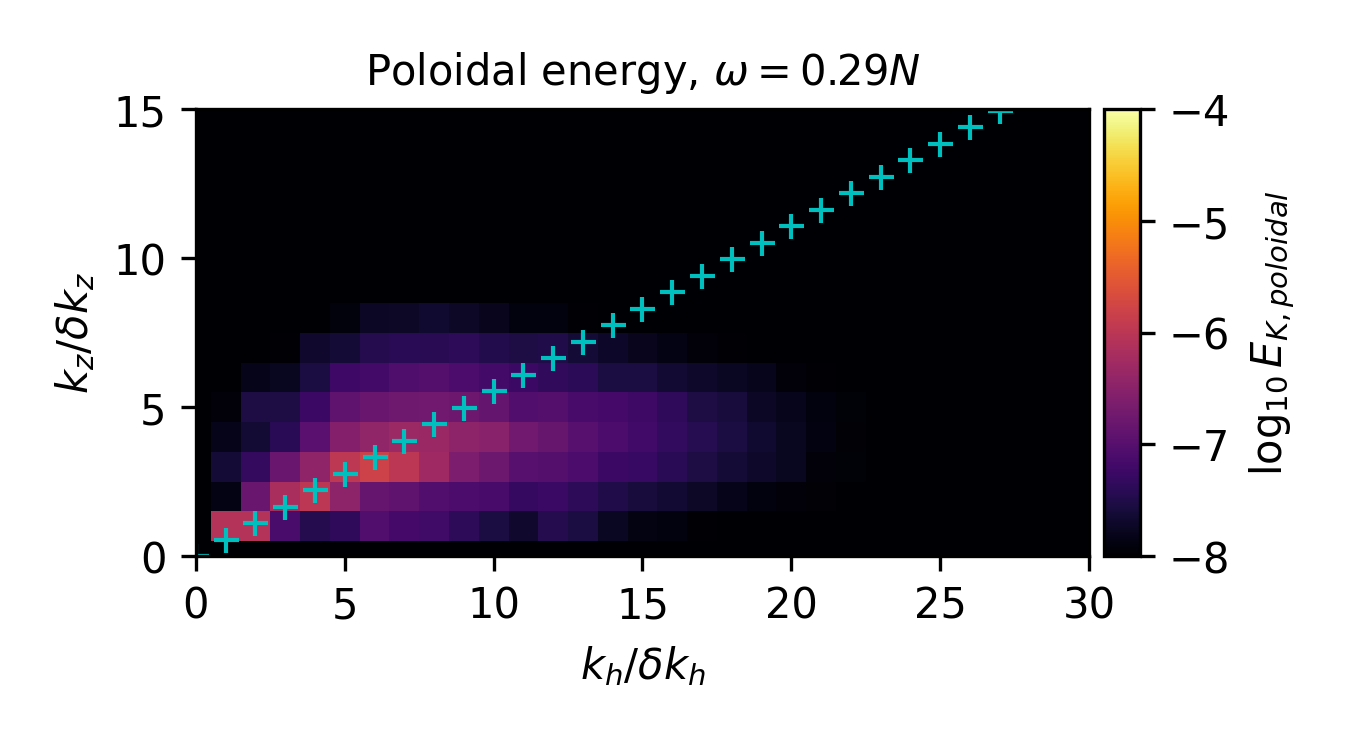}
}
\vspace{-4mm}
\centerline{
\includegraphics[width=0.5\textwidth]{%
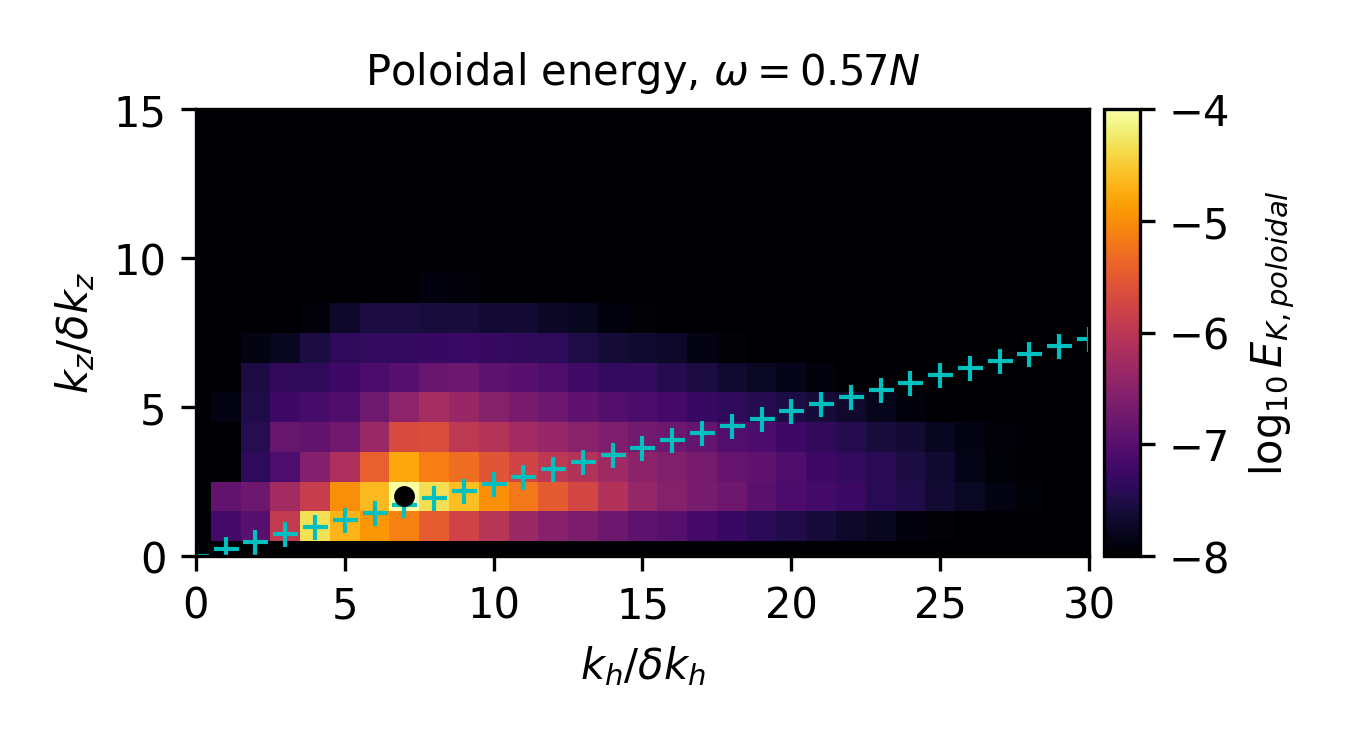}
\includegraphics[width=0.5\textwidth]{%
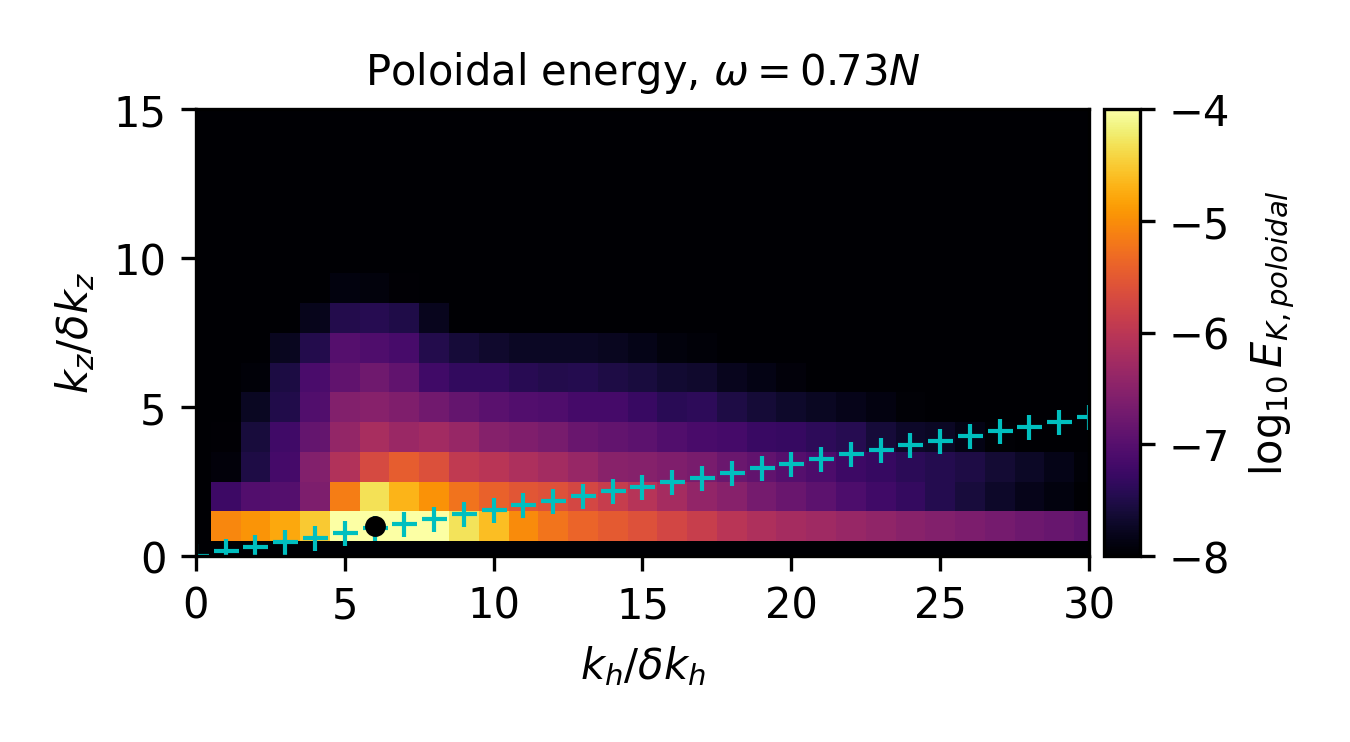}
}
\vspace{-2mm}
\caption{Spatiotemporal spectra of poloidal kinetic energy in the $k_z-k_h$ plane at
fixed $\omega$, for $F = 0.73$ and $a=5$\,cm. Each figure corresponds to an arrow in
figure \ref{fig:temporal-spectra}. Top left: $\omega=0.17N$. Top right: $\omega=0.29N$.
Bottom left: $\omega=0.57N$. Bottom right: $\omega=\omega_f=0.73N$. All axes are
normalized by the grid spacing in its respective direction in Fourier space. Colors are
mapped to the logarithm of the poloidal kinetic energy spectrum, and cyan crosses
represent the internal gravity waves dispersion relation. Black dots represent the
location of the energy maximum. \label{fig:spatiotemporal-spectra-omega}}
\end{figure}

Because the solver used for simulations is pseudospectral, the spatiotemporal analysis
is more conveniently performed in spectral space rather than physical space for the
spatial dimensions. In order to give a complementary point of view of the correlations
picture, we performed such an analysis in $\kk-\omega$ space. In order to do so, we use
the same technique as for the temporal spectra, which consists in saving a reduced
portion of Fourier space at low $k$, with a sufficient time resolution. Contrary to
temporal spectra however, the obtained spectra are not averaged over the whole Fourier
space, so that the spatial information is not lost. Because of the anisotropic nature
of stratified turbulence and internal waves, we rather perform a cylindrical average
around the $z$ axis in spectral space, in order to reduce the number of dimensions to
analyze. The results of this cylindrical average are $(2+1)$-dimensional spatiotemporal
spectra of the variables $(k_h,k_z,\omega)$.

Figure \ref{fig:spatiotemporal-spectra-omega} shows cuts of the spatiotemporal spectra
at constant $\omega$, which corresponds directly to the spectral equivalent of the
picture of figure \ref{fig:spatiotemporal-correlations}. The four panels show the
poloidal energy spectra at frequencies $\omega/N=0.17,0.29,0.57,0.73$, which correspond
to the four arrows of figure \ref{fig:temporal-spectra}. The frequency $\omega=0.29N$
(top right) falls between peaks of the temporal spectra, whereas the three other
correspond to peaks that are engaged in a triadic resonance frequency relation, with
$\omega=0.73N$ (bottom right) being the forcing frequency and $\omega=0.17N$ (top left)
the peak at the frequency of mode $(n_x=0,n_y=1,n_z=1)$. We see that for each
frequency, the energy is concentrated around the internal wave dispersion relation,
which is signaled by cyan plus signs. For the off-peak frequency (top right), the
energy is two to three orders of magnitude lower than for the other frequencies, which
is compatible with the height of the peaks that is observed in figure
\ref{fig:temporal-spectra}. We also note that for this frequency, the energy is
distributed in a rather homogeneous manner along the dispersion relation, with no
preferred wave numbers. This is different for all three on-peak frequencies, where
bright peaks can be observed at rather small wavenumbers. The pixel with maximum energy
is signaled by a black dot for those three frequencies. For $\omega=0.17N$ (top left)
at $(n_h=1,n_z=1)$, for $\omega=0.57N$ (bottom left) at $(n_h=7,n_z=2)$ and for
$\omega=0.73N$ at $(n_h=6,n_z=1)$. It should be kept in mind that as the result of the
cylindrical average around axis $z$, the information about the direction of $\kk_h$ is
lost, and the simple values of $n_h$ are not enough to validate a triadic resonance
relation in $\kk$ space. However, a more thorough study of the spatiotemporal spectra
in the $(k_x,k_y,k_z,\omega)$ space (data not shown) shows that the maxima that are
signaled by black dots actually correspond to 2D modes, namely with $n_y=0$. In that
case, $\kk_h$ coincides with $\kk_x$ and $n_h$ with $n_x$. We thus see that the three
frequencies that are engaged in a triadic resonance relation in frequency are also
engaged in such relation for wavevectors, suggesting that triadic instability plays a
major in the dynamics of the flow, as observed in experiments \cite{Savaro2020}.

\begin{figure}
\centerline{
\includegraphics[width=0.48\textwidth]{%
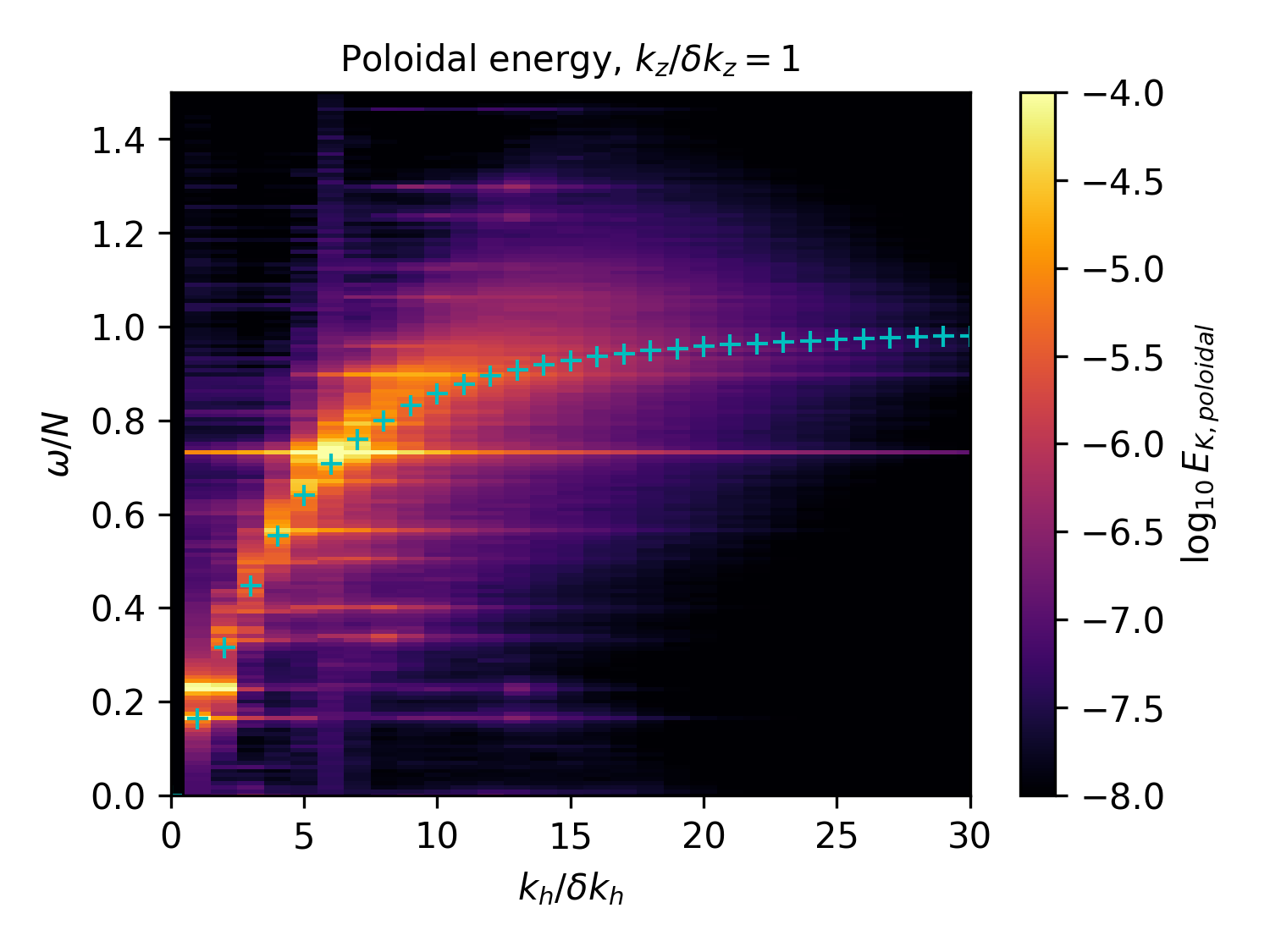}
\includegraphics[width=0.48\textwidth]{%
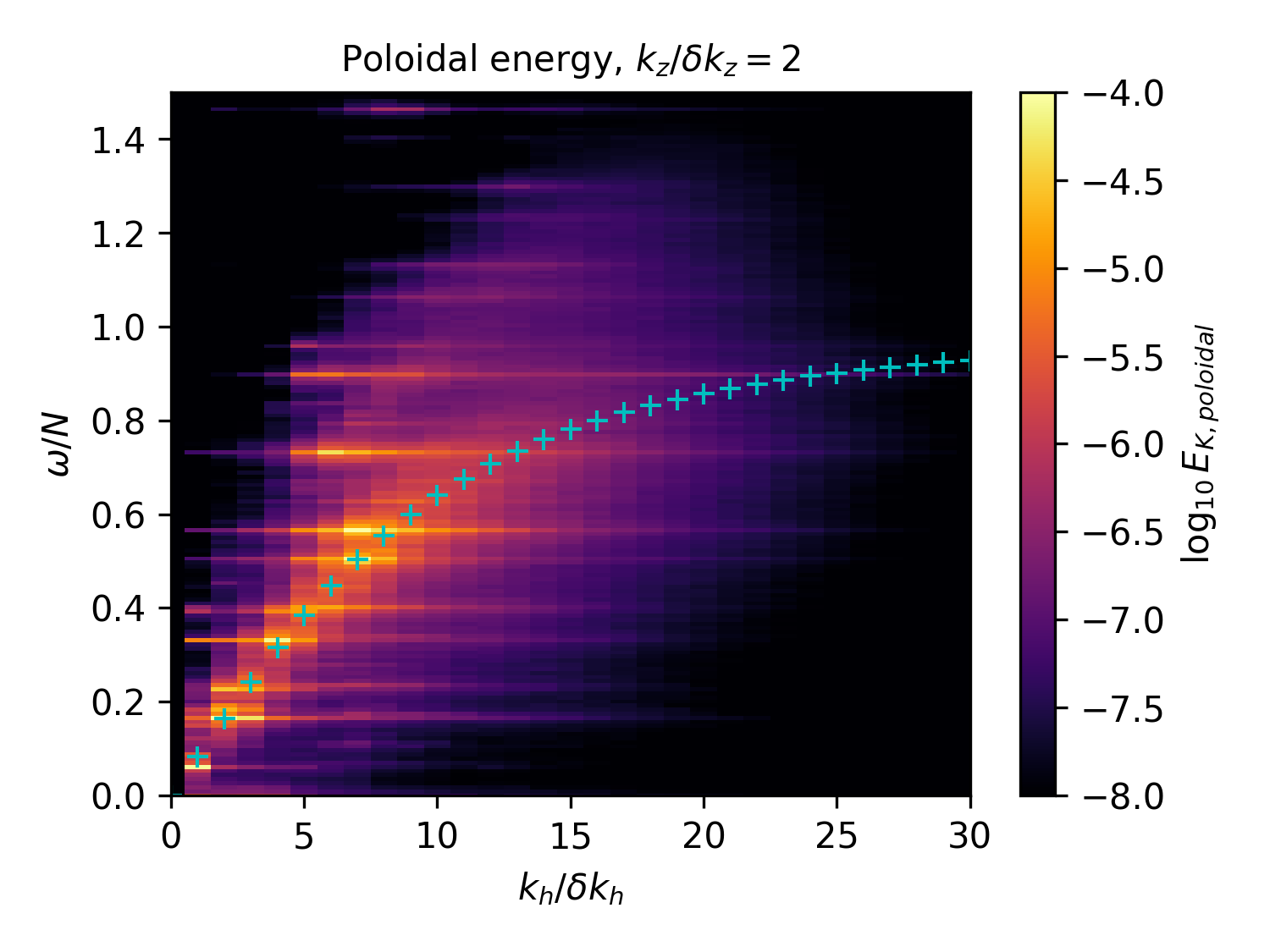}
}
\centering
\includegraphics[width=0.48\textwidth]{%
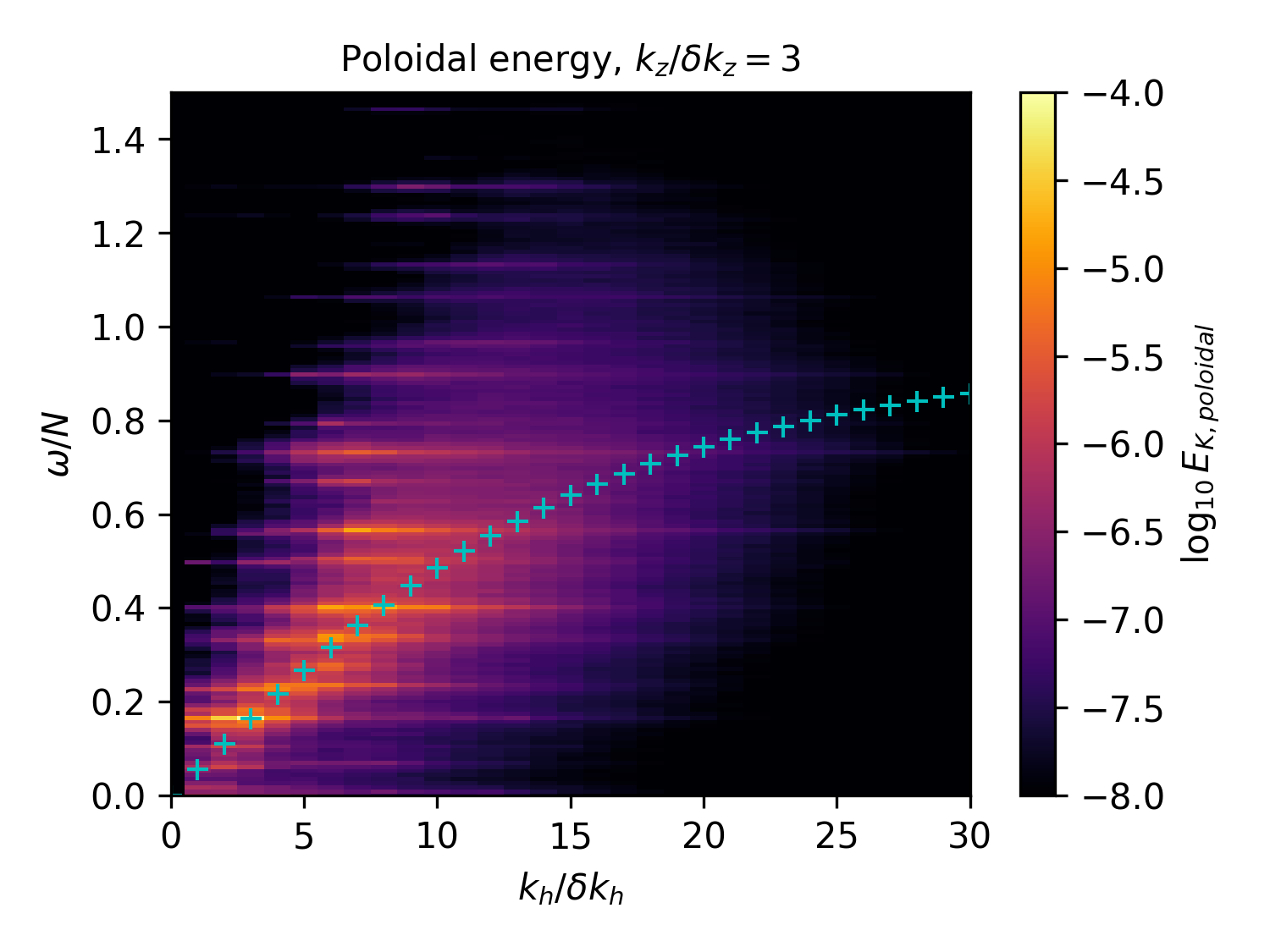}
\caption{Spatiotemporal spectra of poloidal energy in the $\omega-k_h$ plane at
$k_z/\delta k_z=1,2,3$, for $F = 0.73$ and $a=5$\,cm, at resolution
$480\times480\times80$. Horizontal axes are normalized by the horizontal grid spacing
in Fourier space $\delta k_h=\delta k_x = \delta k_y$, and vertical axes are normalized
by $N$. Colors are mapped to the logarithm of the poloidal kinetic energy spectrum, and
cyan crosses represent the internal gravity waves dispersion relation.
\label{fig:spatiotemporal-spectra-ikz}}
\end{figure}

Cuts of the spatiotemporal spectra at fixed $k_z$ can be used to further probe the
waves dispersion relation. Figure \ref{fig:spatiotemporal-spectra-ikz} show the
poloidal energy spectra in the $(k_h,\omega)$ for the three first nonzero values of
$k_z$ (i.e.\ $k_z/\delta k_z=1,2,3$, with $\delta k_z = 2\pi/L_z$). The energy is
clearly concentrated around the dispersion relation, which is signaled again by cyan
plus signs. Along the dispersion relation, frequencies that correspond to the peaks in
the temporal spectra appear more energetically. When $k_z$ increases, the dispersion
relation is less visible as the energy is distributed in a wider area around it. This
behaviour with $k_z$ was also observed in \cite{Savaro2020}, and is consistent with the
hypothesis that small scales are more nonlinear, making the dispersion more diffuse as
wavenumbers increase.

\begin{figure}
\centerline{
\includegraphics[width=0.48\textwidth]{%
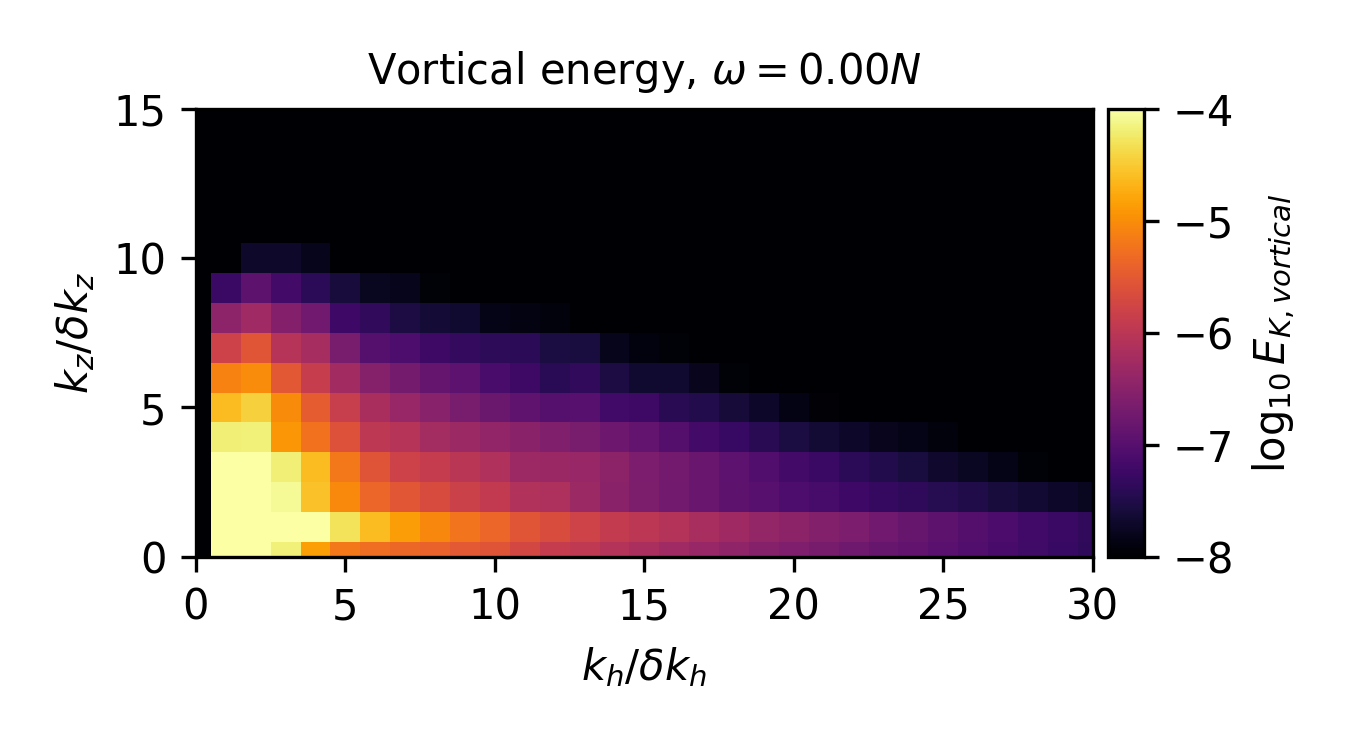}
\includegraphics[width=0.48\textwidth]{%
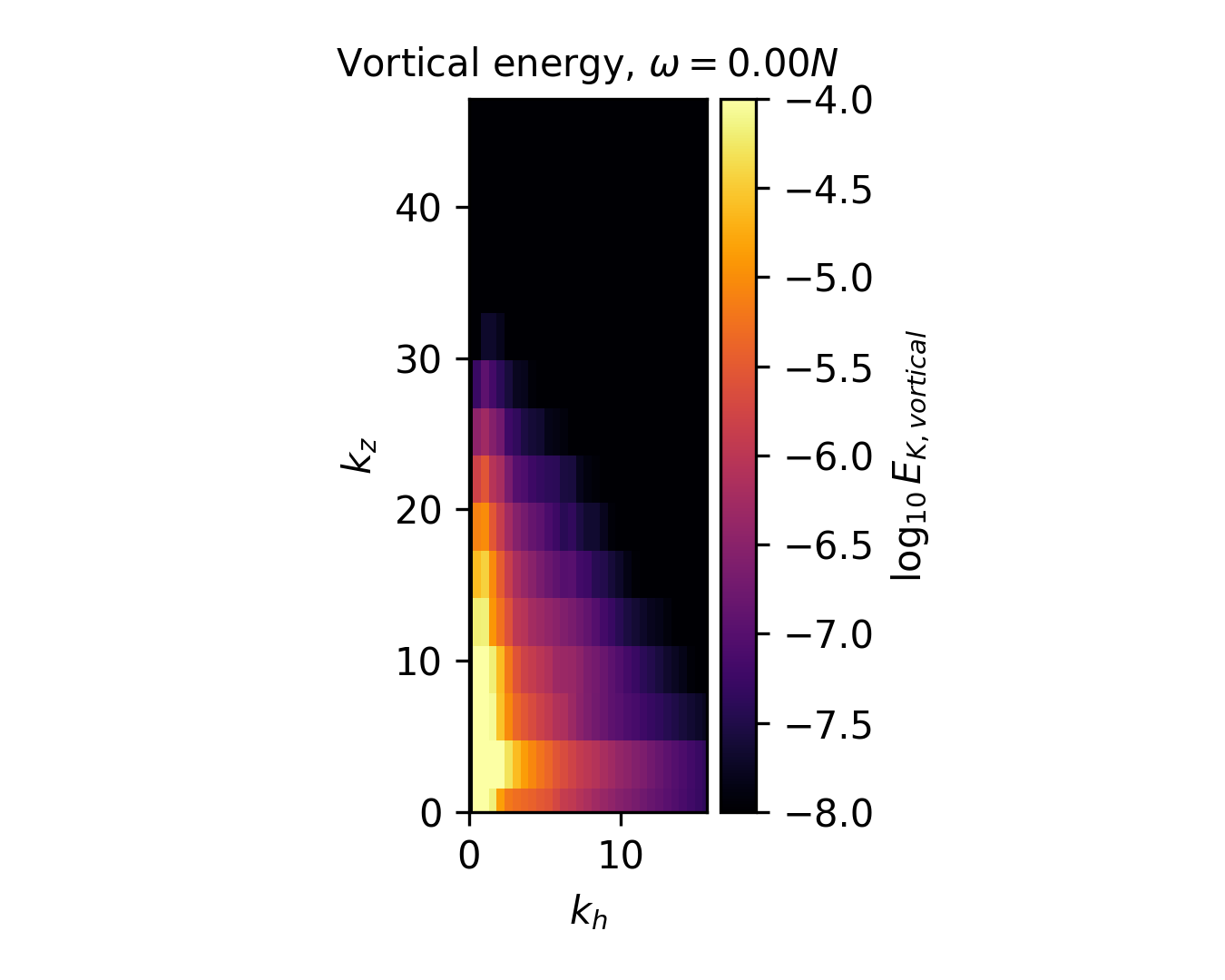}
}
\caption{Spatiotemporal spectra of vortical kinetic energy in the $k_z-k_h$ plane at
$\omega=0$\,rad/s, for $F = 0.73$ and $a=5$\,cm, at resolution $480\times480\times80$.
Left: axes are normalized by the grid spacing in Fourier space. Right: axes in
dimensional units, with the same scale on the $k_z$ and $k_h$ axes. On both figures,
the colors are mapped to the logarithm of the vortical kinetic energy spectrum.
\label{fig:spatiotemporal-spectra-omega-0}}
\end{figure}

Finally, the study of vortical energy in the $(k_h,k_z)$ plane at zero frequency, as
shown in figure \ref{fig:spatiotemporal-spectra-omega-0}, shows high levels of energy
at very large horizontal scales, and with a more extended range of vertical scales.
This is compatible with the observed vortex mode in the mean flow in figure
\ref{fig:vortex}, and in its influence on vertical energy spectra as shown in figure
\ref{fig:spatial-spectra}.

\subsection{Spectral energy budget}

\begin{figure}
\centerline{
\includegraphics[width=0.48\textwidth]{%
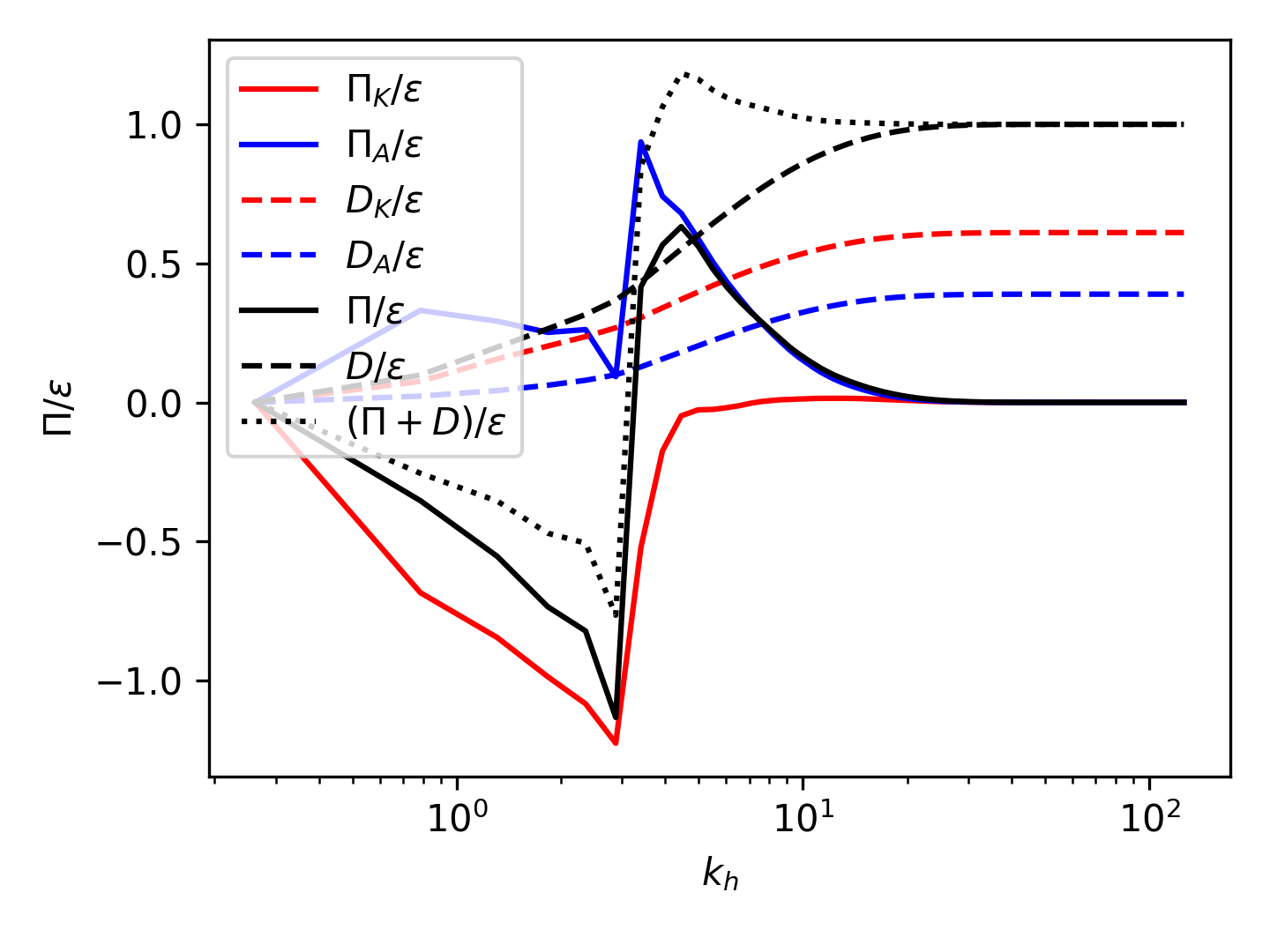}
\includegraphics[width=0.48\textwidth]{%
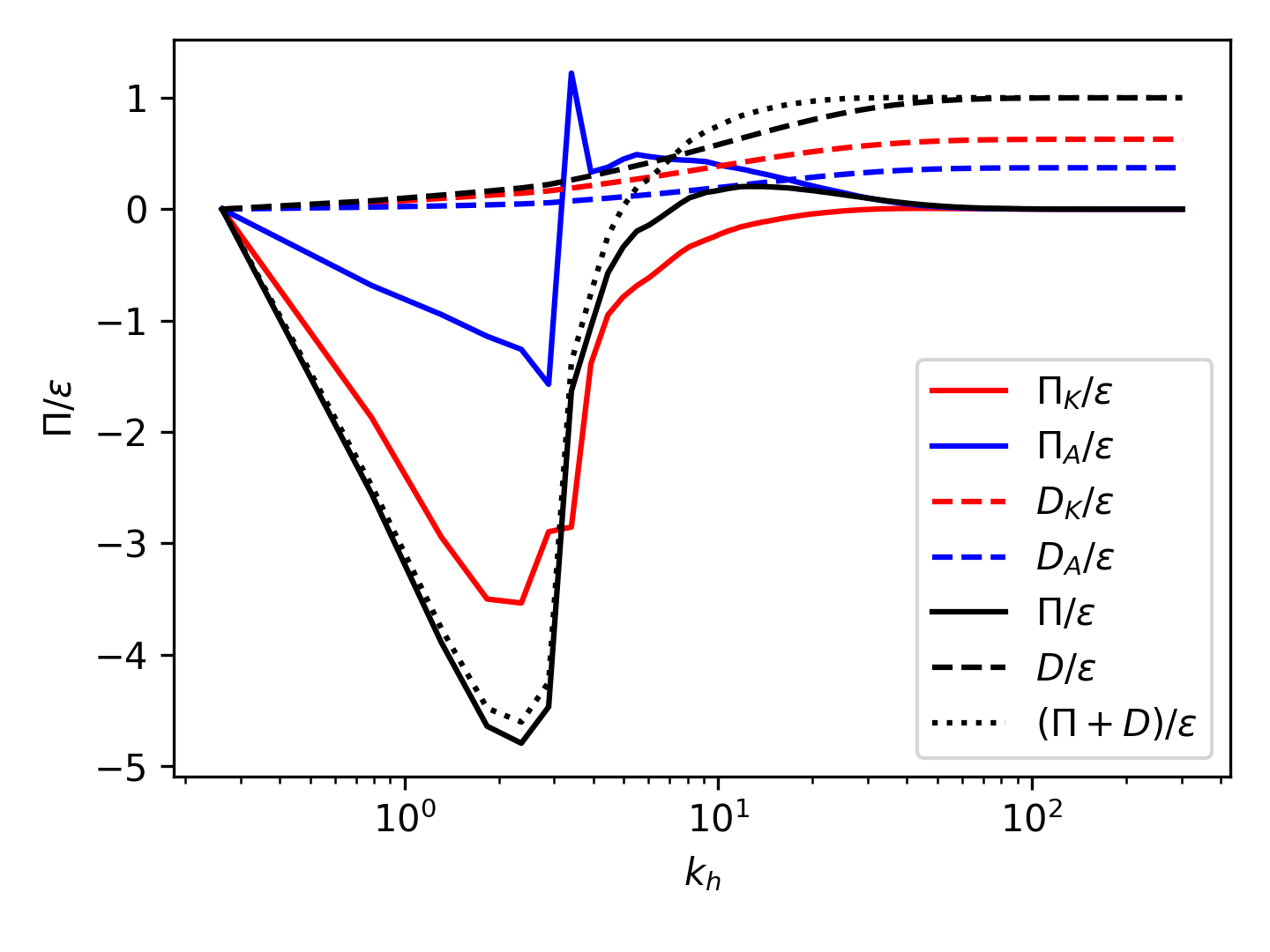}
}
\centerline{
\includegraphics[width=0.48\textwidth]{%
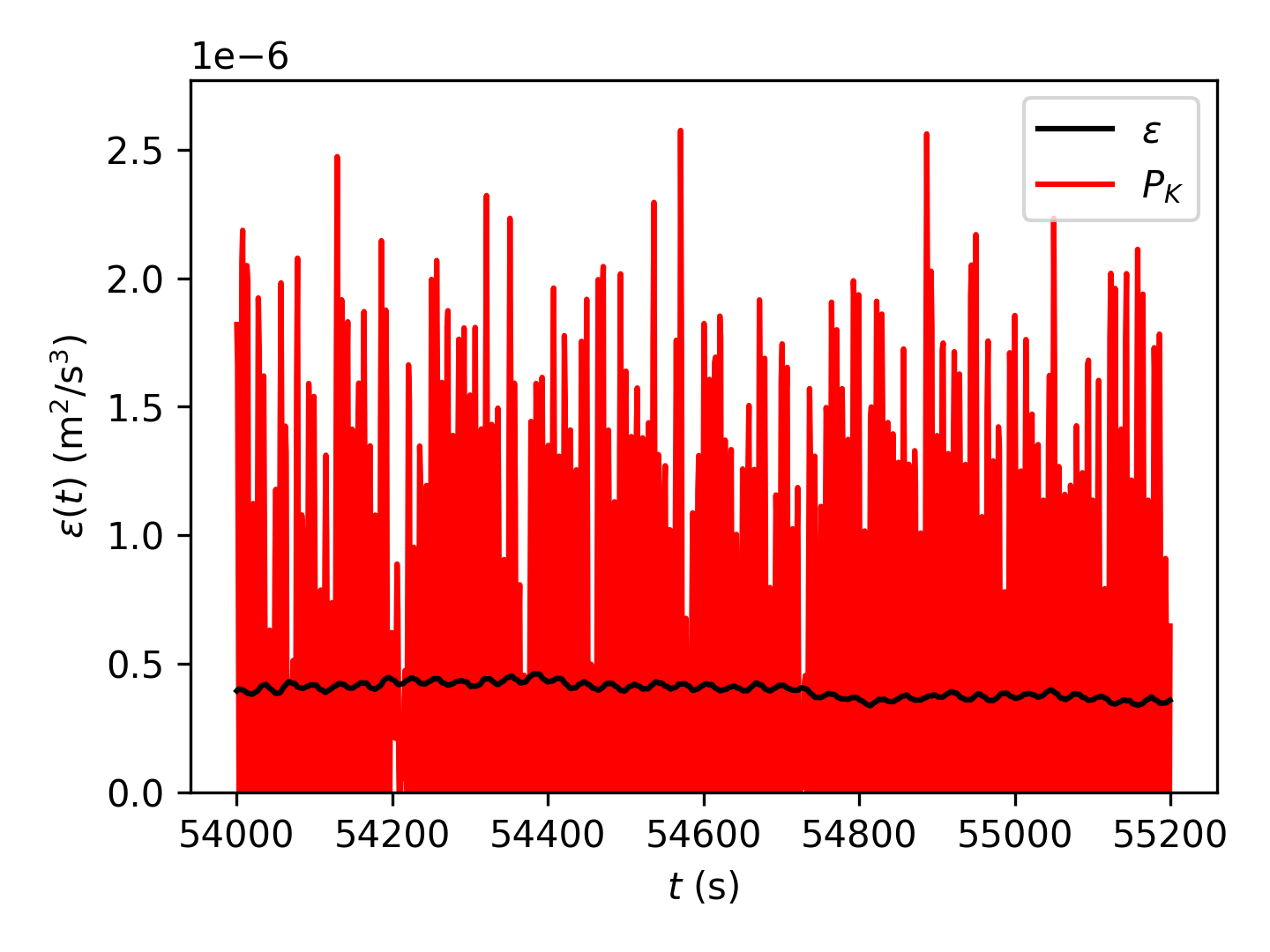}
\includegraphics[width=0.48\textwidth]{%
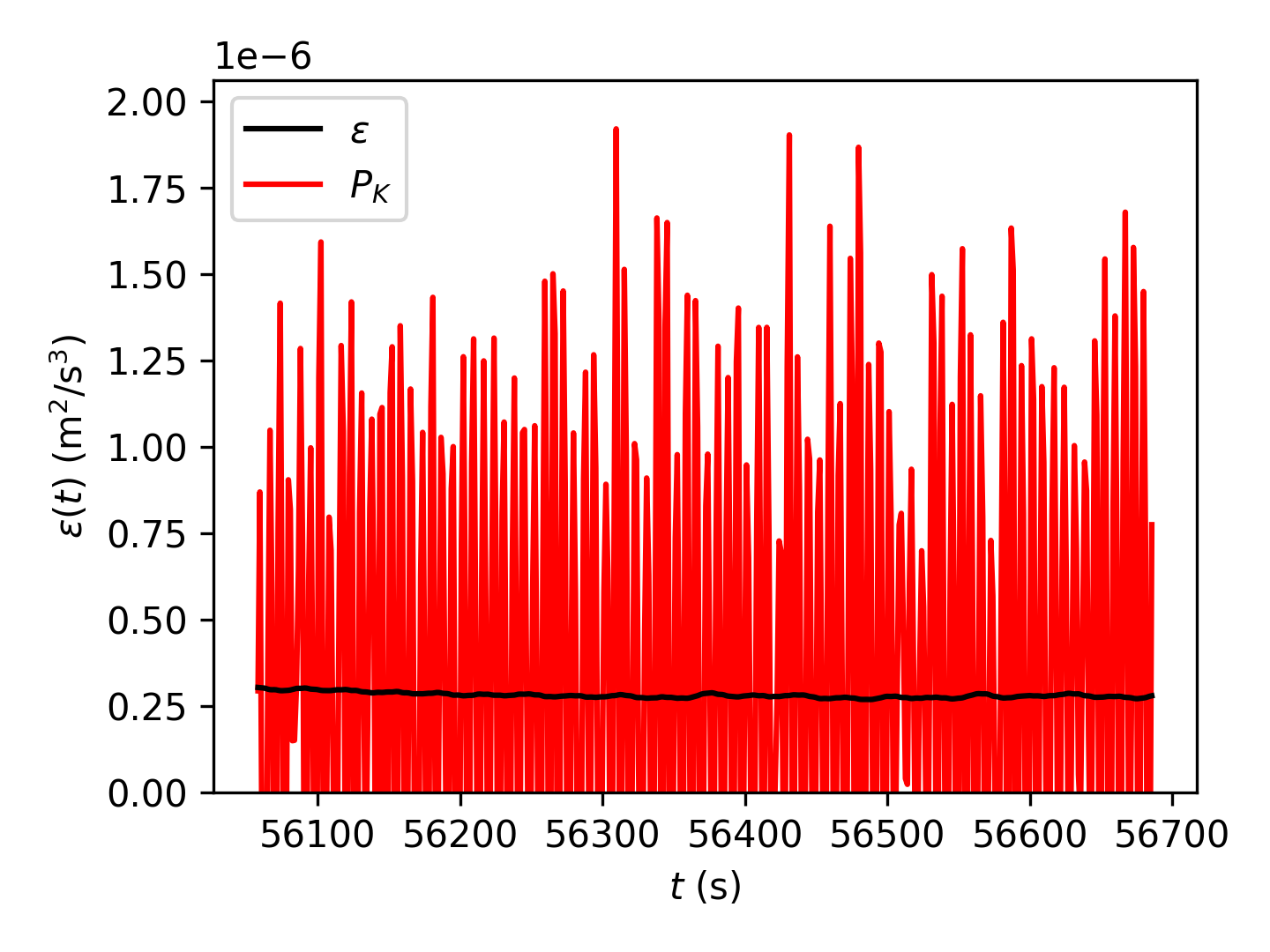}
}
\caption{Top: spectral energy budget for $F = 0.73$ and $a=5$\,cm, at resolutions
$480\times480\times80$ (left) and $1152\times1152\times192$ (right). Solid red, blue
and black lines are respectively the kinetic, potential and total energy fluxes.
Positive values represent fluxes towards the small scales, and negative values towards
the large scales. Dashed red, blue and black lines are respectively the kinetic,
potential and total cumulative dissipations. The dotted black line represent the sum of
the total energy flux and total cumulative dissipation. Bottom : time evolution of the
total energy dissipation (solid black) and injection (dashed red) for the same
simulations. \label{fig:seb}}
\end{figure}

Finally, we study a last aspect of the simulated flow in this first set of parameters,
which is the structure of energy transfers and dissipation in spectral space. These
quantities are extremely difficult to obtain experimentally. For this, we must derive
the evolution equations for the kinetic and potential energy spectra $E_K(\kk,t)$ and
$E_A(\kk,t)$. This is achieved by taking the spatial Fourier transform of the equations
of motion \eqref{ns} and \eqref{buoy}, multiplying them by $\hat{\vv}^*(\kk,t)$ and
$\hat{b}^*(\kk,t)$ respectively and taking the real part. The obtained equations are:
\begin{align}
\partial_t{E_K}(\kk,t) = T_K(\kk,t) - \CKA(\kk,t) - d_K(\kk,t) + P_K(\kk,t)\label{sebK}
\\
\partial_t{E_A}(\kk,t) = T_A(\kk,t) + \CKA(\kk,t) - d_A(\kk,t) \label{sebA},
\end{align}
where $E_K(\kk,t)=|\hat{\vv}|^2/2$ and $E_A(\kk,t)=|\hat{b}|^2/(2N^2)$. The transfer
terms across scales are:
\begin{align}
T_K(\kk,t) = \R\left[\hat{\vv}^* \cdot \mathbb{P}_\bot(\widehat{\vv\times\bomega})\right]%
\label{TK} \\
T_A(\kk,t) = -\frac{1}{N^2} \R\left[\hat{b}^* \widehat{\vv \cdot \bnabla{b}} \right],%
\label{TA}
\end{align}
where $\R$ denotes the real part and $\mathbb{P}_\bot$ is the orthogonal projector onto
the plane orthogonal to $\kk$. By definition, we have
$\sum{T_K(\kk,t)}=\sum{T_A(\kk,t)}=0$. The conversion from kinetic to potential energy
reads:
\begin{equation}
\CKA = -\R\left[{\hat{v}_z}^* \hat{b}\right].
\label{CKA}
\end{equation}
Finally the power injection term is:
\begin{equation}
P_K(\kk,t) = \R\left[\hat{\vv}^* \cdot \hat{\ff}\right],
\label{PK}
\end{equation}
and the dissipation terms are:
\begin{align}
d_K(\kk,t) = f_d(\kk) E_K(\kk,t) \label{DK} \\
d_A(\kk,t) = f_d(\kk) E_A(\kk,t), \label{DA}
\end{align}
where $f_d(\kk)=\nu_2 k^2 + \nu_4 k^4$ is the dissipation frequency, equal for both
types of energy as we chose $\text{Pr}_2=\text{Pr}_4=1$. The equations \eqref{sebK} and
\eqref{sebA} are called the spectral energy budget equations. In stationary state, the
time derivatives of both energy types are zero in average, and the statistics of the
spectral quantities in the left hand side of equations \eqref{sebK} and \eqref{sebA} do
not depend on time. We therefore compute the transfer and dissipation terms of the
spectral energy budget and average them over the length of the time window for the run
at resolution $1152\times1152\times192$. The results are also summed over $k_z$, and
integrated along $k_h$ in order to get the horizontal spectral fluxes
$\Pi_{K,A}=\int_0^{k_h}{T_{K,A}(k_h')}\diff{k_h'}$ and cumulative dissipations
$D_{K,A}=\int_0^{k_h}{d_{K,A}(k_h')}\diff{k_h'}$.

The obtained quantities are shown in figure \ref{fig:seb} (left). Contrary to what we
would expect in the picture of a direct energy cascade towards the small scales, which
would be a constant energy flux between the forcing and dissipative scales, we rather
observe a peculiar shape of the fluxes terms. Indeed, the negative kinetic energy flux
(solid red line) at large scales indicates an average transfer towards the largest
scales, whereas the positive flux above the forcing indicates a slight transfer towards
the small scales. Such shape of the energy fluxes, exhibiting two separated flux loops,
were already observed in 2D simulations of stratified turbulence \cite{boffetta2011,
Linares2020}. In addition, we show the behaviour of the sum of the total flux and
dissipation (black dotted line), which should be constant in stationary state out of
the range of scales where the forcing acts. We see here that this sum is hardly
constant above the dissipative scales, indicating either that the system is not
perfectly stationary or that the forcing is still acting at those scales. Note that the
forcing being oscillatory, and not random, the system cannot be strictly stationary. A
more thorough study of the time evolution of spatial spectra (data not shown) shows
that the statistical stationarity is satisfying, thus suggesting that the forcing term
is the more probable cause of the peculiar shape of the total flux and dissipation.
Indeed, it must be recalled that the forcing scheme used here is local in space and has
no preferred horizontal scale. We therefore do not expect the influence of the forcing
terms to be local in spectral space. The picture drawn by the left panel of figure
\ref{fig:seb} rather suggests that the forcing term used here acts in a complicated
manner over a large range of scales, which is still not well separated from the
dissipative scales at this resolution. In particular, the decrease of the sum of the
total flux and dissipation at large scales indicates that the forcing actually removes
energy in average at those scales.

A time series of the power injection, shown on the right panel of figure \ref{fig:seb}
actually shows frequent negative excursions that are large compared to the mean
injected power, which shows that the dissipative action of the forcing plays an
important role in this type of flow. It should be noted that this fluctuating aspect of
the energy injection rate is typical of experimental setups such as the one used in
\cite{pinton1999power, cadot2010linear,aumaitre2013fluctuations, Savaro2020}, where the
power injection is not a controlled experimental parameter. In such experimental
systems the fluctuation of injected power can overcome the average by several orders of
magnitude, which makes statistical convergence extremely challenging. In contrast, in
numerical simulations for turbulence, the forcing scheme can easily be designed in
order to impose a constant energy injection rate. However the goal of the present study
was to mimic the experimental setup of \cite{Savaro2020}, in order to try and reproduce
the results and give a complementary description of the studied flows. As a
consequence, we do not control the power injection here, thus it is left to fluctuate
as shown by the right panel of figure \ref{fig:seb}.

\section{Exploration of other realistic parameters}

\subsection{Stronger forcing for $F = 0.73$: $a=10$\,cm}

The simulations that are presented in section \ref{sec:like-exp}, though close to the
experiments in terms of parameters, correspond to a regime of low buoyancy Reynolds
numbers, as both second and fourth order buoyancy Reynolds numbers stay slightly
smaller than unity. In order to explore regimes at higher buoyancy Reynolds number, we
performed another set of simulations with a stronger forcing, namely using a forcing
amplitude of ten centimeters. As shown on table~\ref{table_simul}, doubling the
amplitude like so increases the nonlinearity while keeping Froude numbers of the order
of $10^{-3}$, allowing for higher buoyancy Reynolds numbers such as $\R_2\simeq5$.
Experimentally, ten centimeters was the maximum amplitude value that was tried during
the study described in \cite{Savaro2020}. However, this experiment was not included in
the article because of the strong mixing mixing of salt leading to the quick
destruction of the average stratification in the tank. In the present simulations the
\bv{} frequency is a fixed parameter, implying that the linear average stratification
is not subject to mixing. As a consequence, the forcing amplitude could practically be
set to values larger than ten centimeters in our simulations, but we choose in the
present study to limit ourselves to $a=10$\,cm in order to stay into regimes that are
comparable with the experiments.


\begin{figure}
\centerline{
\includegraphics[width=0.48\textwidth]{%
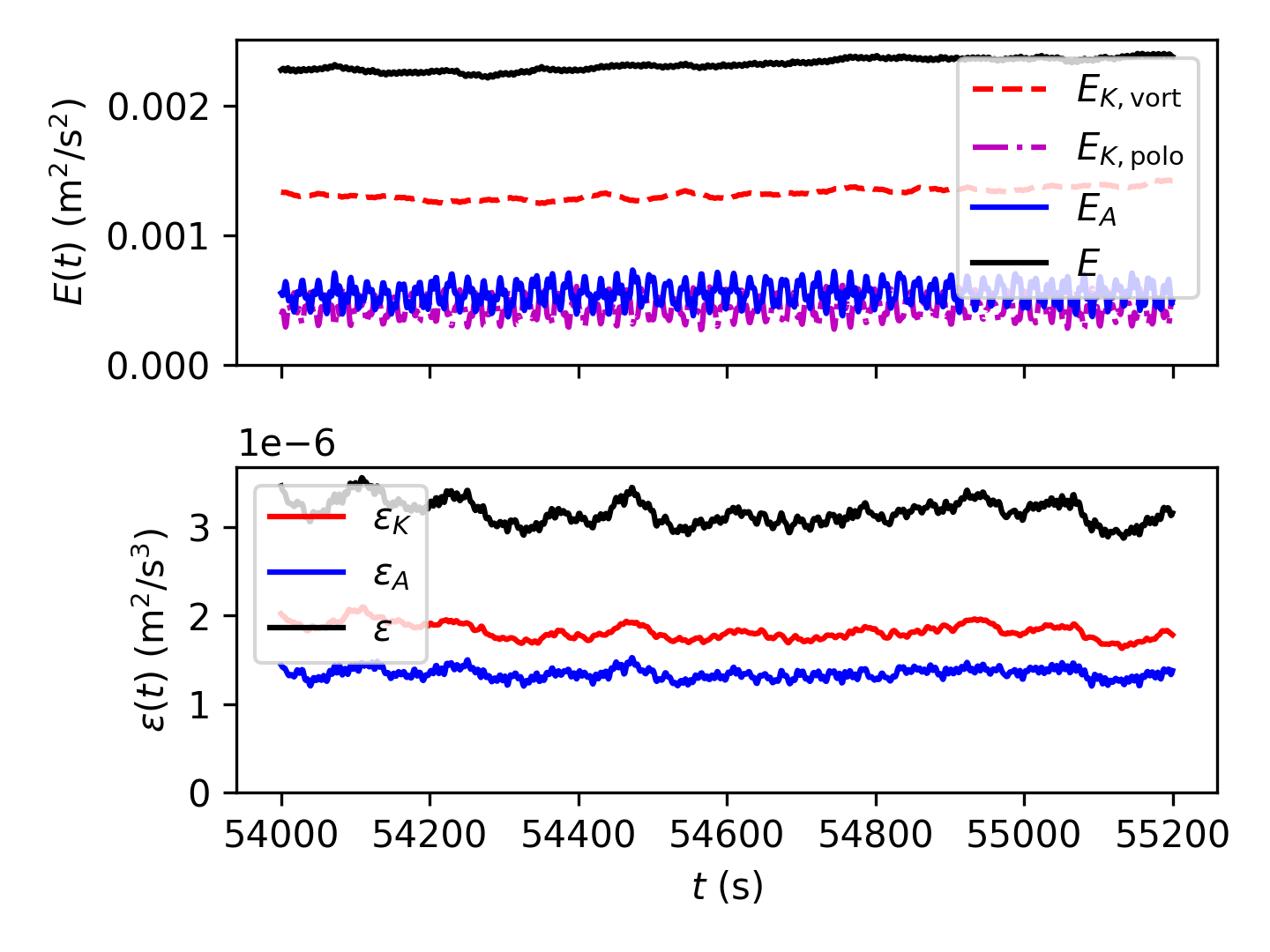}
\includegraphics[width=0.48\textwidth]{%
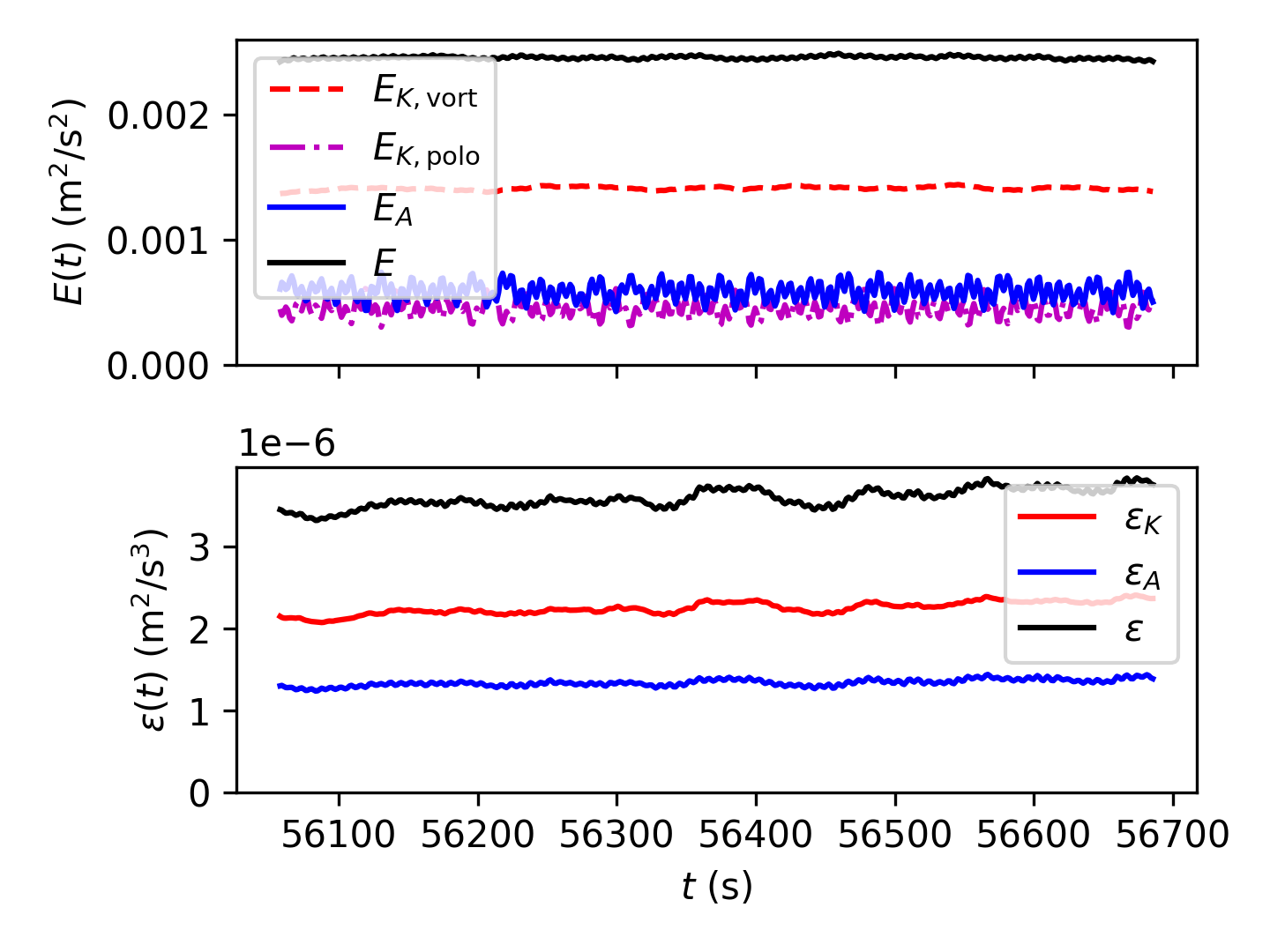}
}
\caption{Time evolution of the energy components (top row) and dissipation (bottom row)
for parameters $F = 0.73$ and $a=10$\,cm. Data are shown for two different resolutions
: $480\times480\times80$ (left column) and $1152\times1152\times192$ (right column),
over a statistically stationary period. \label{fig:spatial-mean-stronger}}
\end{figure}

Figure \ref{fig:spatial-mean-stronger} shows the evolution of energy components and
dissipation for two runs at $a=10$\,cm, at resolutions $480\times480\times80$ (left)
and $1152\times1152\times192$ (right). We note that total energy levels are roughly
three times higher than when forcing at $a=5$\,cm, and total dissipation levels are one
order of magnitude higher. This should be expected, as forcing with a larger amplitude
means injecting more energy into the flow. However, the repartition of energy between
components is more unexpected, as vortical energy levels now represent a much higher
share of the total kinetic energy, with up to 80\% of kinetic energy being vortical
energy, compared to around 50\% for $a=5$\,cm. As shown in table~\ref{table_simul}, the
simulations for this stronger forcing are still very strongly stratified ($F_h \simeq
2\times10^{-3}$) but compared to the weaker forcing ($\R \simeq 0.5$) the buoyancy
Reynolds number is now larger than 1 ($\R \simeq 6$). Therefore, we can expect to see
hints of a transition which could happen for $\R \gtrsim 10$, as in strongly stratified
turbulence forced by vortices \cite{brethouwer_billant_lindborg_chomaz_2007}.


\begin{figure}
\centering
\includegraphics[width=0.65\textwidth]{%
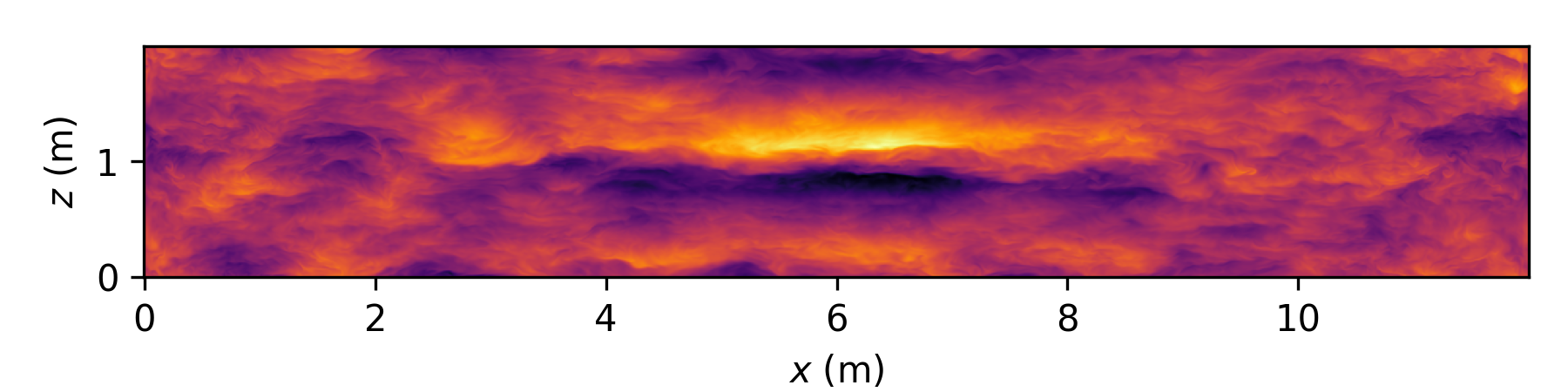}
\includegraphics[width=0.65\textwidth]{%
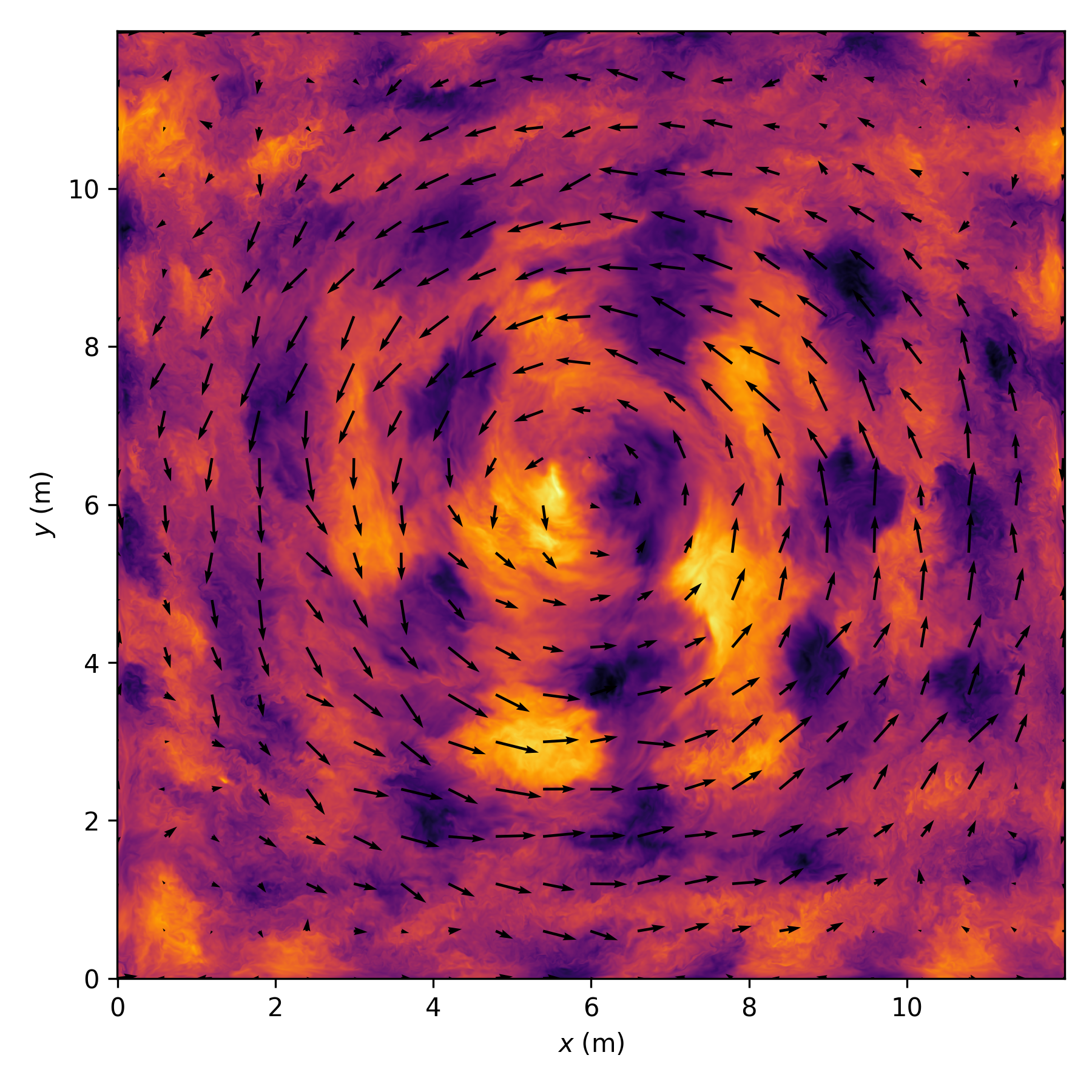}
\caption{Vertical (top) and horizontal (bottom) cross-sessions. Same as
figure~\ref{fig:phys-fields-horiz} ($F =0.73$ ) but for $a=10$\,cm. The resolution is
$2304\times2304\times384$.\label{fig:phys-fields-horiz-stronger}}
\end{figure}

Snapshots of the buoyancy field are displayed in
figure~\ref{fig:phys-fields-horiz-stronger}. Compared to
figure~\ref{fig:phys-fields-horiz}, the small scale structure appear much stronger, in
particular at the periphery of the numerical domain. The velocity field is even more
dominated by the horizontal vortices.


\begin{figure}
\centerline{
\includegraphics[width=0.5\textwidth]{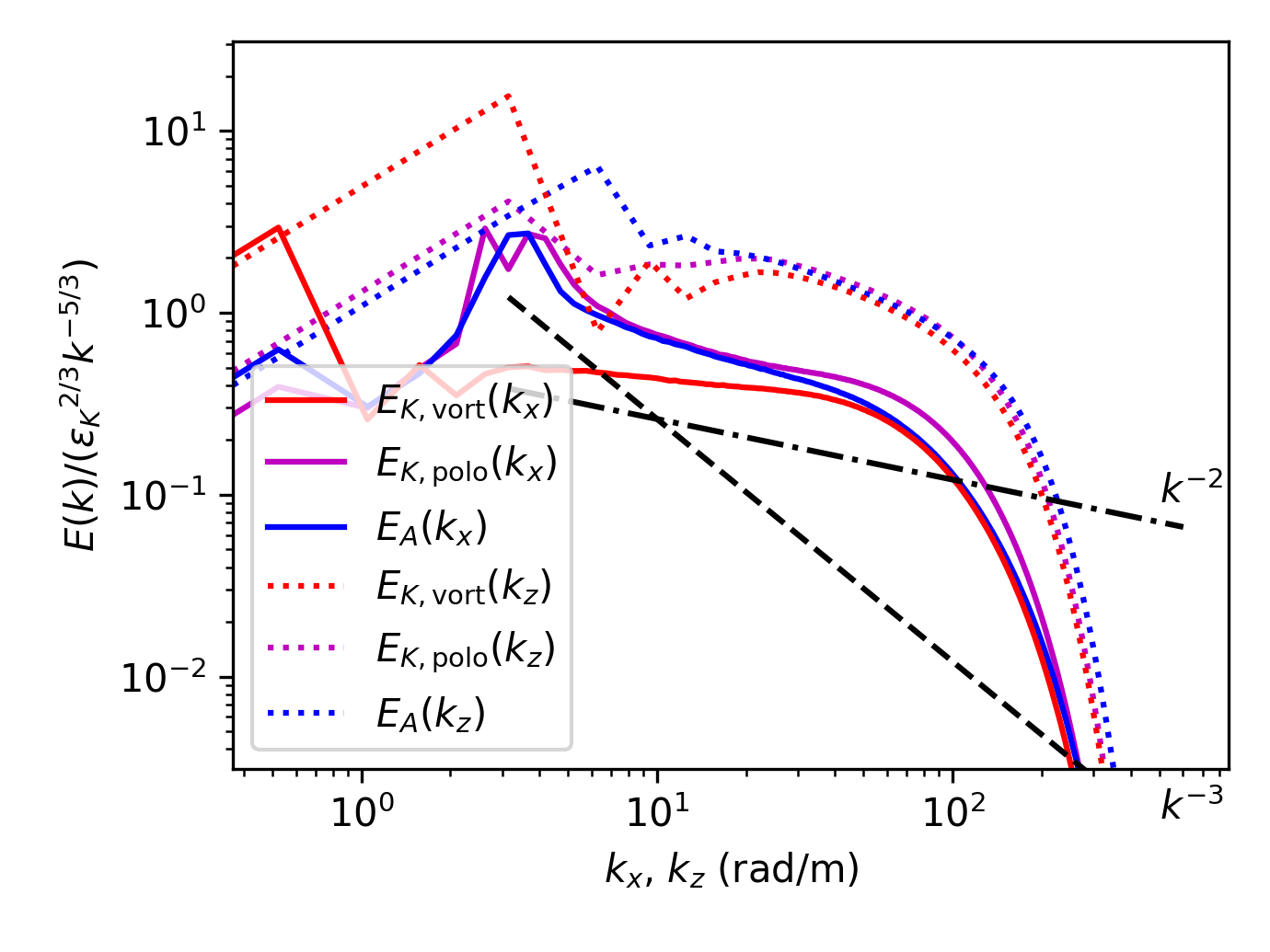}
}
\caption{One-dimensional compensated spatial spectra. Same as
figure~\ref{fig:spatial-spectra} ($F = 0.73$) but for $a=10$\,cm, at resolution
$2304\times2304\times384$ ($\eta\kmax = 0.46$). \label{fig:spatial-spectra-stronger}}
\end{figure}

The compensated spatial spectra are shown in figure~\ref{fig:spatial-spectra-stronger}.
The horizontal and the vertical spectra are also very different, corresponding to
strongly anisotropic flows. The horizontal spectrum of vortical energy (red solid
curve) is nearly flat over more than one decade. This ${k_h}^{5/3}$ scaling law is
consistent with a transition to a regime similar to the LAST regime at higher buoyancy
Reynolds number. However, the LAST regime is associated with equal poloidal and
vortical horizontal spectra, which is not observed here. This might be a finite
buoyancy Reynolds number effect since these two spectra are indeed nearly equal at the
smallest scales.


\begin{figure}
\centering
\includegraphics[width=0.75\figwidth]{%
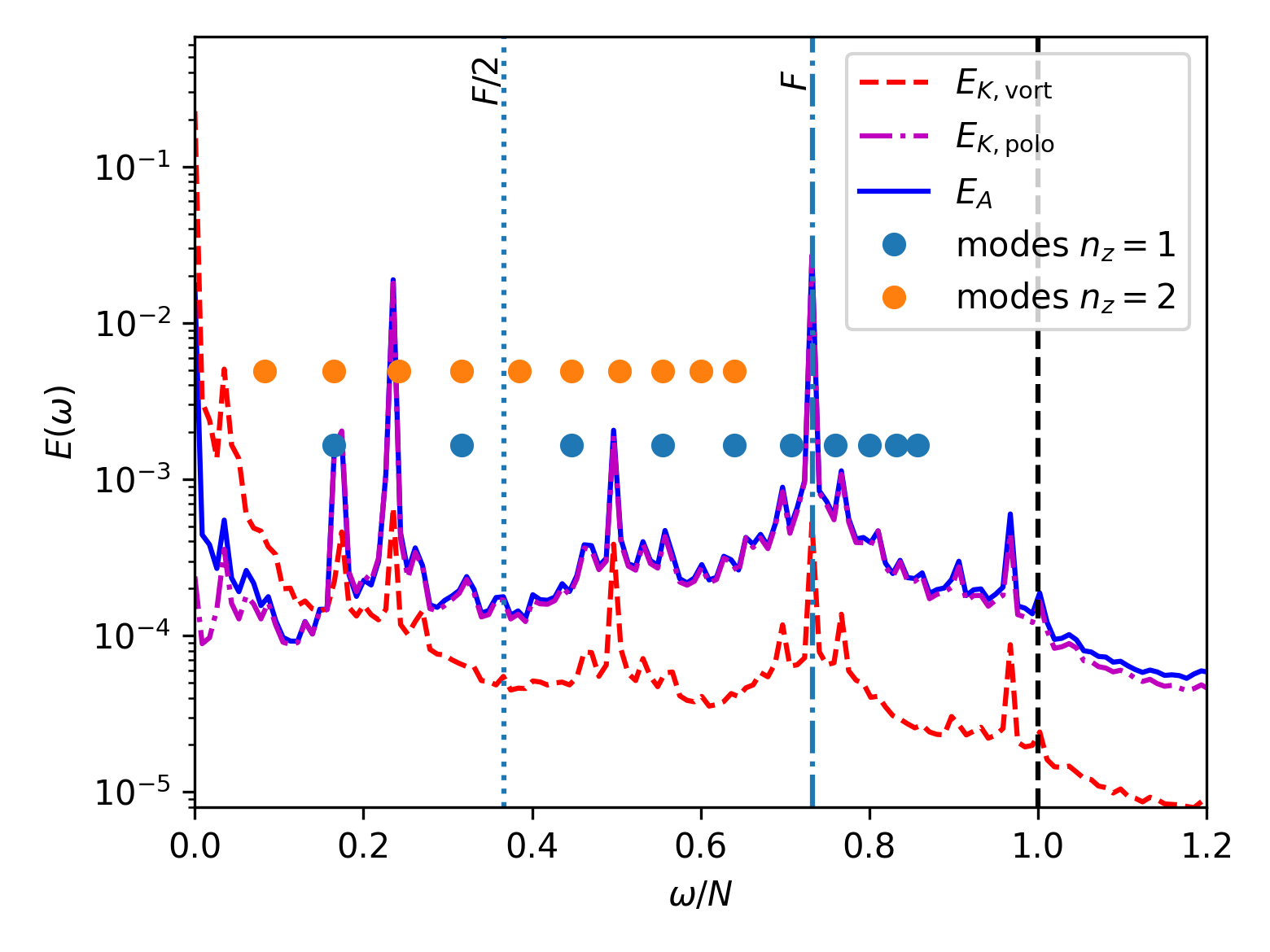}
\caption{Temporal spectra as a function of the normalized frequency $\omega/N$. Same as
figure~\ref{fig:temporal-spectra} but for parameters $F = 0.73$ and $a = 10$\,cm, at
resolution $480\times480\times80$. \label{fig:temporal-spectra-stronger}}
\end{figure}

We show in figure \ref{fig:temporal-spectra-stronger} the temporal spectra for the run
with $a=10$\,cm and resolution $480\times480\times80$. The overall aspect of temporal
spectra is similar to what was observed for $a=5$\,cm (figure
\ref{fig:temporal-spectra}), with poloidal and potential energy dominating above
$\omega=0.2N$, both spectra staying quite close to each other and showing rather
constant levels below the forcing frequency, except for a few sharp peaks. Differences
can be seen first in the number of peaks, as fewer peaks are observable than for
$a=5$\,cm. The off-peak energy continuum of axisymmetric waves increases faster than
the peaks level, as observed in experiments when increasing the forcing amplitude
\cite{Savaro2020}. A second difference lays in the slight offset between poloidal and
potential energy around $F/2$ (dotted vertical line), which was almost not observable
for $a=5$\,cm. This larger offset suggests that waves dynamics are departing from
mostly linear interaction. Finally, the vortical energy levels at very low frequency
are also different from what was observed in figure \ref{fig:temporal-spectra}: the
extension of the frequency range where vortical energy dominates is larger, with
vortical energy remaining non-negligible up to the first peak in the poloidal/potential
energy spectra. These higher vortical energy levels at very low frequencies suggests
that the increase in the share taken by vortical energy corresponds to a relative
growth of the slow vortex mode that was observed in figure \ref{fig:vortex}.


\begin{figure}
\centerline{
\includegraphics[width=0.48\textwidth]{%
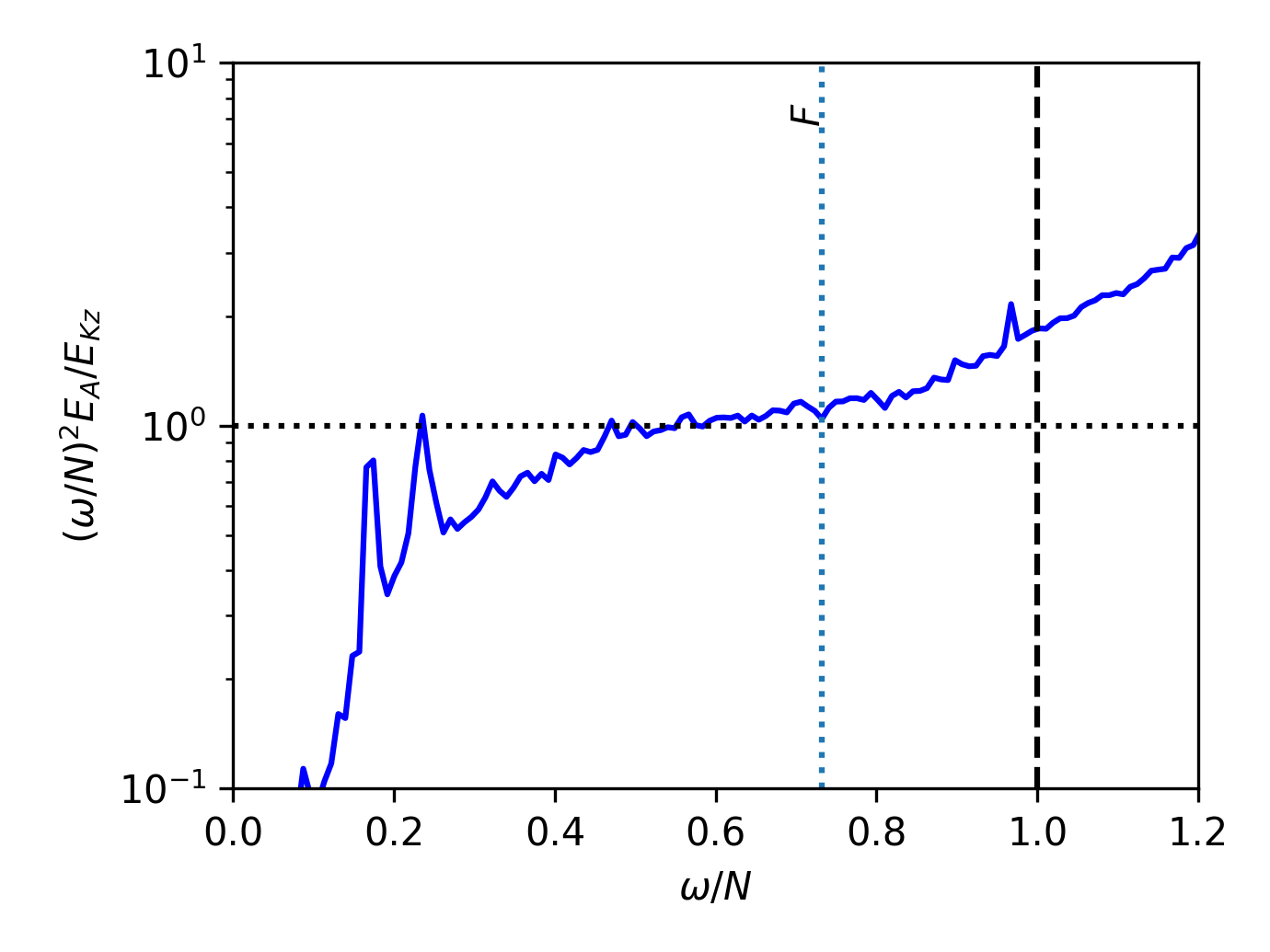}
\includegraphics[width=0.48\textwidth]{%
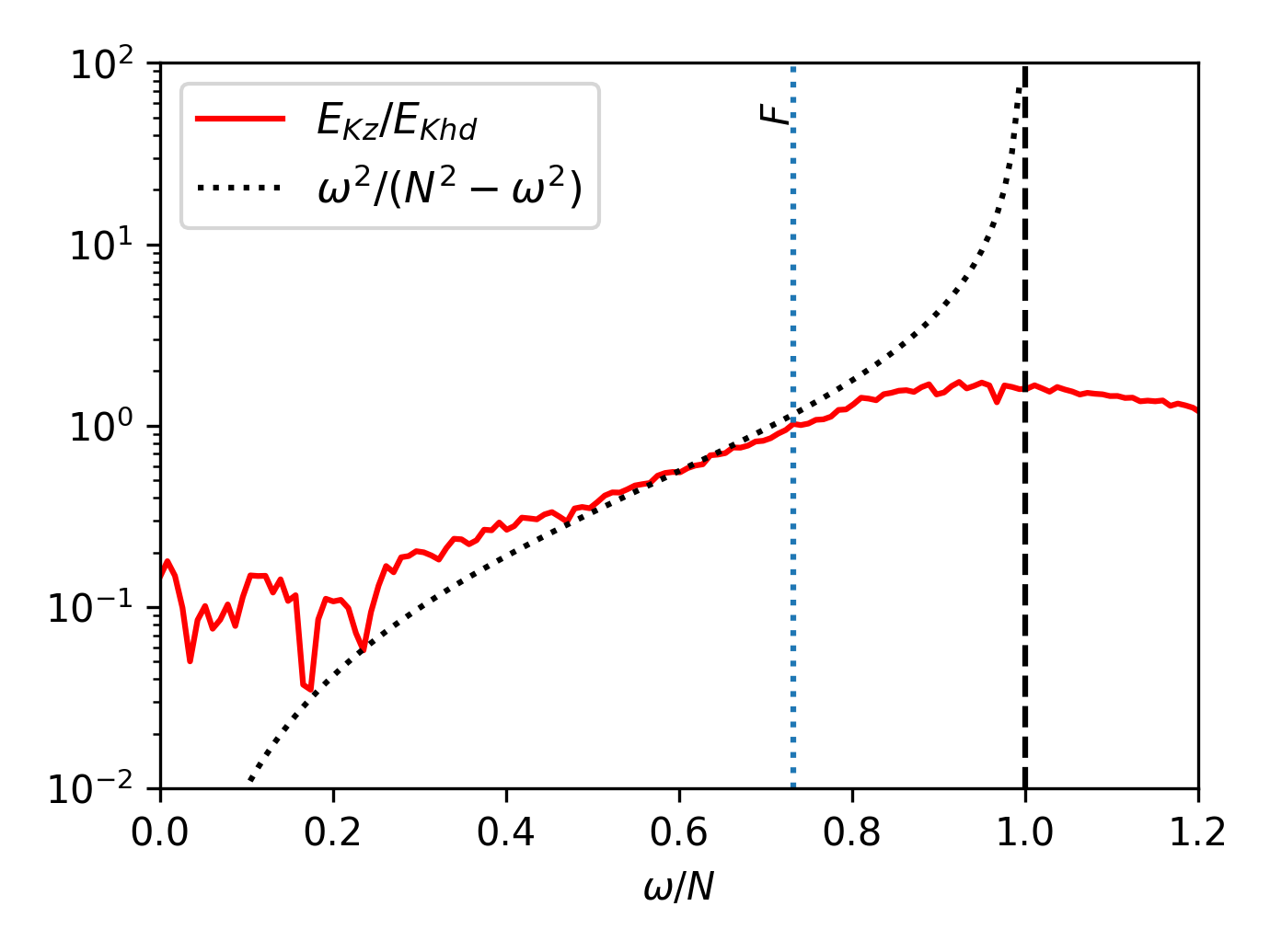}
}
\caption{Linear waves relations in $\omega$-space. Same as
figure~\ref{fig:linear-waves} but for $F = 0.73$ and $a=10$\,cm, at resolution
$480\times480\times80$. \label{fig:linear-waves-stronger}}
\end{figure}

The study of characteristic relations \eqref{ratio_AKz} and \eqref{ratio_Kzh}, as
presented on figure \ref{fig:linear-waves-stronger}, shows that linear wave dynamics
are less at play here than for $a=5$\,cm, which is consistent with higher levels of
nonlinearity due to a stronger energy injection into the system.


\begin{figure}
\centering
\includegraphics[width=0.48\textwidth]{%
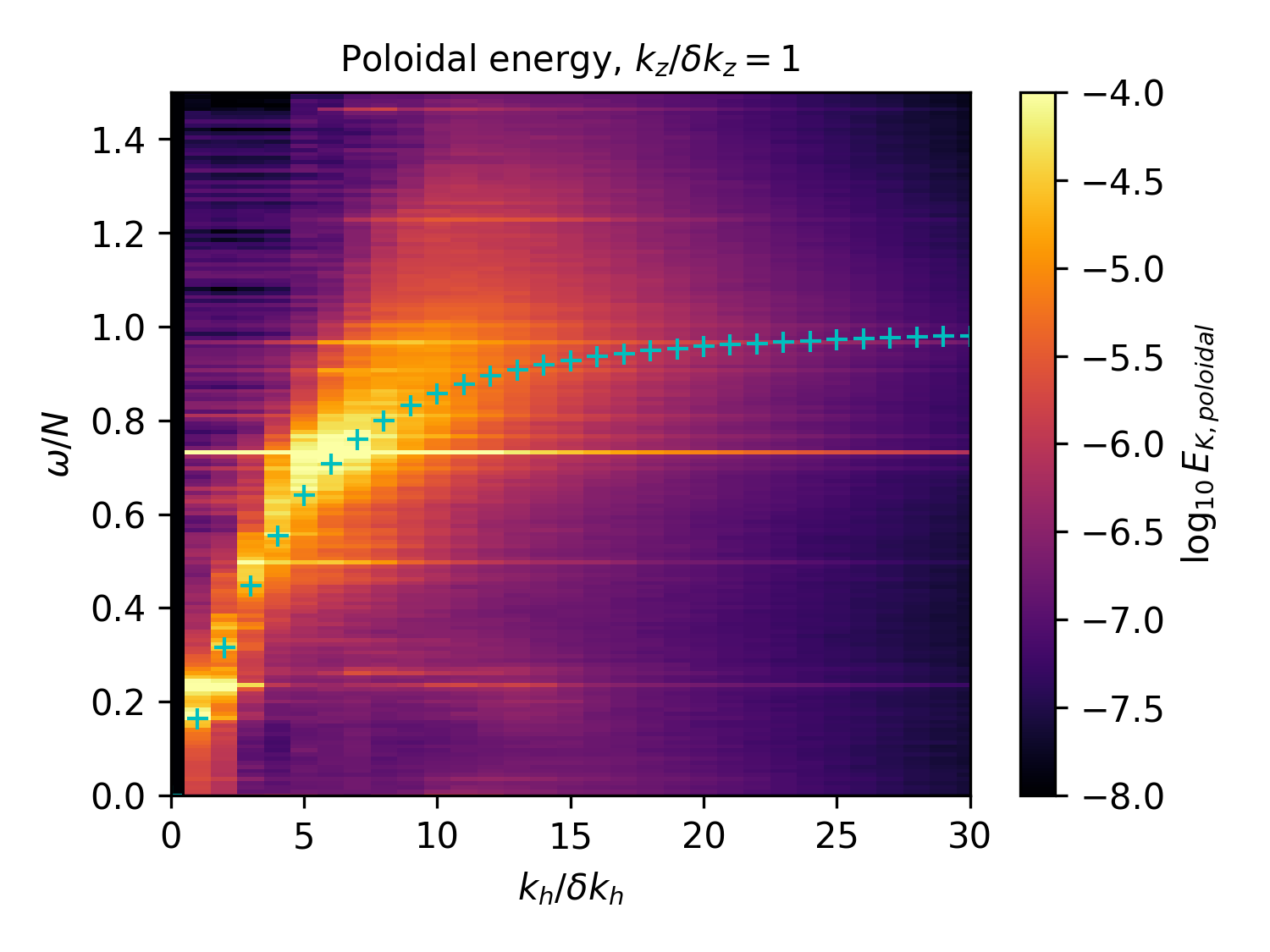}
\caption{Spatiotemporal spectrum of poloidal kinetic energy. Same as
figure~\ref{fig:spatiotemporal-spectra-ikz}, but for $F = 0.73$ and $a = 10$\,cm, at
resolution $480\times480\times80$. The spectrum is shown in the $\omega-k_h$ plane at
$k_z/\delta k_z=1$. \label{fig:spatiotemporal-spectra-stronger}}
\end{figure}

Figure \ref{fig:spatiotemporal-spectra-stronger} (left) shows a cut of the poloidal
energy spatiotemporal spectrum in the $k_h-\omega$ plane, for the first value of $k_z$
(i.e.\ for $k_z/\delta k_z = 1$). We see that the energy remains maximal on the
dispersion relation (cyan plus signs), though it is distributed in a wider region
around the relation. This is compatible with the higher levels of nonlinearity that
were suggested by the temporal analysis. Note also that the spread of energy around the
dispersion relation is wider at higher $k_h$, suggesting that the flow is strongly non
linear at small scales.

\subsection{Slower forcing}
\label{sec:slower}

In all simulations presented in previous sections, we observe the generation of
internal gravity waves at frequencies below the forcing. In order to also study the
behaviour of waves above the forcing, we perform a final set of simulations with a
lower normalized forcing frequency, namely $F=0.4$. Because the velocity of the
oscillating panel of the forcing mechanism is proportional to the motion amplitude $a$
and frequency $F$, the reduction of the forcing frequency induces a decrease of the
power injection into the flow. Simulations with $F=0.4$ and $a=5$\,cm have been
performed (data not shown), though the power injection in this setup was too
inefficient to obtain regimes of dynamics that were comparable to the previous
parameters sets. In order to compensate the decrease of the forcing frequency, the
amplitude is also increased to $a=10$\,cm.


\begin{figure}
\centerline{
\includegraphics[width=0.48\textwidth]{%
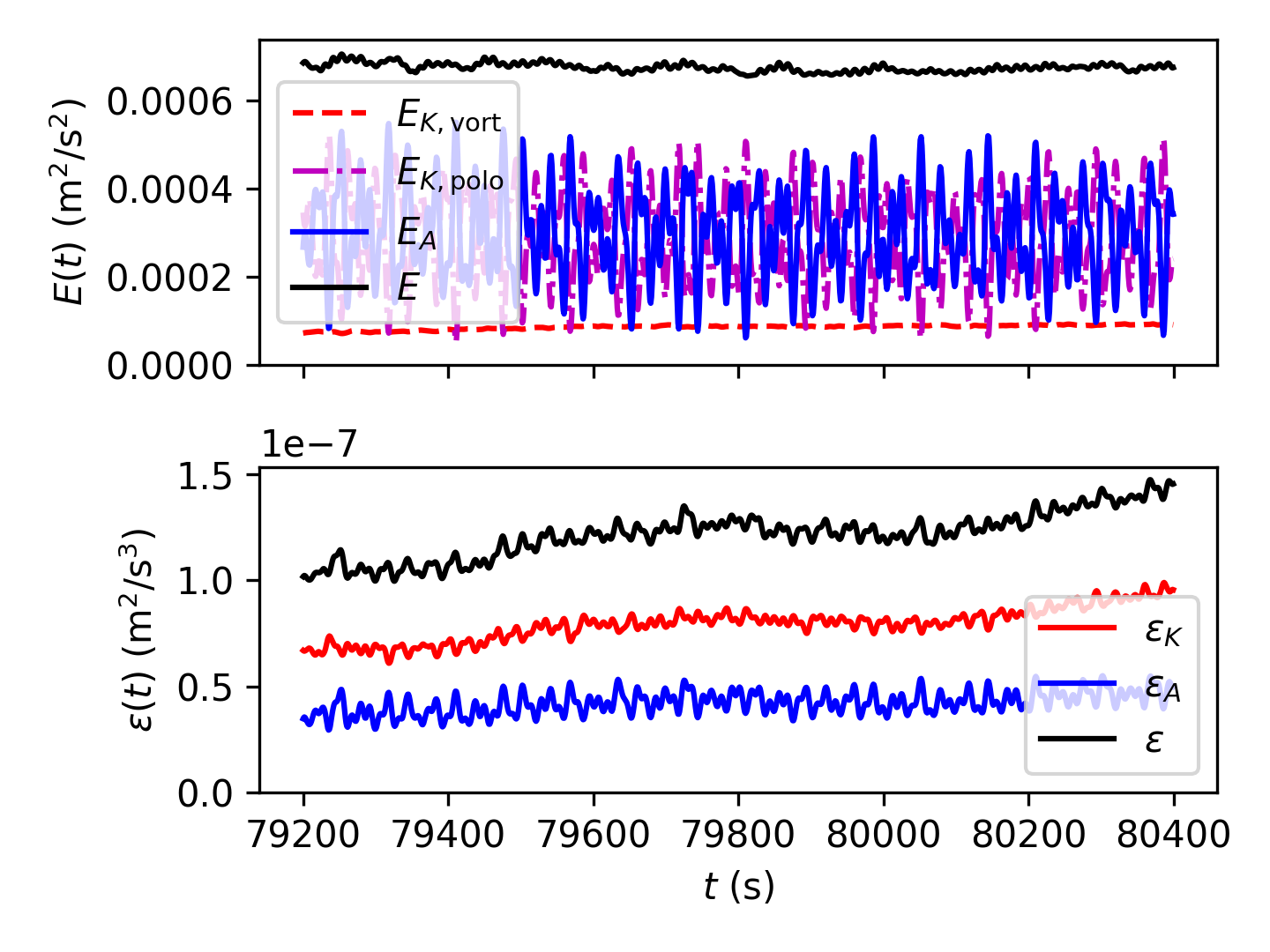}
\includegraphics[width=0.48\textwidth]{%
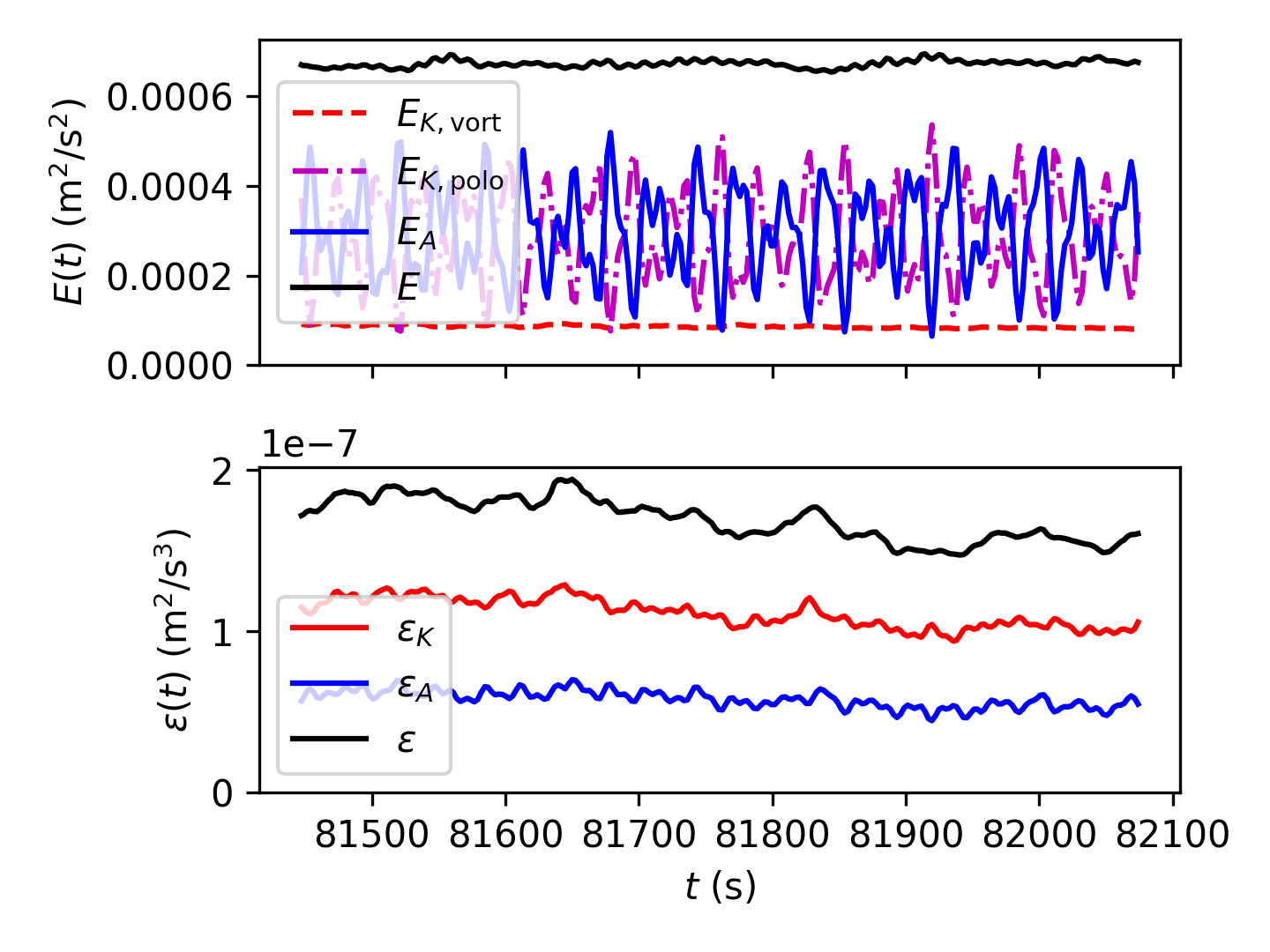}
}
\caption{Time evolution of the energy components (top row) and dissipation (bottom row)
for parameters $F = 0.73$ and $a=10$\,cm. Data are shown for two different resolutions
: $480\times480\times80$ (left column) and $1152\times1152\times192$ (right column),
over a statistically stationary period. \label{fig:spatial-mean-slower}}
\end{figure}

Figure \ref{fig:spatial-mean-slower} shows the evolution of energy components and
dissipations for resolution $480\times480\times80$ and $1152\times1152\times192$. Total
energy and dissipation levels are slightly lower than for $F=0.73$ and $a=5$\,cm,
indicating that the forcing mechanism is injecting slightly less power into the flow.
We also note that the relative level of vortical energy is smaller, with vortical
energy accounting for around 30\% of all kinetic energy.


\begin{figure}
\centerline{
\includegraphics[width=0.5\textwidth]{%
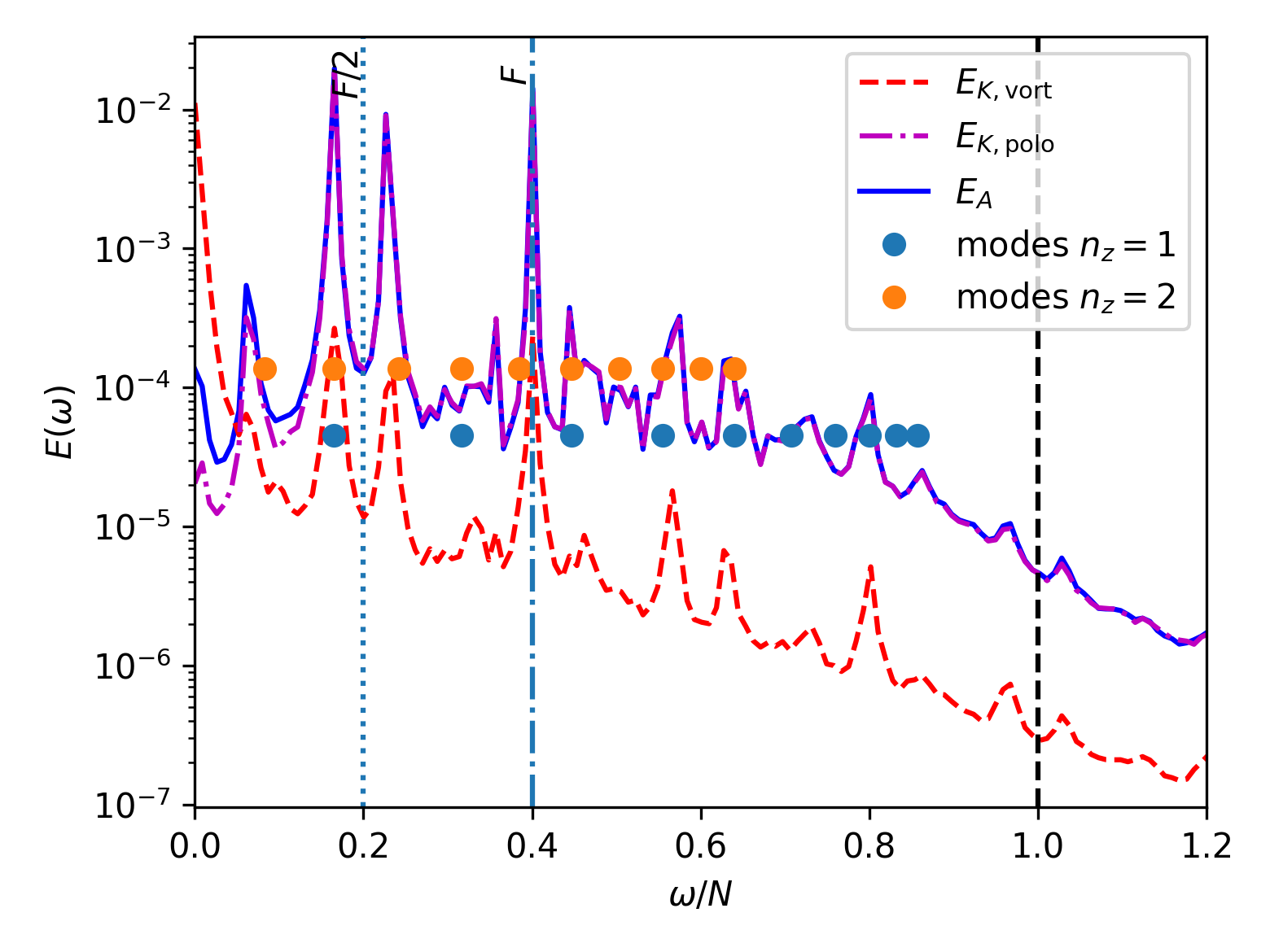}
\includegraphics[width=0.5\textwidth]{%
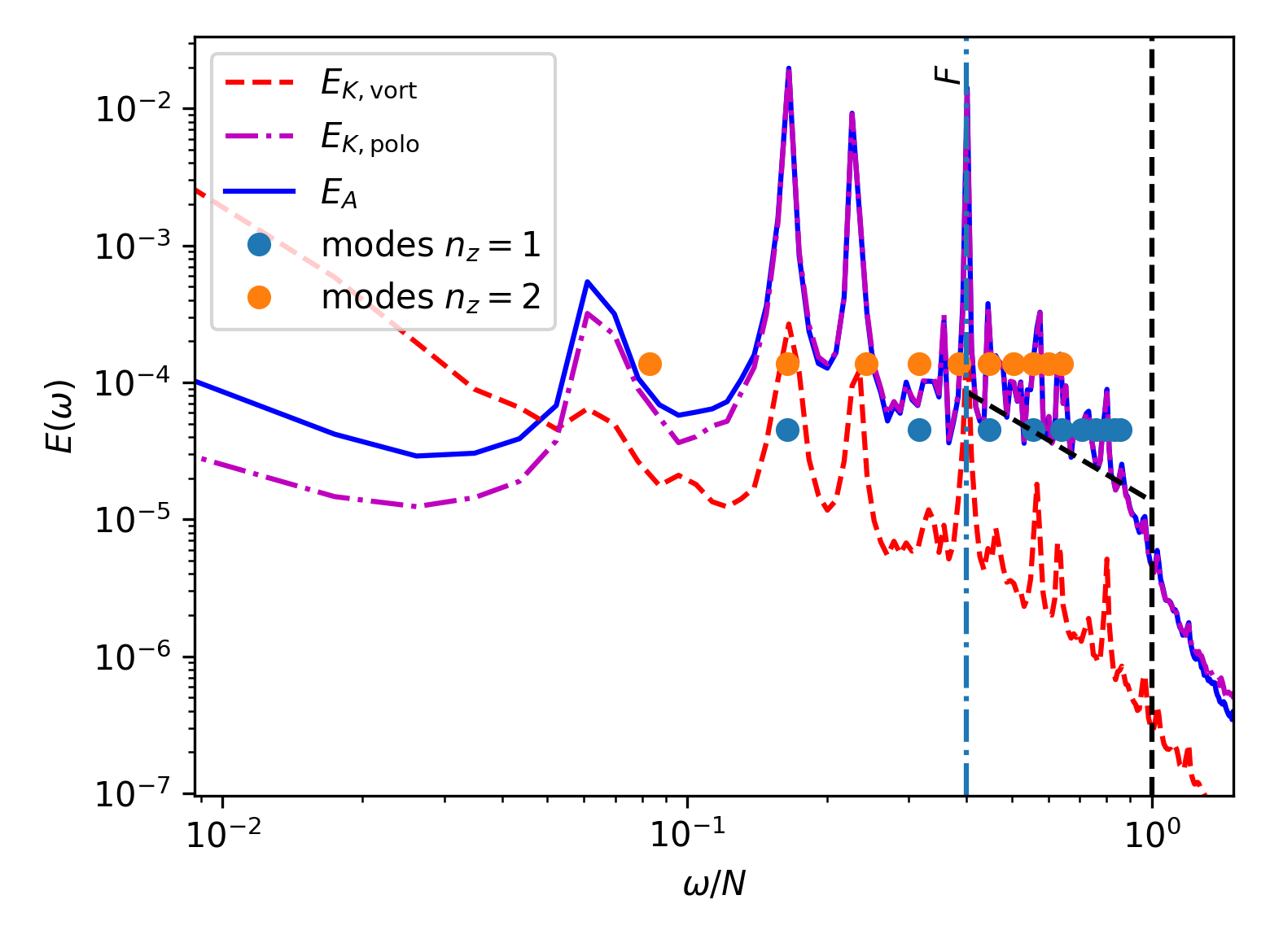}
}
\caption{Temporal spectra as a function of the normalized frequency $\omega/N$. Same as
figure~\ref{fig:temporal-spectra} but for parameters $F = 0.40$ and $a = 10$\,cm, at
resolution $480\times480\times80$. Right: logarithmic scale, with the $\omega^{-2}$
scaling plotted as a black dashed line between $\omega=\omega_f$ and $\omega=N$.
\label{fig:temporal-spectra-slower}}
\end{figure}

Frequency spectra are shown in figure \ref{fig:temporal-spectra-slower}. The overall
aspect of spectra is similar to what was observed in the previous cases, though the
frequency range between $F$ and $N$ is now larger. Below the forcing, poloidal (dashed
red) and potential (solid blue) energy levels are very close to each other, suggesting
linear waves. Toroidal energy is negligible down to very low frequencies below
$\omega=0.1N$, which represents a lower limit than for $a=5$\,cm, showing that the
lower levels of vortical energy are associated with a smaller frequency range of
action. In all spectral components, fewer peaks can be observed than for $a=5$\,cm,
though the three most visible peaks correspond to frequencies that are engaged in a
triadic resonance relation $\omega_a+\omega_b=\omega_c$, where $\omega_a=0.17N$ is
again the frequency of 2D resonant modes such that $n_x=0$, $n_y=n_z$, $\omega_b=0.23N$
and $\omega_c=\omega_f=0.40N$. We note that a triadic resonance relation involving both
the forcing frequency and the frequency of 2D modes such that $n_h=n_z$ is always
observable in all simulations that were performed. This was confirmed by using a
slightly faster forcing frequency $F=0.45$ with the same amplitude (see
table~\ref{table_simul}), showing that the same triadic relation was verified between
the forcing, the first resonant 2D mode and a third frequency $\omega_b=0.28N$ (data
not shown). Above the forcing frequency, poloidal and potential energy are again very
close, indicating that linear internal waves are also generated at those frequencies.
We observe less marked peaks in this range, and a logarithmic scale representation
(left panel of figure \ref{fig:temporal-spectra-slower}) shows that the spectral
behaviour at those frequencies is compatible with a $\omega^{-2}$ Garrett and Munk
spectrum \cite{Garrett1979}. A qualitatively similar spectrum could be described in
experiments \cite{Rodda2022experimental}.


\begin{figure}
\centerline{
\includegraphics[width=0.48\textwidth]{%
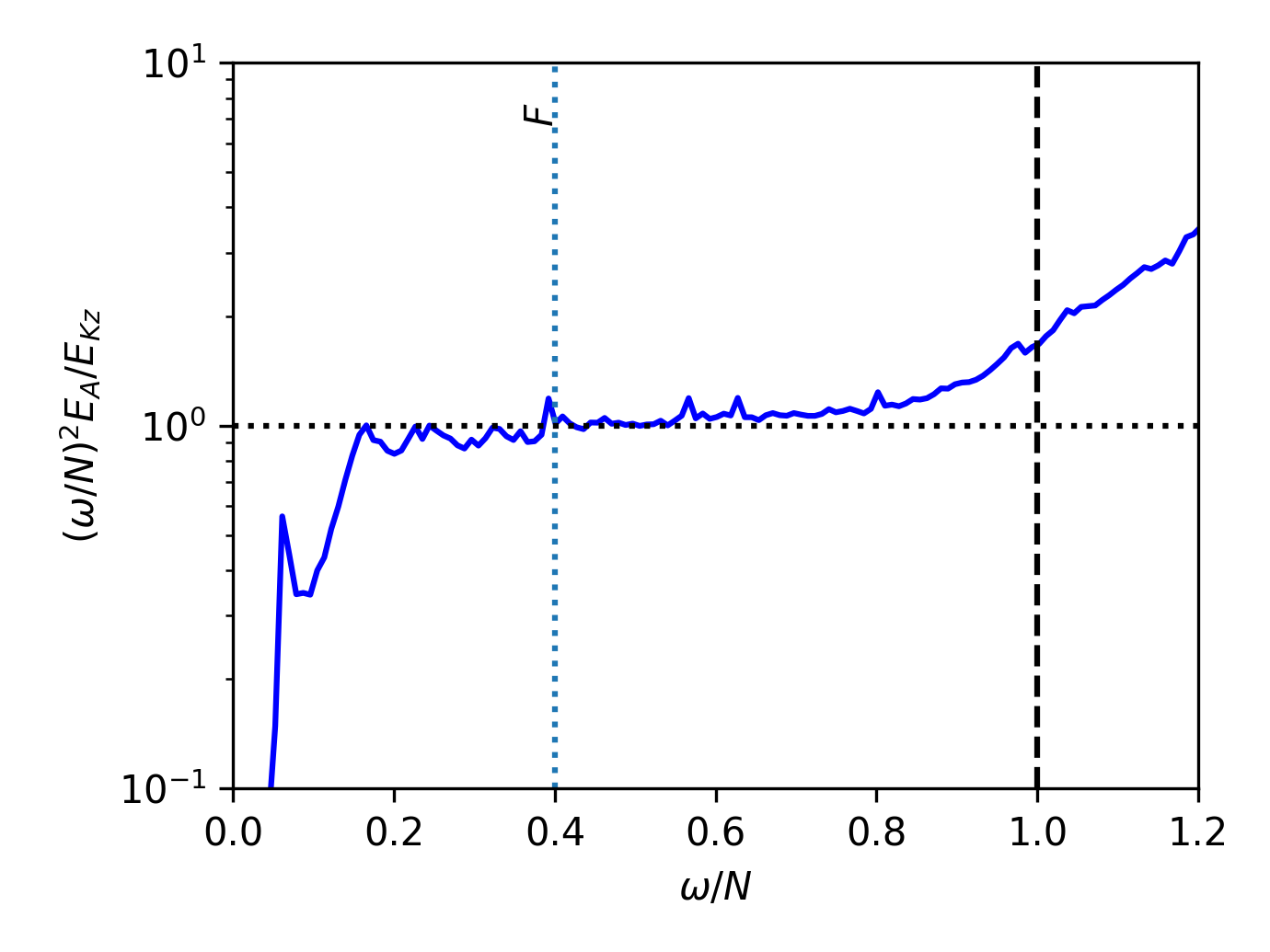}
\includegraphics[width=0.48\textwidth]{%
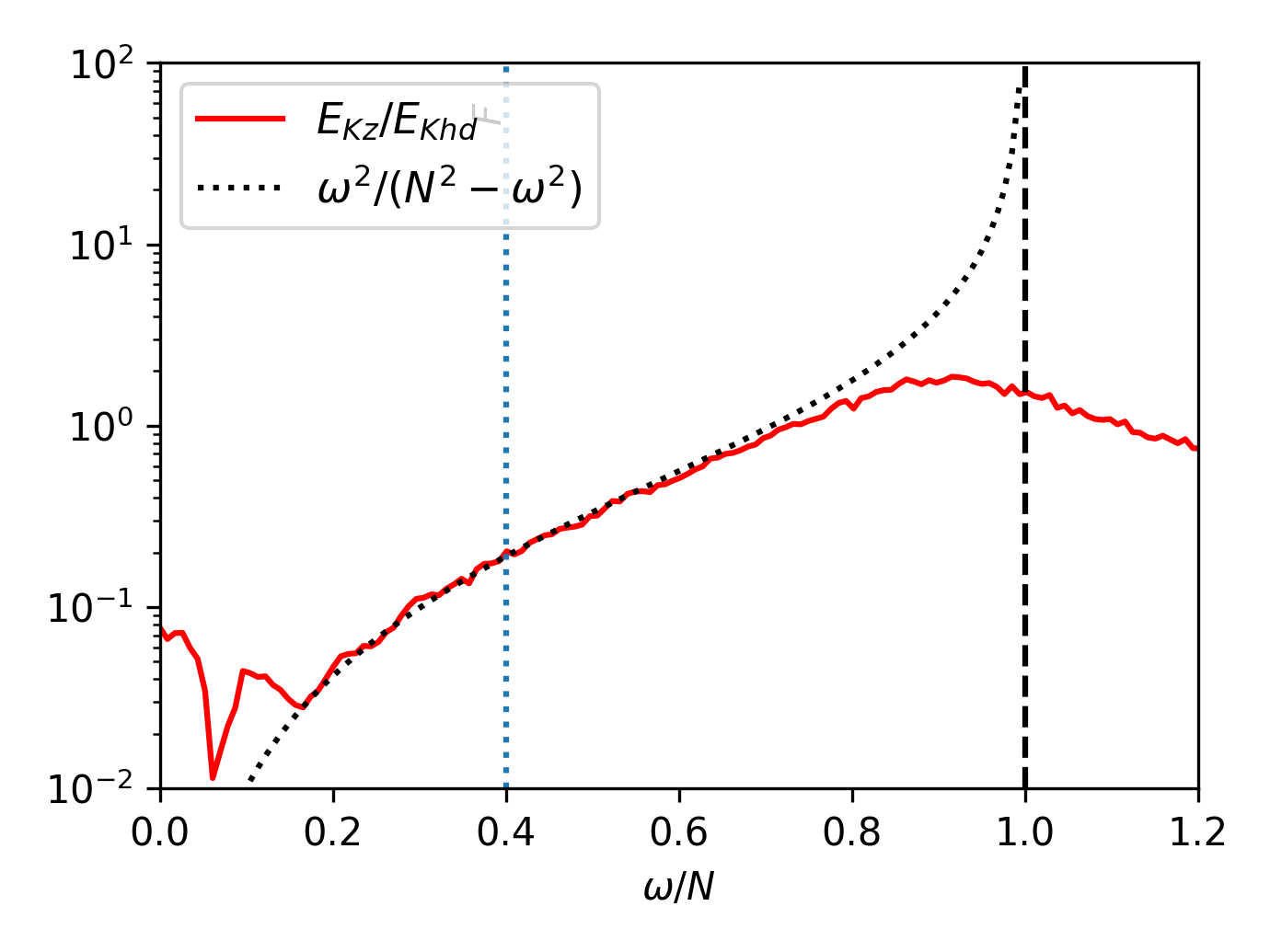}
}
\caption{Linear waves relations in $\omega$-space. Same as
figure~\ref{fig:linear-waves} but for $F = 0.40$ and $a=10$\,cm, at resolution
$480\times480\times80$. \label{fig:linear-waves-slower}}
\end{figure}

Figure \ref{fig:linear-waves-slower} shows the study of linear waves relations
\eqref{ratio_AKz} and \eqref{ratio_Kzh}. We observe a good agreement with linear waves
dynamics in an extended frequency range, which is consistent with a less energetic flow
than for $F=0.73$ and $a=5$\,cm.


\begin{figure}
\centerline{
\includegraphics[width=0.48\textwidth]{%
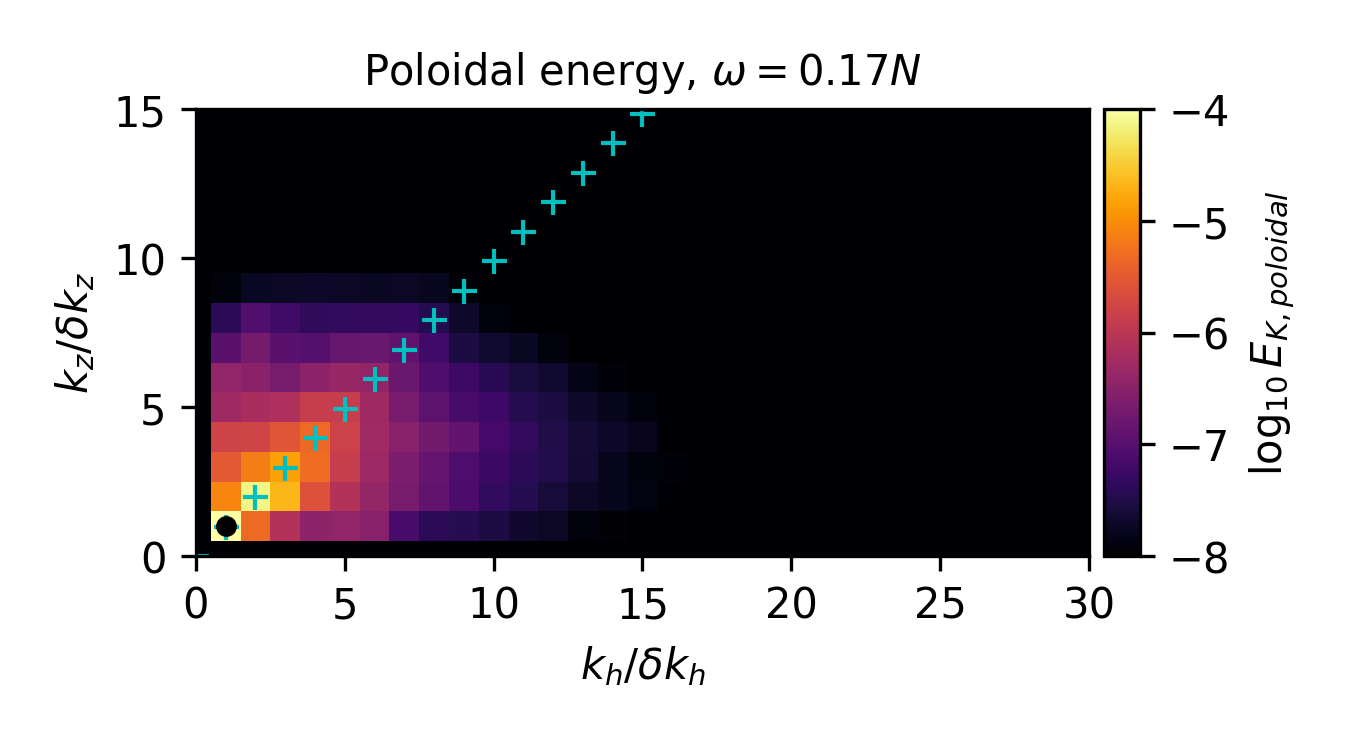}
\includegraphics[width=0.48\textwidth]{%
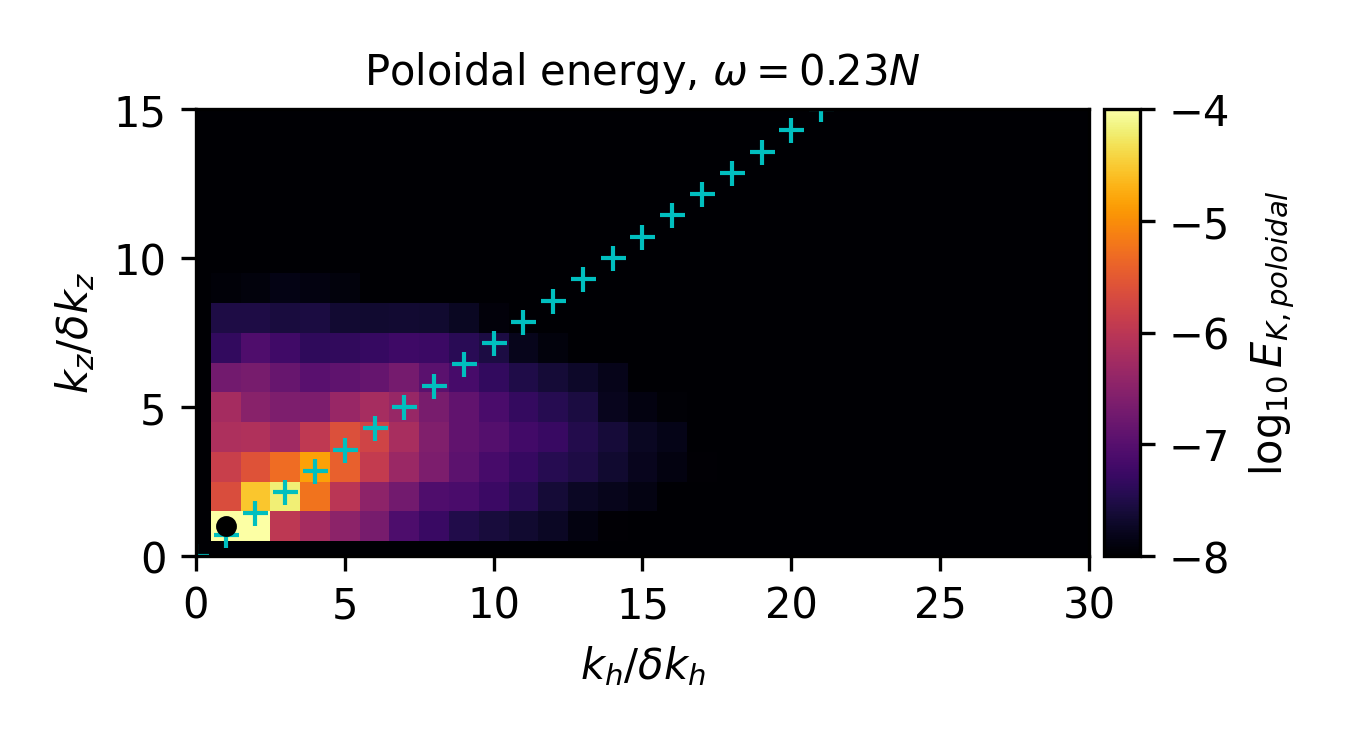}
}
\centerline{
\includegraphics[width=0.48\textwidth]{%
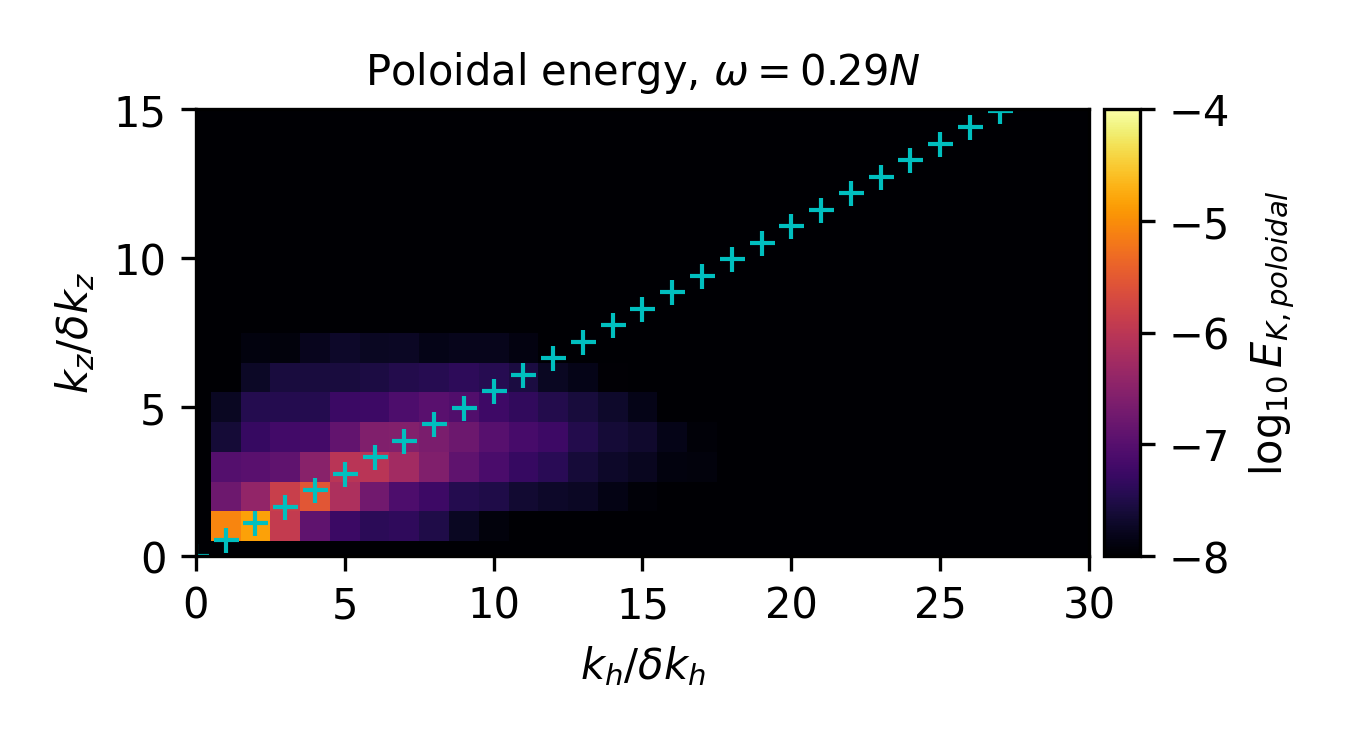}
\includegraphics[width=0.48\textwidth]{%
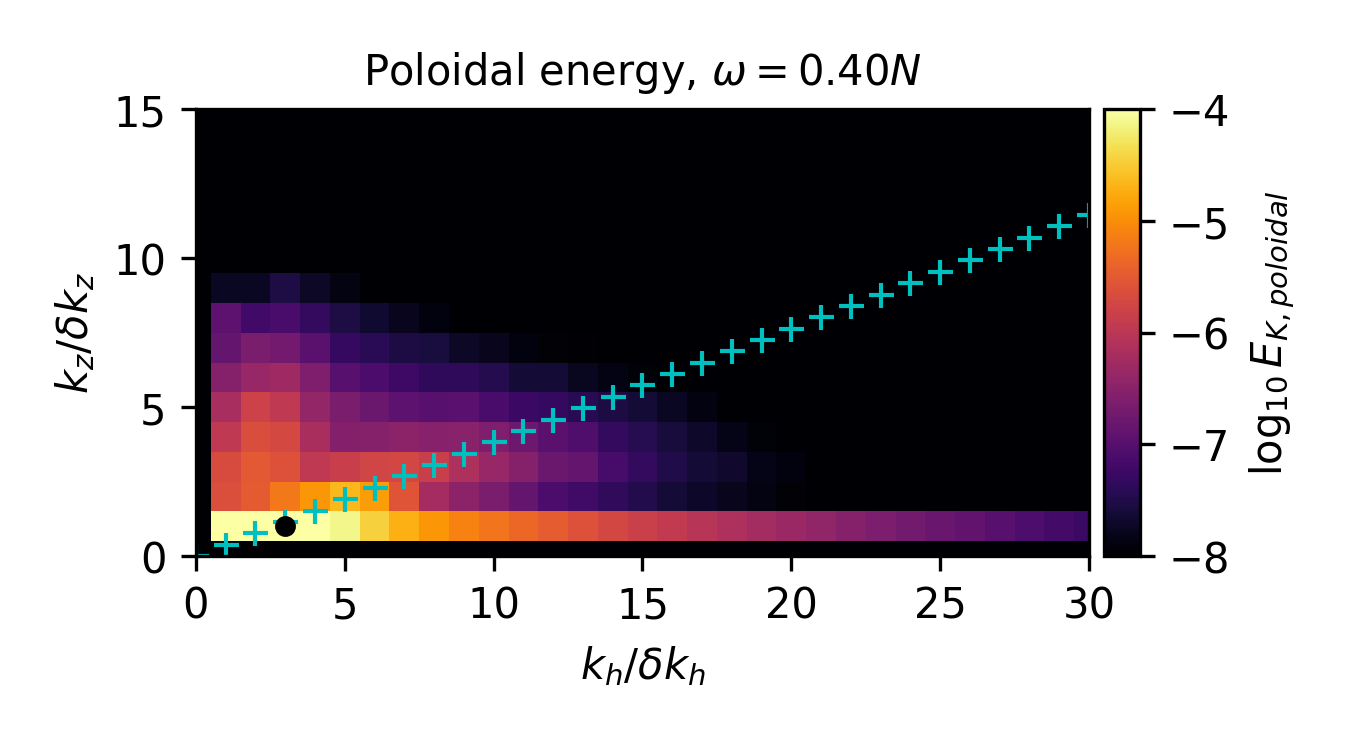}
}
\caption{Spatiotemporal spectra of poloidal kinetic energy in the $k_z-k_h$ plane at
fixed $\omega$. Same as figure~\ref{fig:spatiotemporal-spectra-omega} but for $F =
0.40$ and $a=10$\,cm. Each figure corresponds to an arrow in figure
\ref{fig:temporal-spectra}. \label{fig:spatiotemporal-spectra-omega-slower}}
\end{figure}

The results of spatiotemporal analysis are shown on figure
\ref{fig:spatiotemporal-spectra-omega-slower}, where the poloidal energy spectra are
shown in the $k_h-k_z$ plane for different frequencies. The three frequencies
corresponding to the triadic resonance relation are shown, as well as one off-peak
frequency at $\omega=0.29N$. For the three on-peak spectra, the maximum of energy is
shown by a black dot. In contrast to the case where $a=5$\,cm, the wavenumber values
corresponding to energy maxima cannot be interpreted as being involved in one triadic
resonance relation. To explain the energy repartition for these 3 frequencies, one
needs to consider more modes and few nearly resonant triads.


\begin{figure}
\centering
\includegraphics[width=0.48\textwidth]{%
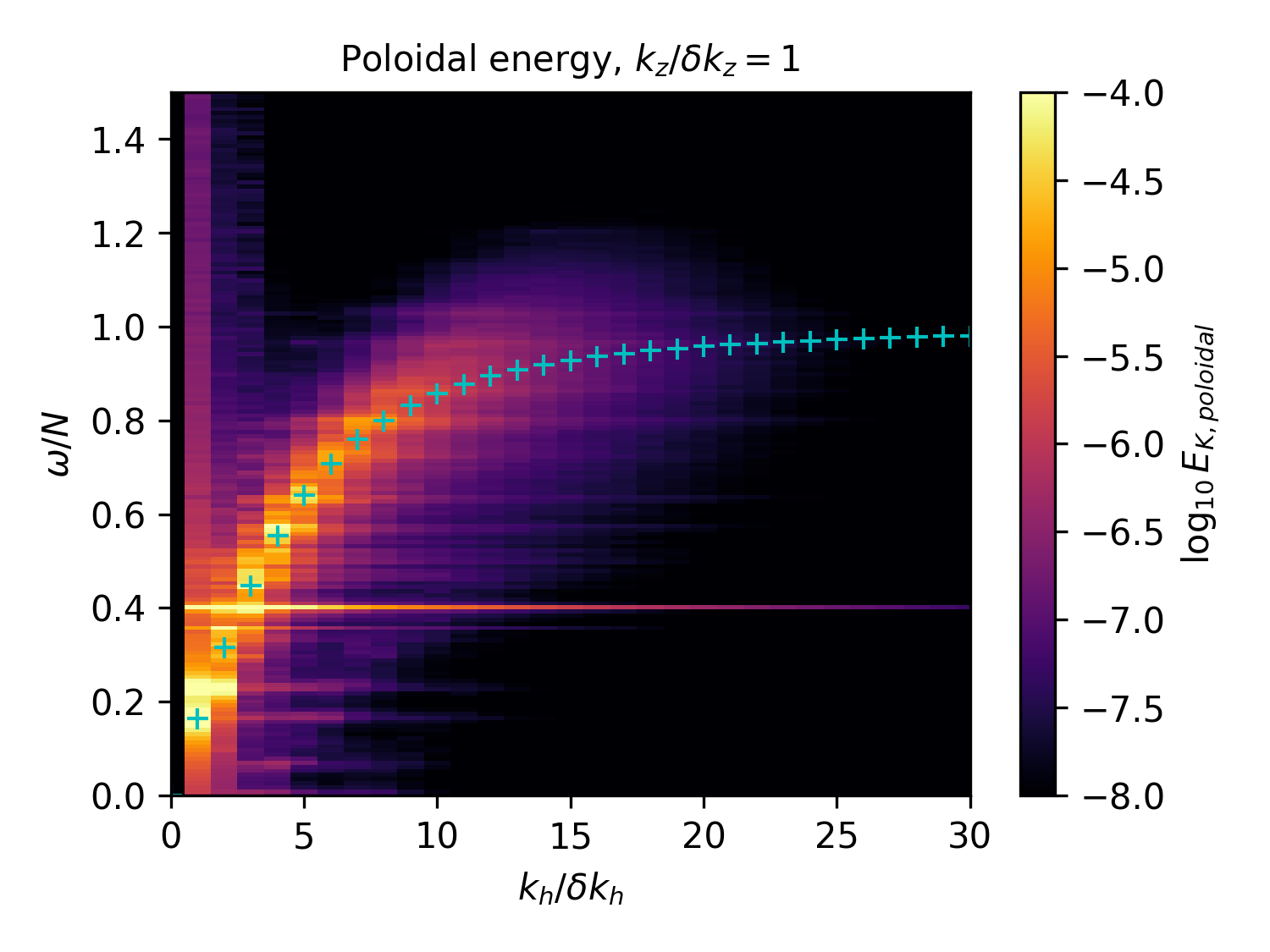}
\caption{Spatiotemporal spectrum of poloidal kinetic energy. Same as
figure~\ref{fig:spatiotemporal-spectra-ikz}, but for $F = 0.40$ and $a = 10$\,cm, at
resolution $480\times480\times80$. The spectrum is shown in the $\omega-k_h$ plane at
$k_z/\delta k_z=1$. \label{fig:spatiotemporal-spectra-slower}}
\end{figure}

Figure \ref{fig:spatiotemporal-spectra-slower} shows a cut of the spatiotemporal
poloidal energy spectrum in the $k_h-\omega$ plane, for the first value of $k_z$. We
see that the energy is concentrated around the dispersion relation, with a smaller
width than for $F=0.73$ and $a=5$\,cm. This closer repartition of energy with respect
to the dispersion relation is consistent with more linear waves dynamics in a regime
where the power injection into the flow is smaller. It is qualitatively consistent with
a regime of weak wave turbulence. Forcing with a more adapted scheme may allow one to
observe a GM spectrum at lower frequency.

\section{Conclusions and perspectives}

We performed numerical simulations of a stratified turbulent flow, using a forcing
mechanism inspired by experiments done in the Coriolis facility \cite{Savaro2020,
Rodda2022experimental}. The generated flows exhibited turbulent Froude numbers around
$10^{-3}$ and buoyancy Reynolds numbers ranging from $0.22$ to $6.3$. Using realistic
parameters, we were able to reproduce some key elements of the experimental flows.
Internal gravity waves are observed at all frequencies $\omega < N$ and mainly at large
scales, where they interact in a weak nonlinear fashion. The increase in power
injection is shown to reduce the range of observable linear waves, which is consistent
with experiments and the standard phenomenology of stratified turbulence. A spectral
analysis enables us to reproduce the spatiotemporal structure of the motion that was
observed in experiments, especially the peculiar shape of temporal spectra below the
forcing frequency, made of a flat continuum and several marked peaks. The discrete
frequencies of the peaks are shown to be involved in triadic resonance relations with
the frequencies of the forcing and a few large scale resonant modes.

A more detailed spatiotemporal analysis has evidenced new aspects of the studied flows.
Vortical energy is shown to condensate at the largest scales in both space and time,
resulting in a strong mean flow that takes the form of a large horizontal vortex the
size of the simulation domain. The relative part of kinetic energy that is taken up by
the vortex, as well as the frequency range at which it acts, are shown to increase
strongly with the power injection from the forcing. This vortical flow could be linked
to the nature of the specific forcing scheme that was used, which was shown to have
complex dynamics in both space and time, injecting and dissipating energy at both large
and intermediate scales.

Finally, using a lower forcing frequency, it was shown that weakly nonlinear waves are
also generated above the forcing frequency, with a structure that is not incompatible
with a $\omega^{-2}$ Garrett and Munk spectrum. However, we also found that some other
key elements from the standard phenomenology of inertial range turbulent cascades, such
as constant energy fluxes through scales or power law spectra behaviour, were absent
from the studied flow. This is coherent with both the presence of waves mainly at large
scales and with the anisotropy of spatial energy spectra. All those observations are
typical of viscosity-affected stratified flows, which are characterized by small
buoyancy Reynolds numbers. Indeed, the flows that we studied here, which were
parametrized to mimic experimental flows, exhibit buoyancy Reynolds numbers of the
order of unity or less, suggesting regimes at the transition between viscosity-affected
stratified flows and strongly stratified turbulence. Regimes at higher buoyancy
Reynolds numbers were explored in 2D in other numerical works \cite{Linares2020}, and
showed the presence of an actual inertial turbulent range.

Further studies should be aimed in the future at a better understanding of the peculiar
forcing that was used in this article and in \cite{Savaro2020, Rodda2022experimental},
as well as its relation to the growth of the large scale vortical flow.

\begin{acknowledgments}

This research was funded, in whole or in part, by the European Research Council (ERC)
under the European Union's Horizon 2020 research and innovation program (Grant No.
647018-WATU). A CC-BY public copyright license has been applied by the authors to the
present document and will be applied to all subsequent versions up to the Author
Accepted Manuscript arising from this submission, in accordance with the grant's open
access conditions. It was also partially supported by the Simons Foundation through the
Simons collaboration on wave turbulence. Part of this work was performed using
resources provided by \href{https://www.cines.fr/}{CINES} under GENCI allocation number
A0080107567.

\end{acknowledgments}


\bibliography{refs}
\end{document}